\documentclass[useAMS,usenatbib]{mnras}
\usepackage[pdftex]{graphicx} 
\usepackage{hyperref}
\usepackage{xcolor} 
\usepackage{lscape}
\usepackage{calc}
\usepackage{amsmath}
\usepackage{afterpage}
\usepackage{bm}
\usepackage{amssymb, physics}
\defcitealias{Jespersen2025a}{J25}

\hypersetup{colorlinks,linkcolor={red!50!black}, citecolor={blue!50!black},urlcolor={blue!80!black} } 

\title[Clustering of high-$z$ galaxies]
{Clustering-based halo mass assignment for high-redshift galaxies: method, validation, and application to JWST}

\author[Lim et al.]
{Seunghwan Lim$^{1,2}$\thanks{E-mail: sl2207@cam.ac.uk},
Sandro Tacchella$^{1,2}$,
Sara Saleh$^{1,2}$,
Roberto Maiolino$^{1,2}$,
Lily Whitler$^{1,2}$,
\newauthor
Joop Schaye$^{3}$,
Evgenii Chaikin$^{4,3}$,
Carlos S. Frenk$^{4}$,
Filip Hu\v{s}ko$^{3}$,
Alexander J. Richings$^{5,6}$,
\newauthor
Brant Robertson$^{7}$
\\
\vspace*{6pt} \\
$^{1}$Kavli Institute for Cosmology, University of Cambridge, Madingley
Road, Cambridge, CB3 0HA, UK \\
$^{2}$Cavendish Laboratory, University of Cambridge, 19 JJ Thomson
Avenue, Cambridge, CB3 0HE, UK \\
$^{3}$Leiden Observatory, Leiden University, PO Box 9513, 2300 RA Leiden, the Netherlands \\
$^{4}$Institute for Computational Cosmology, Department of Physics, University of Durham, South Road, Durham, DH1 3LE, UK \\
$^{5}$Centre for Data Science, Artificial Intelligence and Modelling, University of Hull, Cottingham Road, Hull, HU6 7RX, UK \\ 
$^{6}$E. A. Milne Centre for Astrophysics, University of Hull, Cottingham Road, Hull, HU6 7RX, UK \\
$^{7}$Department of Astronomy and Astrophysics, University of California, Santa Cruz, 1156 High Street, Santa Cruz, CA 95064, USA
\
}

\begin{document} 

\pagerange{\pageref{firstpage}--\pageref{lastpage}}

\date{\today}
\pubyear{2026}

\maketitle

\label{firstpage}

\begin{abstract}
We present a clustering-based method for inferring halo masses of high-redshift galaxies, validated using the COLIBRE simulations combined with the HOMA empirical model, which assumes star formation to be proportional to halo accretion rate and is calibrated to JWST observations. Our approach matches the two-point correlation function of galaxies in stellar mass bins to reference halo clustering, establishing the stellar-to-halo mass relation over an optimal radial range $0.5<r_{\rm p}/\mathrm{cMpc}<1.0$. Validation against true halo masses shows minimal bias, with scatter below $\sim0.3$\,dex for volumes down to $(50$\,cMpc$)^3$ and redshifts to $z=12$, even for photometric data. Cross-validation using the native COLIBRE galaxy population shows that the method is robust to the different galaxy–halo prescriptions. Survey volume dominates the error budget: field-to-field variations in the clustering amplitude vary by factors of $\sim3$ ($\sim2$) for photometric (spectroscopic) samples, translating to 0.3--0.5\,dex uncertainty in halo masses for $\log M_{\rm h}/{\rm M}_\odot \sim 10$--$12$. Splitting the sample by properties such as SFR, colour, and age before applying our clustering matching mitigates assembly bias, offering a key advantage over abundance matching. Application to JWST (JADES) samples at $z\simeq6$ and $10$ yields halo masses of $\log M_{\rm h}/{\rm M}_\odot\simeq 10.52_{-0.21}^{+0.12}$ and $9.91_{-0.34}^{+0.19}$ for $M_{\rm UV}<-17$ galaxies, with linear biases of $b_{\rm h}=4.2_{-0.41}^{+0.26}$ and $7.7_{-1.16}^{+0.77}$, respectively. Current data cannot distinguish between star formation models. Our \textit{Roman} Deep Tier forecasts indicate that at $z\sim10$, the inferred halo masses for $M_{\rm UV}\lesssim-21$ galaxies differ by $\sim0.5$ dex between bursty and non-bursty models, but the expected number of pairs limits the constraining power. Our framework provides robust, empirically-calibrated halo masses essential for interpreting JWST observations and constraining galaxy formation during reionization.
\end{abstract}


\section[intro]{Introduction}
\label{sec_intro}

Establishing the connection between galaxies and their host dark matter haloes is a cornerstone of modern galaxy formation theory \citep[e.g.,][]{WhiteRees1978, Davis1985, MovdBWhite2010, WechslerTinker2018}. This galaxy-halo connection dictates how galaxies form, accrete gas, and evolve within the gravitational scaffolding of the Universe \citep[e.g.,][]{WhiteFrenk1991, ConroyWechsler2009}. At high redshifts ($z\gtrsim5$), during the crucial epoch of reionization and cosmic dawn, this link is particularly critical. It encodes the efficiency of early star formation, the mode of gas accretion, and the impact of the first feedback mechanisms in pristine environments \citep[e.g.,][]{Tacchella2018, Behroozi2019}. 

Furthermore, understanding how galaxies trace the underlying dark matter distribution at these early times may provide the missing puzzle piece to interpret a growing number of challenges revealed by the James Webb Space Telescope (JWST), including a high abundance of ultraviolet (UV)-bright or massive galaxies \citep[e.g.,][]{Donnan2023, Harikane2023a, Finkelstein2024, Robertson2024, Weibel2024, Shuntov2025, Harvey2025, Whitler2025}, the presence of massive, quiescent systems \citep[e.g.,][]{Carnall2023, Carnall2024, Glazebrook2024, Nanayakkara2024, Baker2025b, Turner2025}, and an unexpectedly high number of protoclusters \citep[e.g.,][]{Helton2024b, Lim2024, WuZ2026} and active galactic nuclei (AGN) at $z>5$ \citep[e.g.,][]{Harikane2023b, Maiolino2024b, Kocevski2025}. Disentangling whether these observations signal a need for new physics in galaxy formation models, or are simply a consequence of how galaxies populate their haloes, requires precise and empirically-calibrated methods for assigning halo masses to observed high-redshift galaxies \citep[e.g.,][]{Boylan-Kolchin2023}.

Significant effort has been invested in recent years to map the galaxy-halo connection at $z>4$. One-point statistics, such as the UV luminosity function (UVLF) and stellar mass function (SMF), have been used extensively to infer the relationship between galaxy light and halo mass via techniques like abundance matching \citep[e.g.,][]{Shuntov2025}. However, a fundamental degeneracy limits these one-point statistics: among the proposed explanations for the high abundance of bright galaxies are (i) a high, constant star-formation efficiency (SFE) in massive haloes, or (ii) significant scatter in the UV luminosities of galaxies, which allows lower-mass haloes to host UV-bright objects for short periods \citep[e.g.,][]{Munoz2023, Shen2023, Sun2023, Gelli2024, KravtsovBelokurov2024}. These two scenarios (a high mean efficiency versus a broad distribution of UV luminosities at fixed halo mass) have different implications for the physics of feedback and star formation, yet they cannot be distinguished by abundance alone \citep[e.g.,][]{Munoz2023, Sun2025}.

A third, physically motivated explanation was proposed prior to the launch of JWST by \citet{Cowley2018} using the GALFORM semi-analytic model. In their model, star formation in starburst episodes is assumed to follow a top-heavy initial mass function (IMF), a modification originally introduced to reproduce the number counts and redshift distribution of submillimetre galaxies \citep{Baugh2005}. This model successfully predicted the UVLFs at $z\gtrsim10$ that were subsequently measured by JWST \citep{LuS2025}. While a detailed comparison with this specific model is beyond the scope of this work, the Cowley et al. model serves as an important reminder that different assumptions about the IMF can introduce significant systematic effects in estimates of halo mass. Our focus here is on the two scenarios that are directly implemented in the HOMA empirical model, namely variations in the mean star formation efficiency and the degree of scatter in the UV luminosity--halo mass relation.

Galaxy clustering, a two-point statistic, offers a powerful avenue to break this degeneracy \citep[e.g.,][]{Mirocha2020, Munoz2026}. The clustering strength of a galaxy population, also quantified by its bias relative to the dark matter, is a direct probe of its host halo mass distribution \citep[e.g.,][]{MoWhite1996, Tinker2010}. A model where bright galaxies reside in the most massive, rare haloes will predict a much stronger clustering signal (higher bias) than a model where they are merely the up-scattered tail of a lower-mass population \citep{Munoz2023, Sun2025}. Recent observational campaigns have begun to leverage this. Using a large sample of $\sim4$ million dropouts from the Subaru Hyper Suprime-Cam, \citet{Harikane2022} combined UVLF and clustering measurements to trace the SFE from $z\sim2-7$. Their results supported an almost constant SFE at fixed halo mass over this period, with a gradual increase towards lower redshifts. JWST has now pushed these clustering measurements to even higher redshifts, with studies in the COSMOS-Web and JADES fields providing the first clustering-based constraints on the halo masses of galaxies at $z>10$ \citep{Dalmasso2024, Paquereau2025, Dalmasso2026}. For instance, \citet{Paquereau2025} used halo occupation distribution (HOD) modelling of angular clustering in COSMOS-Web to find that galaxies at $z>10.5$ of $M_\ast\sim 10^{9}\,{\rm M}_\odot$ reside in haloes of $M_{\rm h}\sim10^{10.5}\,\mathrm{M}_\odot$, implying an SFE a full dex higher than at $z\sim1$. Similarly, \citet{Dalmasso2026} presented clustering analysis of JADES galaxies, reporting a significant decline in the average halo mass from $z\sim5.5$ to $z\sim10.6$ for rest-frame UV-selected samples. 

While HOD modelling of clustering is a powerful tool, it often relies on parametric models for the satellite and central galaxy populations, and can be sensitive to assumptions about the halo bias on non-linear scales \citep{Jose2013, Jose2016, Jose2017, Paquereau2025}. In this paper, we introduce an empirically-calibrated approach. Our method matches the measured projected correlation function of a galaxy sample to a library of templates constructed directly from haloes of known mass, thereby establishing a direct stellar mass–halo mass relation (SHMR). A key advantage of this approach is its flexibility. By matching the clustering signal, which integrates information from the entire galaxy population, we can account for the dependence of halo mass on properties beyond stellar mass, such as star formation history or assembly bias, and explicitly mitigate their effects. This is a significant improvement over methods like abundance matching, which assume a monotonic relation between galaxy light (or stellar mass) and halo mass and cannot easily incorporate secondary properties to correct for environmental dependencies. The ability to capture secondary dependencies is crucial for subsequent analyses of environmental effects and correlated galaxy properties within the assigned haloes. 

Our method is built upon a large, $(400\,\mathrm{cMpc})^{3}$ volume from the state-of-the-art COLIBRE hydrodynamical simulation \citep{Schaye2026, Chaikin2026}, populated with realistic galaxy properties using the empirical model HOMA calibrated to reproduce high-redshift UVLFs (Saleh et al. in prep.). This allows us to rigorously quantify the systematic biases and uncertainties introduced by survey volume, redshift errors, and selection effects. Importantly, the relatively small survey volumes probed by JWST have been suggested to be highly susceptible to cosmic variance \citep[e.g.,][]{Lovell2023, Jespersen2025a, Jespersen2025b, Lim2025b, Huang2026}. As we demonstrate in this work, the impact of cosmic variance on high-redshift clustering measurements for typical volumes of JWST surveys is indeed significant (see also \citealt{NewmanDavis2002}, \citealt{Somerville2004}, \citealt{TrentiStiavelli2008}, and \citealt{Robertson2010}). We find that field-to-field variations in the projected correlation function can propagate into systematic uncertainties of up to $\sim$0.5\,dex in the inferred halo masses, making cosmic variance a dominant source of uncertainty when using galaxy clustering to establish the galaxy-halo connection at $z\gtrsim5$. 

The structure of this paper is as follows. In Section \ref{sec_method}, we describe the COLIBRE simulation, the HOMA empirical model for generating mock galaxy catalogues, and the methodology for measuring clustering and inferring halo masses. We present a detailed validation of our technique in Section \ref{sec_result}, exploring its dependence on redshift, survey volume, luminosity cut, and redshift uncertainty. In Section \ref{sec_discussion}, we discuss the profound impact of cosmic variance on our inferences and apply our method to recent JWST observations from the JADES survey, comparing our derived halo masses and biases to independent measurements. Finally, we summarize our findings and their implications for high-redshift galaxy formation in Section \ref{sec_summary}. Throughout this work, we assume a \citet{Chabrier2003} IMF and a flat $\Lambda$CDM cosmology consistent with the Dark Energy Survey year 3 results \citep{Abbott2022}, with parameters $\Omega_{\mathrm{m}}=0.306$, $\Omega_{\mathrm{b}}=0.0486$, $h=0.681$, and $\sigma_8=0.807$.

\section[methods]{methods}
\label{sec_method}

\subsection{The COLIBRE simulations}
\label{ssec_COLIBRE}

We utilize the COLIBRE suite of cosmological hydrodynamical simulations \citep{Schaye2026, Chaikin2026}, a state-of-the-art project designed to model galaxy formation and evolution with unprecedented physical fidelity in large, statistically representative cosmological volumes. COLIBRE represents a significant advancement over previous generations of simulations by directly modelling the multi-phase interstellar medium (ISM) without imposing an artificial pressure floor, and by including a sophisticated on-the-fly model for dust grain evolution. 

The simulations were performed with the \textsc{swift} code \citep{Schaller2024} using the \textsc{sphenix} formulation of smoothed particle hydrodynamics \citep{Borrow2022}. A defining characteristic is the use of four times more dark matter (DM) particles than baryonic particles in the initial conditions, which helps suppress spurious two-body heating and improves the effective resolution for the baryonic component \citep{Ludlow2019, Ludlow2021}. Radiative cooling and heating are tracked element-by-element down to $\simeq 10\,\mathrm{K}$ using a non-equilibrium chemistry network for hydrogen and helium species, with metal cooling tabulated in chemical equilibrium \citep{Ploeckinger2025, Richings2014a, Richings2014b}. The cooling rates are coupled to the dust grain abundances tracked explicitly in the simulation \citep{Trayford2026}. Star formation is implemented stochastically for gravitationally unstable gas, following a Schmidt law with an efficiency of 1\% per free-fall time \citep{Nobels2024}. Stellar mass loss and chemical enrichment from asymptotic giant branch stars, core-collapse supernovae (CCSNe), and type Ia supernovae are tracked using up-to-date nucleosynthetic yields \citep{Correa2026}. Pre-CCSN feedback from stellar winds, radiation pressure, and HII regions is included \citep{Benitez-Llambay2026}, while CCSN feedback is implemented as a combination of stochastic thermal heating \citep{DallaVecchiaSchaye2012} and low-velocity kinetic kicks \citep{Chaikin2023}. Supermassive black holes are seeded in massive haloes and grow via gas accretion and mergers, with AGN feedback implemented thermally in the fiducial model \citep{Bahe2022, BoothSchaye2009}. The subgrid parameters were calibrated to reproduce the observed $z\simeq0$ galaxy SMF, galaxy size-mass relation, and black hole masses in massive galaxies \citep{Chaikin2026}.

For our study, we use the L400m7 COLIBRE simulation, its largest run, with a box size of $400\,\mathrm{cMpc}$ on a side. It provides a statistically representative sample of galaxies while resolving those with halo masses $M_{\rm h}\gtrsim10^{9}\,\mathrm{M}_\odot$, making it well suited for investigating halo properties over a wide mass range. The large simulation volume allows us to examine how the uncertainty in clustering measurements and the reliability of the methods depend on the survey volume. As we will demonstrate later (Sect.~\ref{ssec_CV}), the volume-to-volume variance in clustering strength is significant (up to a factor of two) even for a relatively large volume of $(200\,\mathrm{cMpc})^{3}$ at high $z$, underscoring the necessity of testing the method on a large simulation box. The mass resolution, which matches the faint end of high-$z$ observations, together with the large volume, enables a sample construction that is directly comparable to observations and allows for reliable clustering measurements to disentangle the impact of uncertainties from potential biases.

The L400m7 simulation contains $5\times3008^{3}$ particles in a $(400\,\mathrm{cMpc})^{3}$ box. It has an initial mean baryonic particle mass of $m_{\mathrm{g}}= 1.47\times10^{7}\,\mathrm{M}_\odot$, with DM particles having almost the same mass, $m_{\mathrm{CDM}}=1.94\times10^{7}\,\mathrm{M}_\odot$. The gravitational softening length is $3.6\,\mathrm{ckpc}$, reaching a maximum physical softening of $1.4\,\mathrm{pkpc}$ at $z<1.57$. The simulation assumes the Dark Energy Survey year 3 cosmology \citep{Abbott2022} ($\Omega_{\mathrm{m}}=0.306$, $\Omega_{\mathrm{b}}=0.0486$, $h=0.681$, $\sigma_8=0.807$) and a \citet{Chabrier2003} IMF. The initial conditions were generated at $z=63$ using second-order Lagrangian perturbation theory with the \textsc{monofonic} code \citep{Hahn2020, Michaux2021}, with large-scale modes fixed to their mean variance following \citet{AnguloPontzen2016}. 

Data products include 128 snapshots between $z=30$ and $0$. Haloes and subhaloes are identified using a friends-of-friends (FoF) algorithm and the hierarchical \textsc{hbt-herons} \citep{ForouharMoreno2025} finders, respectively. Galaxy and halo properties are then computed for a wide range of apertures using the Spherical Overdensity and Aperture Processor (SOAP) \citep{McGibbon2025}. For our main analysis, we only use haloes identified by the algorithms, and subhaloes are excluded. Only haloes (subhaloes) with more than 32 (20) particles are identified by the algorithms, and thus used for our analysis. As detailed in Sect.~\ref{ssec_SHMR}, our sample construction for the main analysis is restricted to haloes with \(M_{\rm h}\gtrsim 10^{10}\,\mathrm{M}_\odot\), which are therefore well resolved in this simulation. In some of our analyses, particularly when exploring the faint end for future survey projections or comparing different star formation models, we extend this range down to \(M_{\rm h}\sim10^9\,\mathrm{M}_\odot\). 

\subsection{The empirical model}
\label{ssec_model}

To populate COLIBRE haloes with realistic galaxy properties, we employ the HOMA empirical model (Saleh et al. in prep.), rather than using the galaxy properties directly predicted by COLIBRE. Using an empirical model offers the flexibility to test several variants of different physical scenarios at low computational cost. Importantly, the model provides sufficient flexibility to tune the burstiness of star formation, as the degree of stochasticity remains a key question in explaining the high-$z$ data. This hybrid approach allows us to populate a large cosmological volume with galaxies that have physically motivated spectral energy distributions (SEDs), which is essential for studying galaxy clustering and its dependence on various selection criteria, such as rest-frame UV luminosity.

While we adopt all observables from HOMA and only take halo masses and positions (i.e. the centre of mass of all bound particles) from COLIBRE, we nevertheless choose to use the COLIBRE hydrodynamical simulation rather than its dark matter-only (DMO) counterpart to populate the simulated haloes with the model. This is because baryonic processes are known to affect both halo masses \citep[e.g.,][]{Schaye2023} and the clustering of satellite galaxies \citep[e.g.,][]{vanDaalen2014}. We tested this using the COLIBRE DMO simulation, which has the same volume and resolution, and found that the two-point correlation function of low-mass haloes, particularly on scales $\lesssim 1\,\mathrm{cMpc}$, exhibits larger uncertainties in the DMO simulation compared to the hydro simulation.

HOMA links the star-formation rate (SFR) of a galaxy to the baryonic accretion rate of its host dark-matter halo, 
\begin{equation} 
\mathrm{SFR} = \varepsilon(M_{\rm h})\, f_{\rm b}\, \dot{M}_{\rm h}\, f_{\rm g}(M_{\rm h},z), 
\label{eq:homa_sfr}
\end{equation} 
where $f_{\rm b}$ is the cosmic baryon fraction, $\dot{M}_{\rm h}$ is the halo mass accretion rate from \citet{Fakhouri2010}, and $\varepsilon(M_{\rm h})$ is a double power-law star-formation efficiency with parameters controlling its normalisation, turnover mass, and asymptotic slopes at low and high masses\footnote{We adopt the \citet{Fakhouri2010} accretion histories rather than those measured directly from COLIBRE because finite mass and time resolution can introduce substantial numerical stochasticity and spurious fluctuations in individual halo growth histories (see \citealt{Tacchella2018}, where we have used $N$-body accretion histories).}. The factor $f_{\rm g} \equiv g_{\rm max}/g_{\rm crit}$ follows the acceleration-based model of \citet{Boylan-Kolchin2025}, in which dense high-redshift haloes can reach gravitational accelerations comparable to, or exceeding, a critical value for efficient star formation. Here, $g_{\rm crit} \approx 5\times10^{-10}\,{\rm m\,s^{-2}}$ (or equivalently $\Sigma_{\rm crit}=g_{\rm crit}/(\pi G)\approx 1000\,{\rm M}_\odot\,{\rm pc^{-2}}$) is the threshold acceleration above which stellar feedback (e.g., momentum injection from massive stars) is unable to overcome self-gravity, allowing gas to collapse and form stars with high efficiency. Since $g_{\rm max}\propto M_{\rm h}^{1/3}(1+z)^2$ at fixed concentration, this enhancement becomes important for relatively massive haloes at early times, approximately $M_{\rm h}\gtrsim10^{10}\,{\rm M_\odot}$ at $z\gtrsim8$--10. We use this prescription as an effective parametrization of enhanced star-formation efficiency in rapidly growing, dense high-redshift haloes, rather than as a resolved model of the multiphase ISM or stellar feedback. Physically, it captures the idea that deep early potential wells and high gas surface densities can allow galaxies to approach feedback-limited or near-maximal star formation, as in analytic models of feedback-free or highly efficient starbursts \citep[e.g.][]{Dekel2023}.

The HOMA model is calibrated to reproduce the UVLF at $z\simeq6$ \citep[e.g.][]{Bouwens2021, Harikane2022}. Specifically, the normalization $\varepsilon_0$ of $\varepsilon(M_{\rm h})$ is adjusted while the shape parameters are held fixed. The model is not separately calibrated at higher redshifts. Its comparison to UVLF measurements at $z\simeq4$--12, as well as to SMFs, H$\alpha$ luminosity functions, and the cosmic SFR density, therefore provides a consistency check on whether the same physical prescription can simultaneously reproduce observables that trace star formation and stellar mass assembly on different timescales. Throughout this work, we use the rest-frame far-UV (FUV) luminosities at $1500\,\text{\AA}$ from the HOMA model, which trace star formation on timescales of $10$--$100\,\text{Myr}$. 

In addition to predicting SFRs and UV luminosities, HOMA also assigns stellar masses to each galaxy. The stellar mass is computed by integrating the star formation rate over the formation history of the galaxy and accounting for the mass returned to the interstellar medium by stellar evolution. Specifically, for each galaxy, HOMA uses the Flexible Stellar Population Synthesis (FSPS) code \citep{Conroy2009, Conroy2010} to convolve the bursty star formation history with a single stellar population (SSP) model, assuming a Chabrier IMF. This procedure ensures that the stellar masses are self-consistently linked to the same star formation and accretion histories that determine the UV luminosities.

In this work, we use several HOMA variants to explore how different assumptions about early star formation affect the mock galaxy population. In addition to the fiducial acceleration-enhanced model, we consider three variants. The first is a variant without the acceleration factor (i.e. $f_{\rm g}=1$ in Eq.~\ref{eq:homa_sfr}), which suppresses the enhanced efficiency in dense high-redshift haloes. The second is a variant with stronger stochastic star-formation variability, which introduces larger scatter in the SFR--halo mass relation to mimic burstier star formation. The third is a variant that assumes no stochasticity in the SFR--halo mass relation, effectively removing the scatter while maintaining the same mean relation as the fiducial model. The strength of stochasticity is parameterized by the amplitude parameter $S_0$ of the power spectral density of fluctuations (Saleh et al. in prep.), with $S_0 = 0$ for the no-stochasticity variant, $S_0 = 0.03$ for the fiducial model, and $S_0 = 0.06$ for the stronger stochasticity variant. All variants are calibrated consistently to the same $z\simeq6$ UVLF constraint, allowing us to explore degeneracies between enhanced mean efficiency, bursty star formation, and the role of scatter in the galaxy--halo connection.

For the presentation of our results, we adopt the dust-corrected rest-frame UV magnitudes from the HOMA model following Saleh et al. (in prep.), although our conclusions are insensitive to the dust correction. We use the HOMA model outputs that assume a Chabrier IMF, consistent with the COLIBRE simulation.

For the mock catalogue construction, we use the full HOMA output distributions. For each COLIBRE halo, we assign stellar masses, SFRs, and UV luminosities by drawing from the corresponding HOMA distribution in bins of halo mass and redshift (detailed in Sect.~\ref{ssec_mock}). This preserves the scatter and correlations predicted by the model, allowing us to test how secondary galaxy properties affect the clustering strength and inferred stellar-to-halo mass relation.

In some of our analyses, we use the galaxy and halo properties directly from the COLIBRE simulation, bypassing the HOMA model. This serves two purposes. First, it provides an independent cross-validation of our hybrid method by comparing results from two distinct baryonic models (Sect.~\ref{ssec_validation} and Appendix~\ref{sec_appA}). Second, it allows us to assess the impact of assembly bias and other environmental effects that are naturally encoded in the hydrodynamical simulation but may not be fully captured by the empirical HOMA model (Sect.~\ref{ssec_properties}). By comparing the clustering signals derived from the HOMA-populated mocks with those from the native COLIBRE galaxy catalogues, we can quantify the extent to which our method is sensitive to the underlying assumptions about the galaxy--halo connection.

\begin{figure*}
\includegraphics[width=0.95\linewidth]{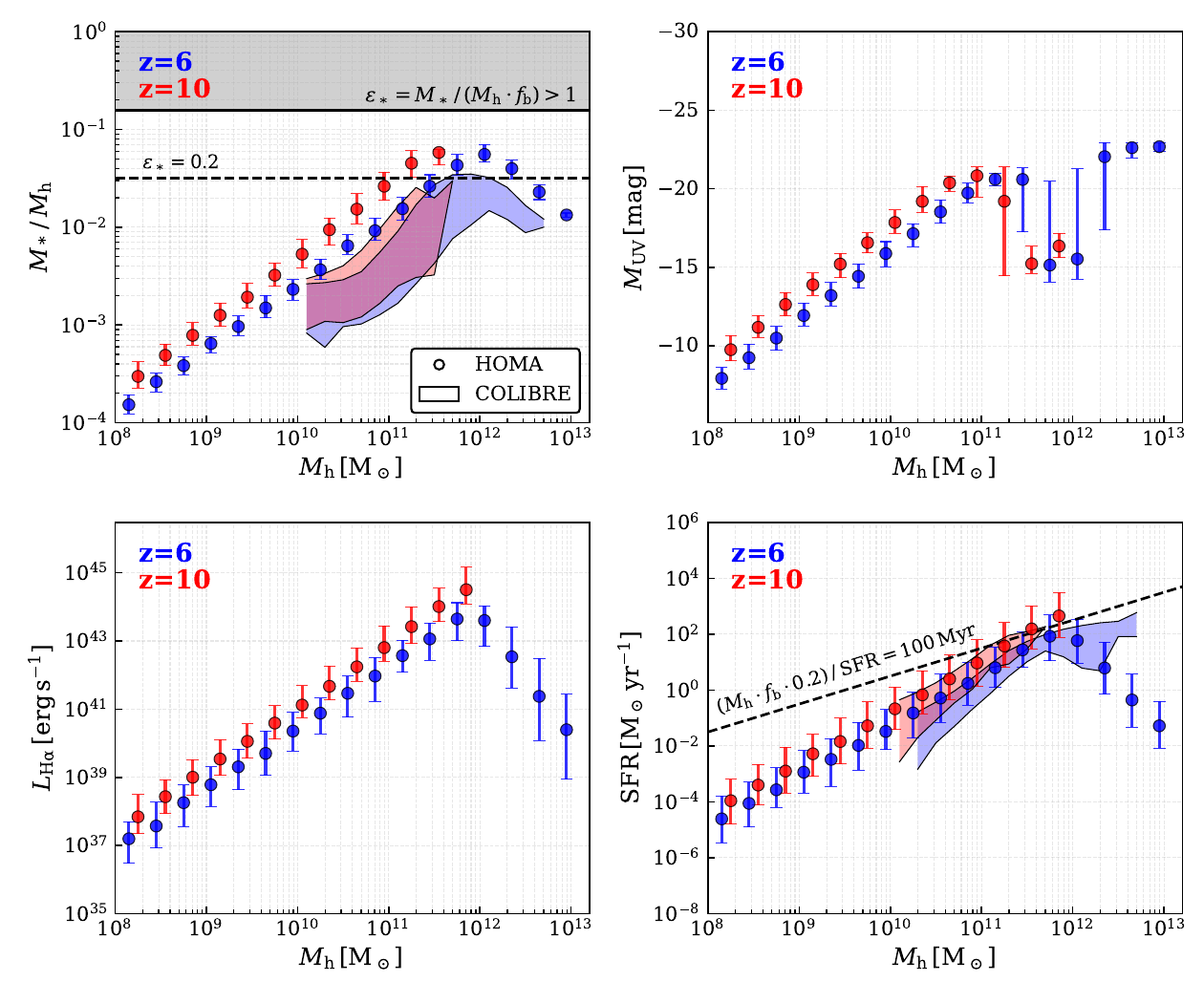}
\vspace{-0.5cm}
\caption{Predictions from the fiducial HOMA model (Saleh et al. in prep.; circles with errorbars indicating the 68 percentile ranges) and the COLIBRE simulation (L400m7; shaded bands indicating the 68 percentile ranges) used for the mock construction and analysis, shown as a function of halo mass at redshifts 6 (blue) and 10 (red). For redshifts 6 and 10, the HOMA model predictions are shown as blue and red circles, respectively. \textit{Upper left}: Stellar-to-halo mass relation. The solid (dashed) lines correspond to $\varepsilon_\ast = M_\ast/(M_{\rm h}\cdot f_{\rm b}) = 1$ ($0.2$). \textit{Upper right}: rest-frame UV magnitude--halo mass relation. \textit{Lower left}: H$\alpha$ luminosity as a function of halo mass. \textit{Lower right}: Star formation rate (SFR) averaged over 1\,Myr (instantaneous SFR for COLIBRE). The dashed line indicates a formation time (or mass-doubling time) of 100\,Myr assuming $\varepsilon_\ast = 0.2$.}
\label{fig_model}
\end{figure*}

\subsection{Construction of mock galaxies}
\label{ssec_mock}

We construct mock galaxy catalogues by combining halo populations extracted from the COLIBRE hydrodynamical simulation with galaxy properties predicted by the empirical model introduced in Sect.~\ref{ssec_model}. The core of the mock construction is the mass-based assignment of model-predicted properties to the COLIBRE haloes. This is accomplished by defining halo mass bins using the unique values from the HOMA model. For each mass bin, we identify all COLIBRE haloes with masses (based on $M_{\rm 200c}$, the total mass within $R_{\rm 200c}$ where the average density enclosed is 200 times the critical density) falling within that bin. For each selected halo in a given mass bin, we randomly select one of the multiple model realizations of HOMA available for that mass bin and assign its predicted properties to the COLIBRE halo. haloes outside the mass range covered by the model are excluded from subsequent analyses where cuts on these properties are applied. This random assignment within each mass bin ensures that the scatter in the mass–property relations is preserved, while maintaining the correct mean trends. The resulting mock galaxy catalogues contain, for each galaxy, the assigned FUV luminosities (from which we compute the rest-frame UV absolute magnitudes $M_{\rm UV}$), H$\alpha$ luminosities, stellar masses, and star formation rates on different timescales. 

We also explored the inclusion of satellites (subhaloes from COLIBRE) in our mock samples, using their (bound) subhalo mass in lieu of \(M_{\rm 200c}\), which is only defined for host haloes. Satellites are known to lose mass dynamically after accretion, and indeed we find that their subhalo masses are typically about 20 per cent lower than those of centrals at fixed stellar mass. To account for this offset, we tested two approaches: (i) scaling their subhalo masses by a factor of 1.2, and (ii) directly assigning them the halo mass that would match centrals of the same stellar mass. In both cases, we populated the subhaloes using the same HOMA model applied to centrals. As we demonstrate in Appendix~\ref{sec_appB}, neither approach yielded a discernible impact on our results (see Fig.~\ref{fig_appB}), with the distributions of residuals $\Delta\log M_{\rm h}$ showing nearly identical standard deviations and negligible median offsets across all three cases. For the samples and scales considered here, this insensitivity arises primarily because satellites constitute only $\sim10\%$ of the sample and our fiducial fitting range ($0.5<r_{\rm p}/\mathrm{cMpc}<1.0$) lies beyond the one-halo regime, where their contribution to the clustering signal is subdominant (see Appendix~\ref{sec_appC}). This conclusion should not be extrapolated to samples with substantially larger satellite fractions or to small-scale clustering analyses, where a more detailed treatment of the central--satellite distinction would be necessary. On these grounds, we opted to exclude satellites from our mock sample construction for the main analysis. 

The original relations from the fiducial HOMA model used for the mock construction are shown in Fig.~\ref{fig_model}. We also overplot the direct COLIBRE predictions for the SHMR and instantaneous SFR. Overall, the HOMA and COLIBRE predictions show good qualitative agreement for the SHMR, with both models exhibiting similar trends. However, we note that at fixed halo mass, the COLIBRE stellar masses are systematically lower than those predicted by HOMA by $\sim 0.5$--$1$\,dex across much of the mass range. This discrepancy reflects the different physical assumptions underlying the two models. COLIBRE self-consistently simulates the complex interplay of gas cooling, star formation, and feedback processes, while HOMA is an empirically calibrated model designed to reproduce the observed UVLF at $z\simeq6$. The difference in stellar mass at fixed halo mass therefore highlights the flexibility of the HOMA model to match observables that may require different star formation efficiencies than those emerging from the hydrodynamical simulation. Importantly, our clustering-based method for inferring the SHMR operates in a self-consistent manner and does not rely on the absolute normalisation of either model, as we will demonstrate in Sects.~\ref{ssec_SHMR} and~\ref{ssec_validation}.

We note that while COLIBRE self-consistently simulates the physics of galaxy formation, it does not match the observed UVLFs at $z\gtrsim10$ when its predicted UV luminosities are compared to JWST observations \citep{LuS2026}. This discrepancy may be related to the assumed IMF, as discussed in \citet{LuS2026}.

\subsection{Measuring clustering via the two-point correlation function}
\label{ssec_2pt}

To characterize the clustering of galaxies, we employ the projected two-point cross-correlation function $w_{\rm p}(r_{\rm p})$, which measures the excess probability of finding a pair of galaxies at a given projected separation $r_{\rm p}$ compared to a random distribution. The projected correlation function is preferred over the full 3D correlation function because it is less sensitive to redshift-space distortions arising from peculiar velocities. It also accounts for redshift errors characteristic of a survey catalog. 

The projected correlation function is derived from the 3D correlation function $\xi(r_{\rm p}, \pi)$, where $r_{\rm p}$ is the projected separation perpendicular to the line of sight and $\pi$ is the separation along the line of sight. The projection is computed by integrating along the line-of-sight (LOS) direction:
\begin{equation}
w_{\rm p}(r_{\rm p}) = \int_{-\pi_{\rm max}}^{\pi_{\rm max}} \xi(r_{\rm p}, \pi) \, d\pi,
\label{eq:wp_integral}
\end{equation}
where $\pi_{\rm max}$ is the maximum LOS separation over which the integration is performed. In practice, $\pi_{\rm max}$ must be chosen to be sufficiently large to include most correlated pairs while avoiding excessive noise from uncorrelated pairs at large separations (see below and Sect.~\ref{ssec_SHMR} for our choice of $\pi_{\rm max}$). 

Throughout this work, we present clustering measurements as the line-of-sight averaged quantity \(w(r_{\rm p}) \equiv w_{\rm p}(r_{\rm p})/(2\pi_{\rm max})\), which is dimensionless and facilitates comparison with observational angular correlation functions \(\omega(\theta)\). In simulations, we compute \(w(r_{\rm p})\) directly from 3D galaxy positions using the \citet{LandySzalay1993} estimator, integrating over the comoving depth corresponding to the assumed redshift uncertainty. For observational data, we convert \(\omega(\theta)\) to \(w(r_{\rm p})\) via the Limber approximation \citep{Limber1953}, assuming the same redshift distribution as used in the mock samples.

To estimate $\xi(r_{\rm p}, \pi)$, we use the \citet{LandySzalay1993} estimator, which minimizes variance and edge effects:
\begin{equation}
\xi(r_{\rm p}, \pi) = \frac{DD(r_{\rm p}, \pi) - 2DR(r_{\rm p}, \pi) + RR(r_{\rm p}, \pi)}{RR(r_{\rm p}, \pi)},
\label{eq:ls_estimator}
\end{equation}
where $DD$, $DR$, and $RR$ are the normalized counts of data-data, data-random, and random-random pairs, respectively, in bins of $(r_{\rm p}, \pi)$. The random catalog consists of uniformly distributed points within the same volume as the data, with the same angular and radial selection functions. For computational efficiency, we employ the \textsc{corrfunc} package \citep{SinhaGarrison2020} to perform the pair counting. Specifically, we use the \texttt{DDrppi} function, which counts pairs as a function of $r_{\rm p}$ and $\pi$. The resulting pair counts are then used to estimate $w_{\rm p}(r_{\rm p})$ for a given $\pi_{\rm max}$.   

Our simulation box is cubic with periodic boundary conditions, which simplifies the clustering analysis as we do not need to correct for complex survey geometries. However, when we divide the full box into subvolumes to study cosmic variance or to mimic finite survey volumes, we must account for edge effects. Galaxies near the boundaries of a subvolume have fewer neighbors in some directions, which biases the pair counts if not properly corrected. To address this, we generate random catalogues that exactly match the subvolume geometry and use them in the estimator. Additionally, we apply the same spatial cuts to both the data and random catalogues to ensure that any edge effects are properly accounted for in the $DR$ and $RR$ terms.

The choice of $\pi_{\rm max}$ in Eq.~\ref{eq:wp_integral} is a critical parameter that affects the amplitude and shape of $w_{\rm p}(r_{\rm p})$. If $\pi_{\rm max}$ is too small, the integral does not capture all correlated pairs, leading to an underestimation of the clustering amplitude. If $\pi_{\rm max}$ is too large, the integral includes uncorrelated pairs, increasing the noise without contributing additional signal. For each redshift under consideration, we compute $\pi_{\rm max}$ values corresponding to different LOS velocity cuts. Specifically, we consider $\Delta z/(1+z)$ values of 0.002, 0.007, 0.02, and 0.07, which translate to comoving distances via:
\begin{equation}
\pi_{\rm max} = |D_{\rm c}(z + \Delta z) - D_{\rm c}(z)|,
\label{eq:pi_max_calc}
\end{equation}
where $D_{\rm c}(z)$ is the comoving distance. This approach ensures that our $\pi_{\rm max}$ values are physically motivated and consistent across different redshifts.

Uncertainties on the measured correlation functions are estimated using jackknife resampling. The simulation volume is divided into 125 spatially contiguous subvolumes, and the correlation function is recomputed 125 times, each time omitting one subvolume. The covariance matrix is then estimated from the variance among these jackknife samples.

These measurements form the basis for comparing the clustering predictions of our model with observational data and for understanding how galaxy properties such as stellar mass and SFR correlate with their large-scale environment across cosmic time.

\subsection{Deriving the stellar mass–halo mass relation from clustering measurements}
\label{ssec_SHMR}

The SHMR is a fundamental constraint on galaxy formation models, encapsulating the efficiency with which baryons are converted into stars as a function of halo mass. In this section, we describe how we combine the clustering measurements obtained in Sect.~\ref{ssec_2pt} with the underlying galaxy populations to infer the SHMR for our mock galaxy catalogues, and to quantify the uncertainties introduced by various observational and methodological choices.

\subsubsection{Overview}

Our approach leverages the fact that galaxies with different stellar masses exhibit different clustering amplitudes, which reflect the masses of their host dark matter haloes. We establish a mapping between stellar and halo mass by comparing two projected cross-correlation functions. Both functions are computed within a fixed mock galaxy sample, selected with a rest-frame UV magnitude cut. The first is the cross-correlation between the sample and its own galaxies binned by stellar mass; the second is the cross-correlation between the same sample and its own galaxies binned by halo mass. This mapping is then validated against the true SHMR directly measured from the simulation, allowing us to quantify systematic biases as well as variances arising from sample selection, volume effects, and redshift uncertainties.

The method proceeds in several steps. First, we measure the projected cross-correlation function $w_{\rm p}(r_{\rm p})$ for galaxies in multiple stellar mass bins, as described in Sect.~\ref{ssec_2pt}, and similarly for haloes in multiple halo mass bins using the same estimator. Second, for a given stellar mass bin, we compare its clustering signal $w_{\rm p}(r_{\rm p})$ to those of the halo mass bins across a range of projected separations, identifying the halo mass whose clustering most closely matches that of the galaxy sample. Third, we repeat this matching process across all stellar mass bins to construct an empirical SHMR. Finally, we validate this clustering-inferred relation against the true SHMR obtained by directly matching galaxies to their host haloes in the mock catalog, quantifying the scatter and bias as a function of redshift, survey volume, luminosity threshold, and redshift uncertainty.

\subsubsection{Construction of reference halo clustering templates}

To establish a reference for matching, we first measure the projected correlation function for haloes in bins of halo mass $M_{\rm h}$. For this purpose, we use the full $(400\,\mathrm{cMpc})^3$ simulation box to minimize cosmic variance and ensure well-converged clustering measurements across a wide dynamic range in halo mass. The halo mass bins are defined in logarithmic space, spanning $\log M_{\rm h}/\mathrm{M}_\odot=10$ to $13$ with a bin width of $0.5$ dex for our fiducial analysis. For some applications, such as exploring faint galaxy populations for future surveys or comparing different star formation model variants, we extend this range down to $\log M_{\rm h}/\mathrm{M}_\odot \sim 9$ (as shown in Sects.~\ref{ssec_obs} and \ref{ssec_constrain}). For each bin, we compute $w_{\rm p}(r_{\rm p})$ using the same estimator and $\pi_{\rm max}$ values as for the galaxy samples. These halo clustering templates provide a one-to-one mapping between halo mass and clustering amplitude, where more massive haloes exhibit stronger clustering. The values of the cross-correlation with UV magnitude cuts are provided in Table~\ref{tab_corr} in Appendix~\ref{sec_appD}.

\subsubsection{Matching galaxy clustering to halo clustering}

For each stellar mass bin, we seek the halo mass whose clustering most closely matches that of the galaxy sample. To enable a continuous mapping rather than being limited to discrete halo mass bins, we construct a fine grid of halo masses by interpolating between the measured halo clustering templates. Specifically, we generate a fine grid in $\log M_{\rm h}$ space with 50 points per bin and interpolate the $w_{\rm p}(r_{\rm p})$ curves in log–log space using linear interpolation. This yields a set of finely sampled reference clustering templates covering a near-continuous range of halo masses.

For a given stellar mass bin with measured clustering $w_{\rm p,gal}(r_{\rm p})$, we compare it to each fine-grid halo template over a fixed range of projected separations, with a fiducial choice of $0.5<r_{\rm p}/\mathrm{cMpc}<1.0$. This range is chosen to balance sensitivity to halo mass, which is maximal on intermediate scales, while minimizing the impact of nonlinear effects on very small scales and noise and cosmic variance on very large scales (see Sect.~\ref{ssec_CV}). For reference, the virial radii of the samples are approximately 0.1$-$0.4\,cMpc. The comparison is performed in log-space, and the goodness of fit is quantified by the reduced chi-squared statistic, where each radial bin is weighted by the inverse of its jackknife variance (as calculated in Sect.~\ref{ssec_2pt}). The halo mass that minimizes $\chi^2$ is adopted as the characteristic mass corresponding to that stellar mass bin. 

\subsubsection{Validation against the true SHMR}

Having obtained a clustering-inferred SHMR for each combination of redshift, stellar mass bin, and survey parameters, we validate the results against the true SHMR directly measured from the mock catalog. For each stellar mass bin, the matching procedure yields a characteristic halo mass, and we fit a linear relation in log-space to these matched mass pairs, weighting each bin by the square root of the number of galaxies it contains. This fitted relation provides a continuous mapping from stellar mass to inferred halo mass, which we then apply to every galaxy in the sample to assign its clustering-matched halo mass $M_{\rm h,CM}$.

To validate the clustering-based inference of halo mass, we compute the residual $\Delta \log M_{\rm h} = \log M_{\rm h,true} - \log M_{\rm h,CM}$, where $M_{\rm h,true}$ is the true halo mass of a galaxy. We compute the distribution of $\Delta \log M_{\rm h}$ across all galaxies in a given sample, characterizing its median, which quantifies systematic bias, and its standard deviation ($\sigma_{\Delta \log M_{\rm h}}$), which quantifies the statistical scatter in the inference. To evaluate the overall performance of our method, we define a combined bias-and-scatter metric $\sigma_{\rm comb} = \sqrt{\mathrm{median}(\Delta \log M_{\rm h})^2 + \sigma_{\Delta \log M_{\rm h}}^2}$. This metric is motivated by the bias-variance decomposition, where the total prediction error is given by the sum of the squared bias and the variance. We use $\sigma_{\rm comb}$ throughout our analysis as a quantitative measure of the combined systematic offset and random scatter in our halo mass estimates.

\subsubsection{Quantifying the impact of observational effects}

A key advantage of our approach is the ability to systematically vary the parameters that affect real observational surveys and quantify their impact on the inferred SHMR. For each combination of parameters, we compute the distribution of $\Delta \log M_{\rm h}$ and extract its median, standard deviation, and $\sigma_{\rm comb}$. 

\begin{figure*}
\includegraphics[width=1.\linewidth]{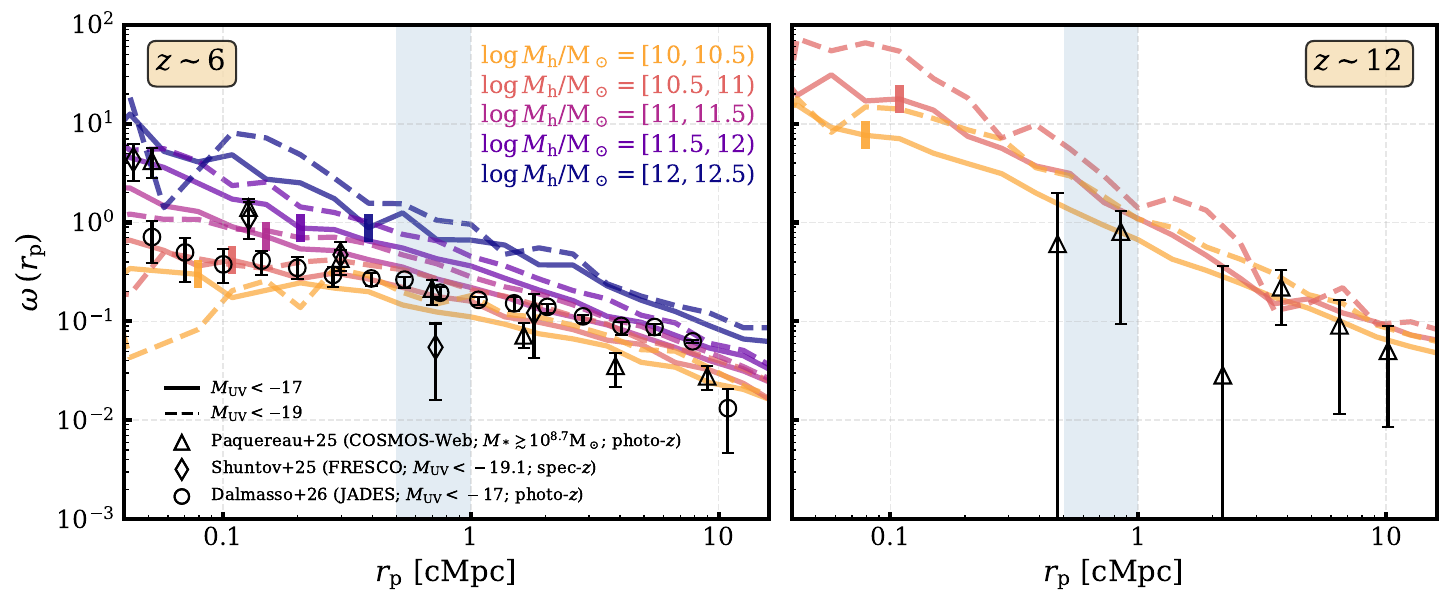}
\vspace{-0.3cm}
\caption{Two-point cross-correlation functions at redshifts 6 (left) and 12 (right) for $\Delta z/(1+z)=0.07$, computed between the full mock galaxy sample (selected with $M_{\rm UV}<-17$ or $-19$, shown as solid and dashed lines, respectively) and subsamples of galaxies in the indicated halo mass bins. The mock galaxies are constructed from the COLIBRE simulation (L400m7; Sect.~\ref{ssec_COLIBRE}) combined with the HOMA empirical model (Saleh et al. in prep.; Sect.~\ref{ssec_model}). The vertical shaded band highlights the radial range $0.5<r_{\rm p}/\mathrm{cMpc}<1.0$, which we adopt as optimal for our halo mass inference (Sect.~\ref{ssec_SHMR}; see also Appendix~\ref{sec_appC}). The vertical ticks indicate the mean virial radii. Comparison with observational data (\citealt{Paquereau2025, Shuntov2025b, Dalmasso2026}) suggests that the high-redshift galaxies revealed by JWST typically reside in haloes with $\log M_{\rm h}/{\rm M_\odot}\simeq 10-11$. The correlation values for the halo mass bins at $r_{\rm p}=0.75$\,cMpc, the median of our fiducial radial range $0.5<r_{\rm p}/\mathrm{cMpc}<1.0$, are provided in Table~\ref{tab_corr}.}
\label{fig_wp_Mh}
\end{figure*}

We investigate four redshifts ($z=$\,4, 6, 10, 12) spanning the epoch of reionization and the early buildup of galaxies, allowing us to assess how the reliability of clustering-based SHMR inference evolves with cosmic time. We apply rest-frame UV magnitude cuts of $M_{\rm UV}<-13, -15, -17, -18, -19, -20$, mimicking the selection of high-redshift galaxy surveys with varying depth. This reveals how luminosity selection biases the inferred SHMR and what corrections may be needed when comparing to observations. We analyze subvolumes corresponding to cubical regions of 25, 50, 100, and 200\,cMpc on a side, obtained by subdividing the full $(400\,\mathrm{cMpc})^3$ simulation box, as well as the entire $(400\,\mathrm{cMpc})^3$ volume itself, to quantify the minimum volume required to obtain converged SHMR constraints and the impact of cosmic variance on the inferred relation. 

Finally, we consider four values of LOS velocity uncertainty, parameterized as $\Delta z/(1+z)=$\,0.002, 0.007, 0.02, 0.07, which translate to different $\pi_{\rm max}$ values in the clustering measurement, to assess how redshift errors affect the clustering signal and consequently the inferred SHMR. We apply top-hat cuts of $\Delta z$ to mimic these uncertainties, but find no significant impact on the performance of the method when assuming the values as Gaussian errors. This insensitivity arises because our method applies the same redshift-error treatment consistently to both halo and galaxy clustering measurements, thereby operating in a self-consistent manner and not biasing the inferred relation. The four chosen values span the typical redshift uncertainties of spectroscopic and photometric data at high redshifts for which clustering measurements have been reported. For example, $\Delta z/(1+z)$ ranges between $0.05$ and $0.09$ for the photometric samples of \citet{Dalmasso2026}, while it is generally $\lesssim 0.001$ for spectroscopic data. Note that we do not correct for peculiar velocities, as we directly map redshift uncertainties to the maximum LOS integration scale $\pi_{\rm max}$. Rather, the purpose of testing multiple integration depths is precisely to account for intrinsic redshift errors, including peculiar velocities, which affect even spectroscopic samples. Our smallest tested value, $\Delta z/(1+z) = 0.002$, is comparable to the typical amplitude of peculiar velocities. The results at this limit therefore remain susceptible to unmodelled peculiar velocity effects.

Note that our implementation of LOS redshift errors always preserves the clustering signal associated with a given halo. The integration depth inevitably includes some uncorrelated population, which increases noise, but it never misses the true clustering signal. In observations, however, redshift errors are typically characterized by a Gaussian width rather than a sharp cut $\Delta z$. Consequently, a target halo may occasionally be misidentified at a redshift bearing no true associated clustering. Such cases contribute nothing to the clustering measurements, thereby underestimating the average signal. Assuming a 10 per cent fraction of such LOS outliers (which is a typical maximum for high-$z$ photometric data; e.g., \citealt{Shuntov2025c}), the mean clustering amplitude is reduced by the same fraction. Even for a halo whose associated clustering lies within the integration depth, adding 10 per cent uncorrelated galaxies likewise reduces the measured clustering by 10 per cent. Thus, the total expected underestimation is approximately 20 per cent, corresponding to about 0.08\,dex. Given the logarithmic slope of $\sim$\,0.5 between halo mass and clustering strength (shown later), this translates into a 0.15$-$0.2\,dex error in halo mass. Indeed, when we randomly scatter 10 per cent outliers across our mock samples, we find a bias of roughly this magnitude. Crucially, this bias appears only if the same outlier fraction is omitted when constructing the reference halo clustering templates. Once outliers are consistently included in both the galaxy and halo templates, the same fractional reduction applies to both, resulting in no net bias. Hence, as long as the redshift error of a survey is accurately known, our method fully accounts for it and introduces no bias in the SHMR mapping. 

By systematically varying these parameters and computing the resulting $\Delta \log M_{\rm h}$ statistics, we build a comprehensive picture of the reliability and limitations of clustering-based SHMR inference at high redshift. These results provide essential guidance for interpreting observational measurements and for designing survey strategies that minimize systematic biases in the inferred galaxy–halo connection.

\section[result]{Results}
\label{sec_result}

In this section, we present the results of our clustering-based method for inferring the SHMR from mock galaxy catalogues. We first establish reference clustering templates as a function of halo mass, then validate the accuracy and precision of our matching technique against true halo masses from the simulation, systematically exploring the impact of survey volume, redshift uncertainty, and UV luminosity threshold. Finally, we investigate how secondary galaxy properties can mitigate assembly bias and environmental effects, improving the accuracy and precision of halo mass estimates. 

\subsection{Clustering strength as a function of halo mass}
\label{ssec_wp_Mh}

As outlined in Sect.~\ref{ssec_SHMR}, the first step of our method is to establish reference templates for the average clustering strength as a function of halo mass. These templates serve as the basis for matching, where we identify the halo mass whose clustering best matches that of a given galaxy sample. The cross-correlation functions $w_{\rm p}(r_{\rm p})$ between the full mock galaxy sample (defined by $M_{\rm UV,cut} = -17$ and $-19$) and halo mass-selected subsamples are shown in Fig.~\ref{fig_wp_Mh}. 

To construct these templates, we use the largest available volume, the full simulation box of $400$\,cMpc on a side. This choice minimizes cosmic variance, which we find can be substantial even for relatively large volumes. For $(200$\,cMpc$)^3$ subvolumes, the measured clustering strength at fixed mass for haloes with $M_{\rm h}>10^{10}\,{\rm M}_\odot$ can vary by up to a factor of $\sim 2$ between different subvolumes, across all redshifts considered. A $(200$\,cMpc$)^3$ volume corresponds to survey areas of approximately 40 and 160\,deg$^2$ at redshift 6 and 10, respectively, when assuming LOS depth of $\Delta z = 0.5$. Given that the relationship between $\log w_{\rm p}$ and $\log M_{\rm h}$ at a given $r_{\rm p}$ has a slope of approximately $0.5$, such variations in clustering can translate into biases of up to $\sim 0.6$ dex in the inferred halo masses. This effect represents one of the largest potential uncertainties when applying our clustering matching technique to real survey volumes, which are necessarily limited in size. We return to a detailed discussion of cosmic variance in Sect.~\ref{ssec_CV}.

To provide context for how these templates can be compared with observational data and used to connect observed galaxies to their host haloes, we compare our predictions with the results from \citet{Paquereau2025}, \citet{Shuntov2025b}, and \citet{Dalmasso2026}. Using a sample of 6,500 Lyman Break Galaxies (LBGs) at $5 < z < 11$ from the JADES survey, \citet{Dalmasso2026} measured the two-point angular correlation function and interpreted it within a HOD framework. For this comparison, we adopt a value of $\Delta z/(1+z)$ corresponding to the redshift range of their sample and, since their selection corresponds to $M_{\rm UV} < -17$, we compare with our templates for that threshold. Based on a photometric redshift catalog from the 0.53\,deg$^2$ COSMOS-Web survey \citep{Casey2023}, \citet{Paquereau2025} constructed stellar mass-limited, complete galaxy samples with a F444W magnitude limit of 27.75, enabling mass-limited angular clustering measurements from $z = 0.1$ to $z \sim 12$. Finally, \citet{Shuntov2025b} reported clustering measurements of H$\alpha$ and [\textsc{Oiii}] emitters from the FRESCO \citep{Oesch2023} and CONGRESS \citep{Egami2023} surveys in 124\,arcmin$^2$ of the GOODS fields at $3.8<z<9$.

\begin{figure*}
\includegraphics[width=1.\linewidth]{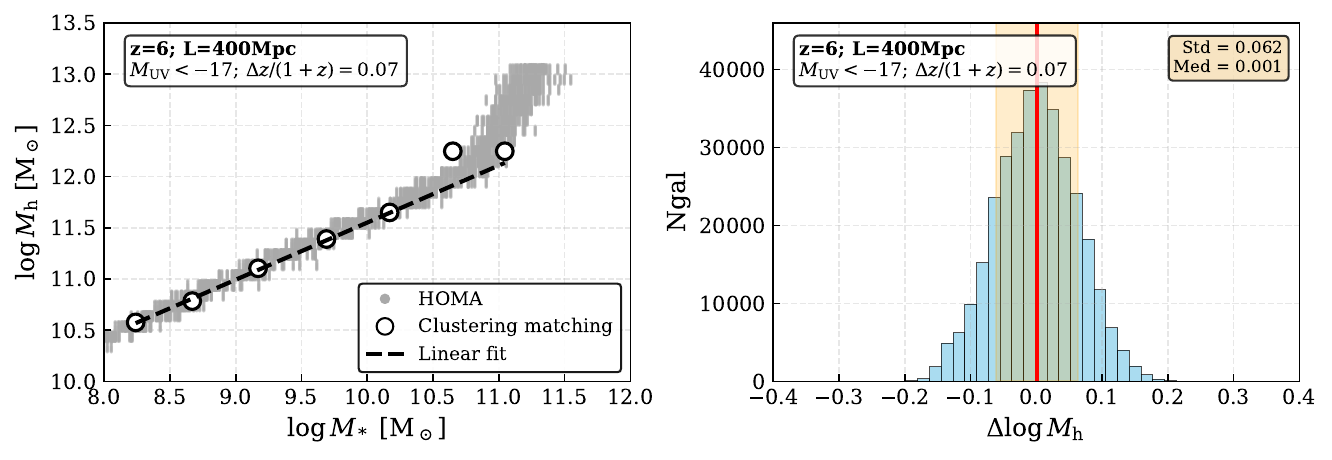}
\vspace{-0.2cm}
\caption{Validation of clustering-based halo mass inference for the mock constructed by populating the COLIBRE haloes with the fiducial HOMA model. \textit{Left:} Stellar-to-halo mass relation (SHMR) inferred by our clustering matching method (circles; power-law fit shown as black dashed line) for stellar mass bins of width $0.5$ dex at redshift $z=6$, for a $(400$ cMpc$)^3$ volume, a UV magnitude cut of $M_{\rm UV, cut}=-17$, and a redshift uncertainty of $\Delta z/(1+z)=0.07$. The gray band indicates the true SHMR distribution from the HOMA model input. \textit{Right:} Distribution of residuals $\Delta \log M_{\rm h} = \log M_{\rm h,true} - \log M_{\rm h,CM}$ between the true halo mass and that inferred from clustering matching (using the power-law fit from the left panel). The red vertical line and orange band indicate the median and $1\sigma$ dispersion of the distribution, respectively. The combined uncertainty $\sigma_{\rm comb} = \sqrt{\mathrm{median}(\Delta \log M_{\rm h})^2 + \sigma_{\Delta \log M_{\rm h}}^2}$ is consistent with the intrinsic scatter of the HOMA model input, confirming that our method recovers the underlying SHMR dispersion without introducing bias or additional uncertainty.}
\label{fig_validation}
\end{figure*}

Over our fiducial fitting range of $0.5 < r_{\rm p}/\mathrm{cMpc} < 1.0$, the best-matching halo mass for the \citet{Dalmasso2026} sample is $\log M_{\rm h} /{\rm M}_\odot  \simeq 10.5$. This is approximately $0.8$\,dex lower than the typical halo mass inferred from their HOD analysis. The origin of this discrepancy is uncertain. Given the relatively shallow slope of the $w_{\rm p}$–$M_{\rm h}$ relation, a bias or uncertainty of a factor $\sim 2$ in the underlying assumptions of either framework could potentially reconcile the two estimates. Indeed, the \citet{Dalmasso2026} measurements have errors of about a factor of $2$ on angular scales of $10{-}20$ arcsec, which correspond roughly to our fitting range. Moreover, their best-fit HOD models show discrepancies of a similar factor relative to the data over the same scales. Similar discrepancies between HOD models and observations in the transition range from one-halo to two-halo terms have been reported by other studies (e.g. \citealt{Harikane2022, Paquereau2025, Shuntov2025b}). These discrepancies likely arise from the assumption of linear bias in their HOD framework. \citet{Paquereau2025} demonstrated that accounting for non-linear halo bias increases clustering strength by a factor of 2${-}$3 on $\sim$1\,cMpc scale at $z\sim6$, significantly improving the fit to observations, and lowering estimated halo mass (see also \citealt{Jose2013, Jose2017, Mead2015, Harikane2018}). 

In \citet{Dalmasso2026}, their tightest constraints come from larger scales near $100$ arcsec ($\sim 4$\,cMpc), where the measurement errors are smallest. However, as we will show in Sect.~\ref{ssec_CV}, on scales $r_{\rm p} \gtrsim 3$ cMpc the two-point correlation function becomes highly sensitive to cosmic variance, varying by up to an order of magnitude even between $(200$\,cMpc$)^3$ volumes. For the JADES survey volume, which is approximately $(50$\,cMpc$)^3$, we find the impact of cosmic variance to be even more severe, causing fluctuations of two orders of magnitude at $r_{\rm p}\gtrsim3$\,cMpc. In fact, when we fit our clustering template to the observational estimate at larger scales of $r_{\rm p} \approx 4-5$ cMpc, we obtain a higher best-fit halo mass of $\log M_{\rm h}/{\rm M}_\odot \simeq 11$, which lies within approximately $0.2$ dex of the value inferred by \citet{Dalmasso2026}. This suggests that much of the apparent discrepancy can be attributed to the different radial scales used in the analysis, rather than to fundamental differences in the underlying galaxy–halo connection.

Another notable difference concerns the satellite fraction. In our mock catalog, approximately $10$ per cent of galaxies satisfying the $M_{\rm UV} < -17$ selection are satellites (where the satellite/central classification comes from the COLIBRE simulation, while the UV magnitudes are assigned by the HOMA model). In contrast, the HOD framework of \citet{Dalmasso2026} infers a satellite fraction of only $0.1$ per cent, effectively negligible, despite yielding much higher host halo masses for their sample. This discrepancy is not easily reconciled by simply invoking a model with a lower satellite fraction, as such a model would tend to associate UV-selected galaxies with systematically lower-mass haloes. It would therefore predict a lower host halo mass alongside a lower satellite fraction, contrary to the higher halo mass and negligible satellite fraction reported by \citet{Dalmasso2026}. This discrepancy highlights degeneracies and different assumptions inherent in the two modelling approaches, and underscores the importance of systematic comparisons between clustering-based methods and HOD analyses.

\begin{figure*}
\includegraphics[width=1.\linewidth]{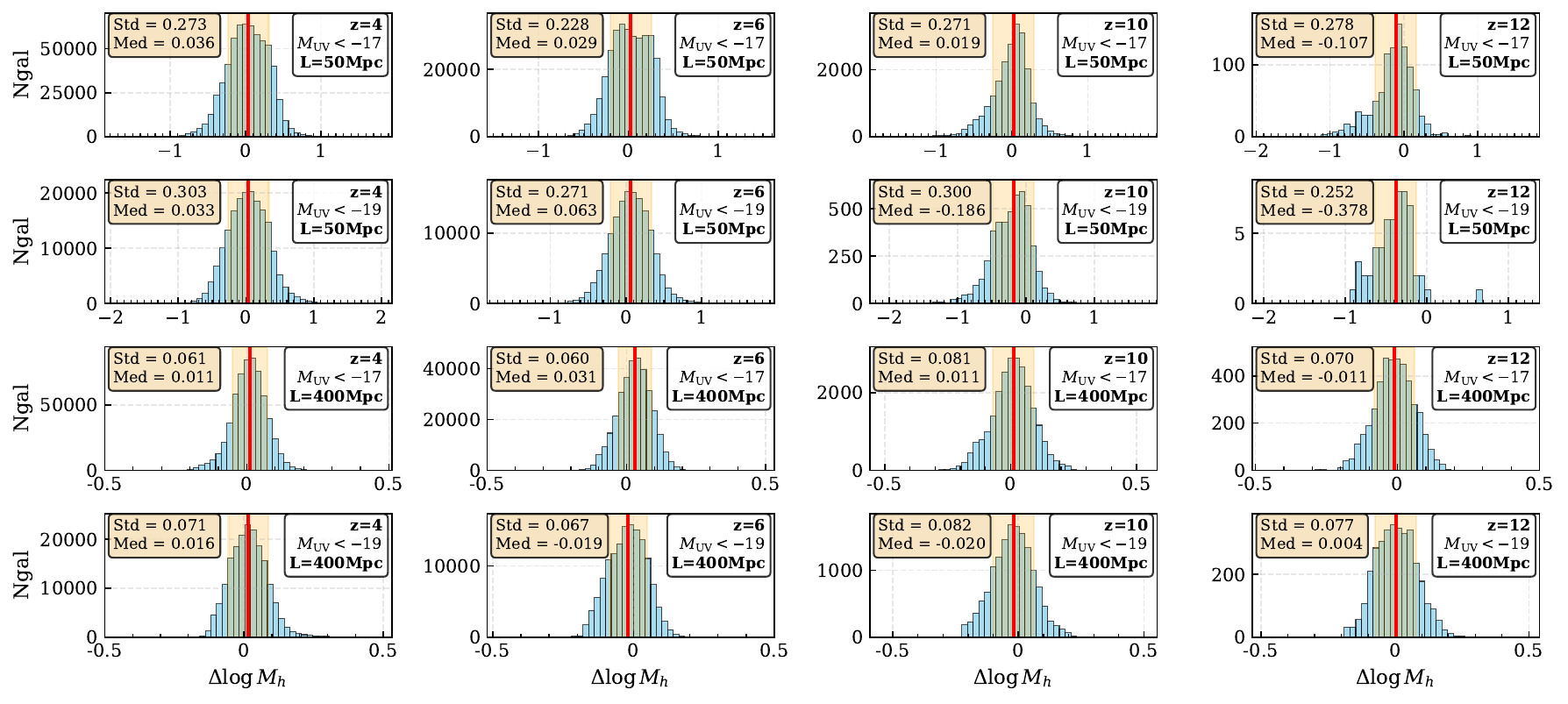}
\vspace{-0.2cm}
\caption{Probability distribution function (PDF) of the difference between the halo mass estimated by our clustering matching method and the true value, for mock galaxies and their two-point correlation functions constructed under various observational configurations as indicated in each panel (for a fixed redshift uncertainty of $\Delta z/(1+z)=0.002$). The halo mass is estimated via a power-law fit to the average values obtained for each stellar mass bin, as shown by the dashed line in Fig.~\ref{fig_validation}. The red vertical line and orange band indicate the median and 1-sigma range of the distribution, respectively. The results demonstrate that our method recovers halo mass and the SHMR to within 0.3\,dex for survey volumes spanning (50\,cMpc)$^3$ to (400\,cMpc)$^3$ and across redshifts 4 to 12. The combined bias and dispersion $\sigma_{\rm comb} = \sqrt{\mathrm{median}(\Delta \log M_{\rm h})^2 + \sigma_{\Delta \log M_{\rm h}}^2}$, in particular, depends strongly on survey volume while showing only weak dependence on redshift and UV magnitude.}
\label{fig_validation2}
\end{figure*}

We also apply our clustering matching method to estimate the best-fitting halo mass for the COSMOS-Web samples of \citet{Paquereau2025} based on their angular measurements, focusing on their $z \sim 6-8$ results where the data are sufficiently constraining; we include their $z \sim 12$ measurement only as a reference, as the observational uncertainties at higher redshift are too large for precise estimates. At $z \sim 6$, we find that COSMOS-Web galaxies with stellar mass limit $M_* \gtrsim 10^{8.7}\,{\rm M}_\odot$ are associated with haloes of $M_{\rm h} \simeq 10^{10.1}\,{\rm M}_\odot$, obtained by fitting our templates over the fiducial $0.5 < r_{\rm p}/\mathrm{cMpc} < 1.0$ range after applying the same stellar mass limit to our mock sample. From our model prediction, this stellar mass limit roughly corresponds to $M_{\rm UV} \lesssim -18.5$, bridging the selection of the two observational datasets. Notably, despite their different selection criteria and survey volumes, the two independent measurements are consistent with each other within our fiducial radial range. The discrepancies, however, become pronounced only at smaller ($r_{\rm p} < 0.2$\,cMpc) and larger ($r_{\rm p} > 1.0$\,cMpc) scales, which we attribute to nonlinear effects and cosmic variance, respectively. The fact that the two datasets agree precisely in the range where our method is designed to operate reinforces our confidence in this choice of radial range as optimal for robust halo mass assignment to high-redshift galaxies. 

We also include the measurements of \citet{Shuntov2025b} in Fig.~\ref{fig_wp_Mh}. However, due to the large uncertainties and sparse coverage within our fiducial range $0.5 < r_{\rm p}/\mathrm{cMpc} < 1.0$ at $z>5$, we do not attempt quantitative halo mass estimation using these data. Their measurements primarily constrain the small-scale regime ($r_{\rm p} < 0.1$\,cMpc), which is sensitive to non-linear effects (Sect.~\ref{ssec_CV}) and therefore less ideal for robust halo mass inference. Nonetheless, the broad consistency among these three independent datasets on the scales relevant to our analysis provides a basis for validating our method and interpreting comparisons with our results (Sect.~\ref{sec_discussion}).

\subsection{Validation of the clustering-based halo mass inference and SHMR}
\label{ssec_validation}

Following the methodology outlined in Sect.~\ref{ssec_SHMR}, we now validate our clustering-based inference of halo masses by comparing the inferred values with the true halo masses directly measured from the mock catalog. The left panel of Fig.~\ref{fig_validation} illustrates this validation for a representative case: redshift $z=6$, a full $(400$ cMpc$)^3$ volume, a UV magnitude cut of $M_{\rm UV,cut} = -17$, and a redshift uncertainty of $\Delta z/(1+z) = 0.07$. For each stellar mass bin of width $0.5$ dex, we plot the mean halo mass inferred from clustering matching, compared against the true input relation indicated by the grey band.

Deviations of up to $0.2$ dex relative to the true values are observed, approximately three times the intrinsic scatter present in the mock catalog. Since both the halo template clustering and the galaxy clustering are measured from the same volume, cosmic variance does not contribute to these deviations. Instead, the primary source of fluctuation is the large line-of-sight integration depth corresponding to $\Delta z/(1+z) = 0.07$, which translates to redshift errors of $\pm 0.5$ or a comoving distance of $\sim 200$\,cMpc at $z \simeq 6$. The clustering signal declines rapidly with distance from a galaxy. Integrating over LOS depths much larger than $\sim 5$ cMpc, beyond which $\xi(r)$ drops below unity at these redshifts, primarily adds noise rather than signal, causing the measured averages to fluctuate.

\begin{figure*}
\includegraphics[width=0.95\linewidth]{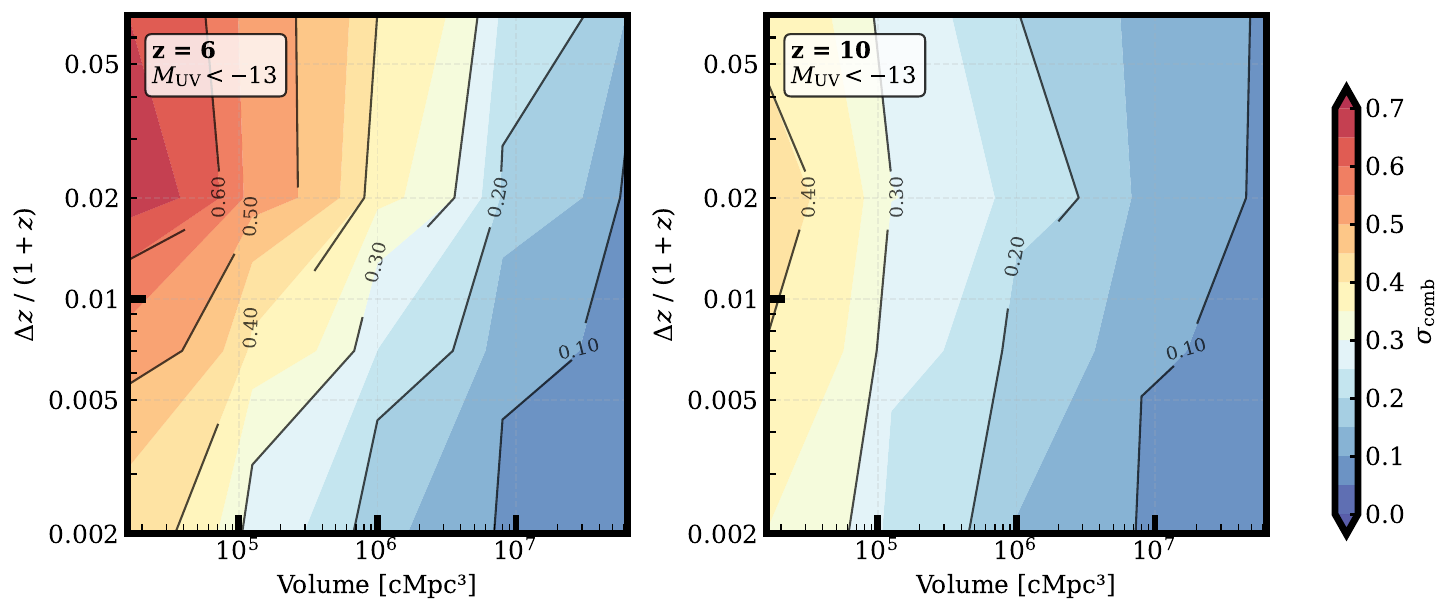}
\vspace{-0.2cm}
\caption{Combined bias and dispersion, $\sigma_{\rm comb} = \sqrt{\mathrm{median}(\Delta \log M_{\rm h})^2 + \sigma_{\Delta \log M_{\rm h}}^2}$, in the halo mass of mock galaxies estimated by our clustering matching method, compared to the true value, as a function of survey volume and redshift uncertainty. Results are shown for redshifts 6 (left) and 10 (right) and for a UV magnitude cut of $M_{\rm UV, cut}=-13$ as an example. The uncertainty depends strongly on survey volume while showing only mild dependence on the redshift error.}
\label{fig_contour1}
\end{figure*}

The dashed line in Fig.~\ref{fig_validation} shows a power-law fit to the results from individual stellar mass bins, which provides a continuous mapping from stellar mass to inferred halo mass as described in Sect.~\ref{ssec_SHMR}. While we adopt a simple power-law fit throughout this analysis for simplicity, more flexible functional forms could be employed to better describe trends present in other mock or observational datasets. The right panel of Fig.~\ref{fig_validation} displays the distribution of residuals $\Delta \log M_{\rm h} = \log M_{\rm h,true} - \log M_{\rm h,CM}$, with a median of $0.001$ and a standard deviation of $0.062$. The combined scatter and bias, $\sigma_{\rm comb} = \sqrt{\mathrm{median}(\Delta \log M_{\rm h})^2 + \sigma_{\Delta \log M_{\rm h}}^2}$, of $\sim 0.07$ dex is comparable to the intrinsic dispersion built into the model, demonstrating that our clustering matching technique can estimate halo masses accurately for high-redshift faint sources even in the presence of photometric redshift uncertainties.

Fig.~\ref{fig_validation2} expands this analysis to a broader range of redshifts, survey volumes, and selection criteria, all assuming a more optimistic redshift error of $\Delta z/(1+z) = 0.002$. Except where the number of galaxies passing the selection becomes too small (e.g., $M_{\rm UV} < -19$ samples at $z=12$), the combined dispersion of scatter and bias remains $\lesssim 0.3$ dex. This holds even for a survey volume comparable to that of the current JADES data out to redshift 12. These precisions are comparable to the predictions by \citet{Endsley2020} for JWST observations of $M_{\rm UV}\sim-19$ galaxies at $z\sim4-10$. The accuracy and precision of clustering matching appear largely insensitive to galaxy magnitude or redshift. Survey volume emerges as the most significant factor: for a $(400$ cMpc$)^3$ volume, the clustering matching recovers the true halo mass to within the intrinsic scatter of the model across all redshifts and selections considered, indicating that cosmic variance is essentially the only significant source of uncertainty aside from observational errors in derived galaxy properties such as stellar mass or UV luminosity.

Fig.~\ref{fig_contour1} summarizes the performance across volumes spanning nearly four orders of magnitude ($10^4$ to $10^8$ cMpc$^3$) and redshift uncertainties $\Delta z/(1+z)$ ranging from $0.002$ to $0.07$, now extended to much fainter galaxies with $M_{\rm UV} < -13$. At both $z=6$ and $z=10$, survey volume dominates over redshift errors in determining the accuracy and precision of halo mass inference. For volumes smaller than $10^5$ cMpc$^3$ (roughly $(50$ cMpc$)^3$), the combined uncertainty $\sigma_{\rm comb}$ increases to $>0.5$ dex at $z=6$ and $>0.3$ dex at $z=10$. At fixed volume and redshift error, the precision improves with increasing redshift: same brightness cuts at higher redshifts select rarer density peaks with stronger clustering, yielding more stable $w_{\rm p}$ measurements and thus more robust clustering matching.

\begin{figure*}
\includegraphics[width=0.95\linewidth]{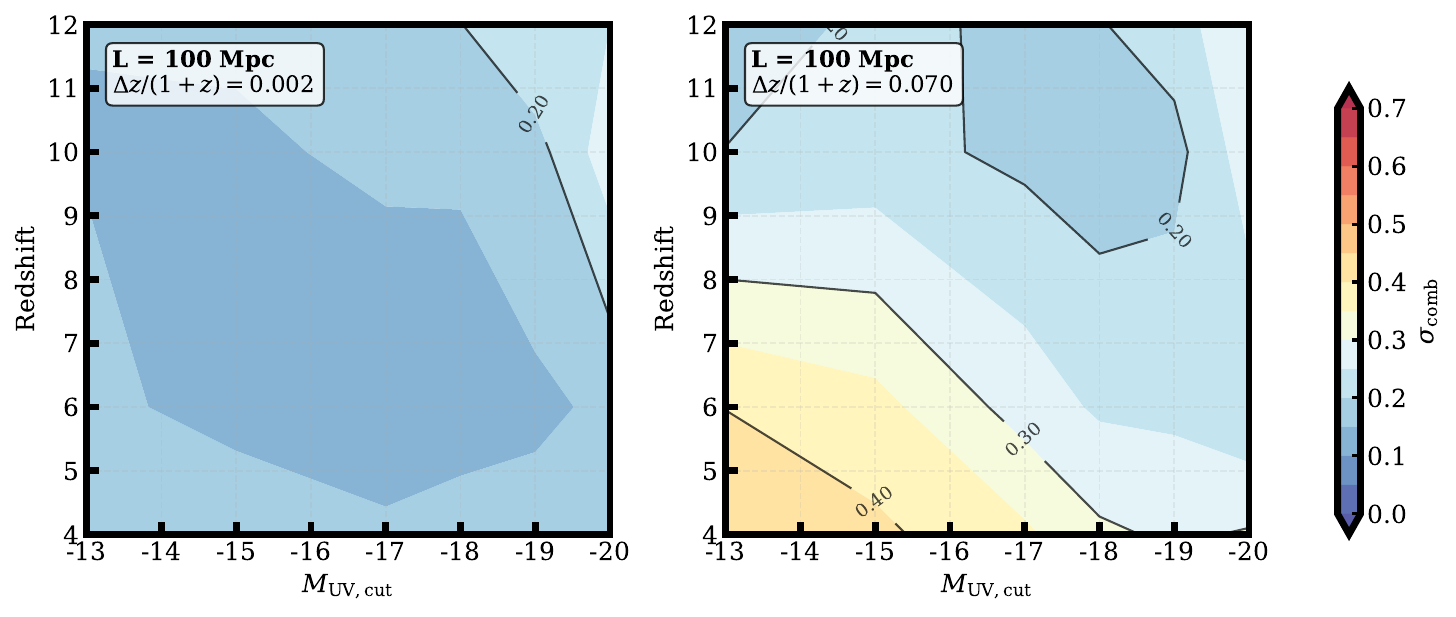}
\vspace{-0.2cm}
\caption{Combined bias and dispersion, $\sigma_{\rm comb} = \sqrt{\mathrm{median}(\Delta \log M_{\rm h})^2 + \sigma_{\Delta \log M_{\rm h}}^2}$, in the halo mass of mock galaxies estimated by our clustering matching method, compared to the true value, as a function of UV magnitude selection cut and redshift. Results are shown for redshift uncertainties of $\Delta z/(1+z)=0.002$ (left) and 0.07 (right) and for a survey volume of (100\,cMpc)$^3$ as an example. }
\label{fig_contour2}
\end{figure*}

Fig.~\ref{fig_contour2} presents a similar performance summary but now as a function of $M_{\rm UV,cut}$ and redshift for a fixed volume of $(100$ cMpc$)^3$ and redshift uncertainties of $\Delta z/(1+z) = 0.002$ and $0.07$. The left panel shows that with accurate redshifts, the uncertainty in inferred halo mass depends only weakly on redshift or magnitude cut, remaining below $\sim 0.2$ dex. As redshift uncertainty increases, however, the combined accuracy and precision degrades more strongly for fainter and lower-redshift samples. This reflects the fact that brighter cuts at higher redshifts select higher density peaks with intrinsically stronger clustering, which are more robust to the diluting effects of photometric redshift errors.

Finally, to verify that our validation results are not specific to the HOMA model, we repeat the same clustering matching analysis using galaxy and halo properties taken directly from the COLIBRE simulation (see Sect.~\ref{ssec_mock}). The COLIBRE SHMR exhibits an intrinsic scatter approximately twice as large as that of the HOMA model at fixed stellar mass. Our method recovers the true COLIBRE halo masses with no significant bias and with dispersions that closely match the input scatter, confirming that the clustering-based inference traces the underlying SHMR dispersion regardless of its amplitude (see Appendix~\ref{sec_appA} for details). This cross-validation demonstrates that the performance metrics derived from the HOMA mock catalogues are representative of the method's ability to recover the true halo masses without introducing model-dependent systematics.

\subsection{Dependence on secondary galaxy properties: mitigating assembly bias and environmental effects}
\label{ssec_properties}

Thus far, our analysis has used stellar mass as the galaxy property when measuring clustering and constraining the SHMR. The UV magnitude cut is only applied to determine which galaxies are considered as surrounding neighbours of the target. This approach, while effective, implicitly assumes that the clustering strength of a galaxy population is fully determined by its stellar mass, and that any residual dependence on other properties such as star formation history, colour, or stellar age is negligible or can be absorbed into the scatter of the SHMR. However, galaxy properties beyond stellar mass are known to correlate with the large-scale environment and assembly history of their host haloes \citep[e.g.,][]{Gao2005, Wechsler2006, Lim2016, Lim2025}. If such secondary properties are correlated with clustering strength at fixed stellar mass, ignoring them could introduce systematic biases in the inferred halo masses.

To investigate this, we use the galaxy and halo properties taken directly from the COLIBRE simulation (L400m7). This approach has two distinct advantages: first, the COLIBRE simulation naturally encodes a broad range of baryonic processes—including feedback, accretion history, and environmental effects—that are not fully captured by the empirical HOMA model. Second, it provides direct access to the true halo masses, allowing us to quantify how the inclusion of secondary properties improves the accuracy and precision of our clustering-based halo mass estimates without introducing additional model-dependent assumptions. In principle, one could attempt to incorporate accretion history and assembly bias into the HOMA framework by populating COLIBRE haloes with HOMA-predicted properties using the halo accretion rates directly measured from the simulation. However, the HOMA model parameters are calibrated to match observed galaxy properties (Sect.~\ref{ssec_model}), and applying them to the accretion histories of a different simulation would require re-tuning these parameters to ensure consistency with the underlying baryonic physics. Moreover, the way accretion rates are defined differ between the HOMA model and the COLIBRE simulation, making an implementation non-trivial. While we plan to explore such a self-consistent implementation in future work, for the present analysis we adopt the direct COLIBRE properties as a robust and physically motivated alternative. 

\begin{figure}
\includegraphics[width=0.98\linewidth]{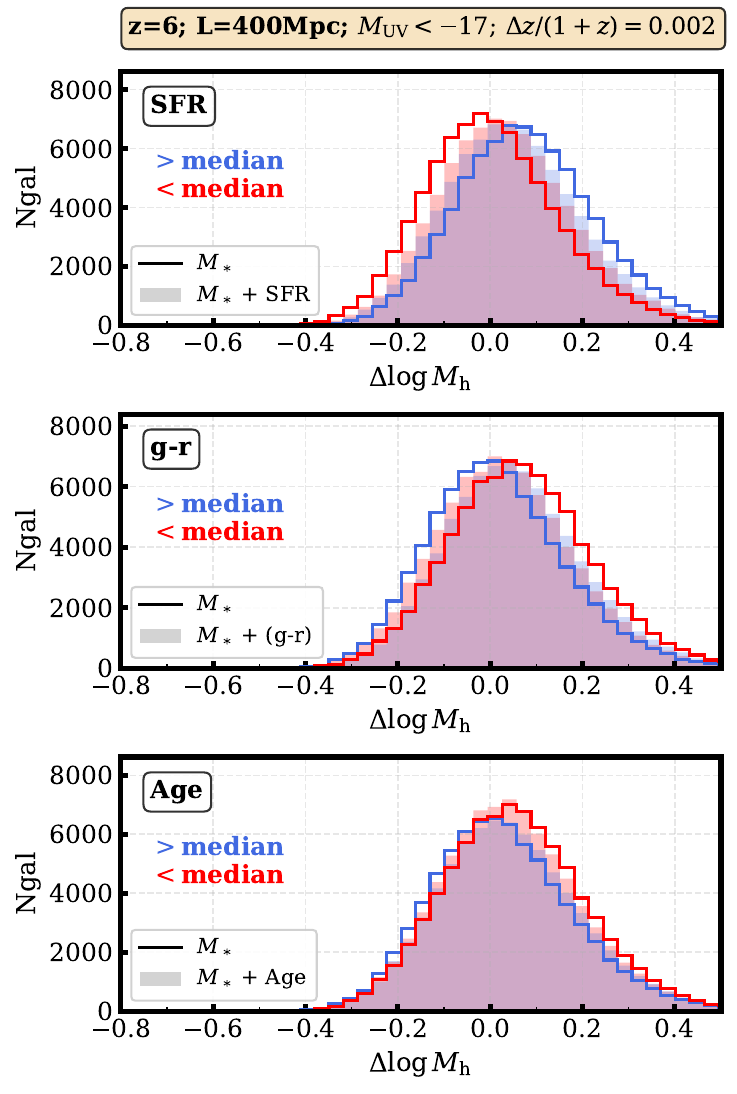}
\vspace{-0.2cm}
\caption{Impact of secondary galaxy properties on clustering-based halo mass inference using galaxy and halo properties taken directly from the COLIBRE simulation (L400m7; Sect.~\ref{ssec_COLIBRE}), with UV magnitudes from HOMA for selection. Results are shown at redshift $z=6$ for a $(400$ cMpc$)^3$ volume, a UV magnitude cut of $M_{\rm UV, cut}=-17$, and a redshift uncertainty of $\Delta z/(1+z)=0.002$. Each panel corresponds to a different secondary property: star formation rate averaged over 100\,Myr (SFR; top), rest-frame colour $(g-r)$ (middle), and stellar age (bottom). In each panel, the histogram shows the distribution of residuals $\Delta \log M_{\rm h} = \log M_{\rm h,true} - \log M_{\rm h,CM}$ between the true halo mass and that inferred from clustering matching. The filled histograms (blue for above-median, red for below-median) show the residuals obtained when using coefficients derived independently for each subsample. The unfilled histograms show the residuals obtained when using a single global coefficient derived from the full sample. The significant reduction of the offsets between the subsamples in the filled histograms demonstrates that our method can effectively mitigate assembly bias and environmental effects, improving the accuracy and precision of halo mass estimates.}
\label{fig_SFR100}
\end{figure}

Fig.~\ref{fig_SFR100} presents the results of this analysis for three secondary properties: the SFR averaged over 100\,Myr, the rest-frame colour $(g-r)$, and the luminosity-weighted stellar age. For the rest-frame colour, we use the $g$ and $r$ band luminosities from the GAMA filter set, which are rest-frame dust-free AB luminosities of the stellar particles computed using the \textsc{GALAXEV} models \citep{BruzualCharlot2003} convolved with the corresponding filter bands. The luminosity-weighted stellar age is computed using the same GAMA $r$-band luminosities as weights. For each property, we split the galaxy sample at each stellar mass bin into two subsamples, those with the property above the median and those below. We then apply our clustering matching method in two ways. First, we use the standard approach with a single global coefficient derived from the full sample (shown as unfilled histograms). Second, we apply the method separately to the above- and below-median subsamples using coefficients derived independently for each subsample (filled histograms). The histograms display the distribution of residuals $\Delta \log M_{\rm h} = \log M_{\rm h,true} - \log M_{\rm h,CM}$, where $M_{\rm h,CM}$ is the halo mass inferred from clustering matching.

The results reveal a clear and systematic pattern. When using the standard approach with a single global relation, the inferred halo masses exhibit offsets that correlate with the secondary property. For SFR, galaxies above (below) the median show a positive (negative) median offset. This indicates that high-SFR galaxies at fixed stellar mass preferentially inhabit more massive haloes, while low-SFR reside in lower-mass haloes. Likewise, galaxies with bluer colours or younger ages are assigned to higher-mass haloes, reflecting the environmental dependence of quenching and star formation histories. These offsets represent systematic biases introduced by ignoring the environmental dependence of galaxy properties.

When we apply our clustering matching method separately to the above- and below-median subsamples using independently derived coefficients, these systematic offsets are largely eliminated. The filled histograms in each panel show that the distributions for the two subsamples become significantly more symmetric and centered near zero, demonstrating that the offsets introduced by assembly bias and environmental effects can be effectively removed by accounting for secondary properties.

The improvement seen across all three secondary properties suggests that galaxies with different recent star formation histories and stellar populations occupy different halo environments at fixed stellar mass, and that these environmental differences manifest as differences in clustering strength. By accounting for these secondary dependencies, our method effectively corrects for assembly bias, yielding halo mass estimates that are both more accurate and more precise. The combined uncertainties after correction are comparable to or smaller than the intrinsic scatter of the COLIBRE simulation itself, indicating that we have recovered essentially all the information available from clustering.

The ability to incorporate secondary properties in a physically motivated way represents a key advantage of our clustering-based method over alternative approaches such as abundance matching, which relies solely on stellar mass or luminosity and cannot easily account for assembly bias or environmental effects. While we have focused on three illustrative properties here—SFR, colour, and age—the same approach can be extended to any galaxy property that correlates with clustering strength at fixed stellar mass. This flexibility is particularly valuable for interpreting the diverse and growing suite of JWST observations, which now probe a wide range of galaxy properties across the first billion years of cosmic history.

\section[discussion]{Discussion}
\label{sec_discussion}

\begin{figure*}
\includegraphics[width=0.95\linewidth]{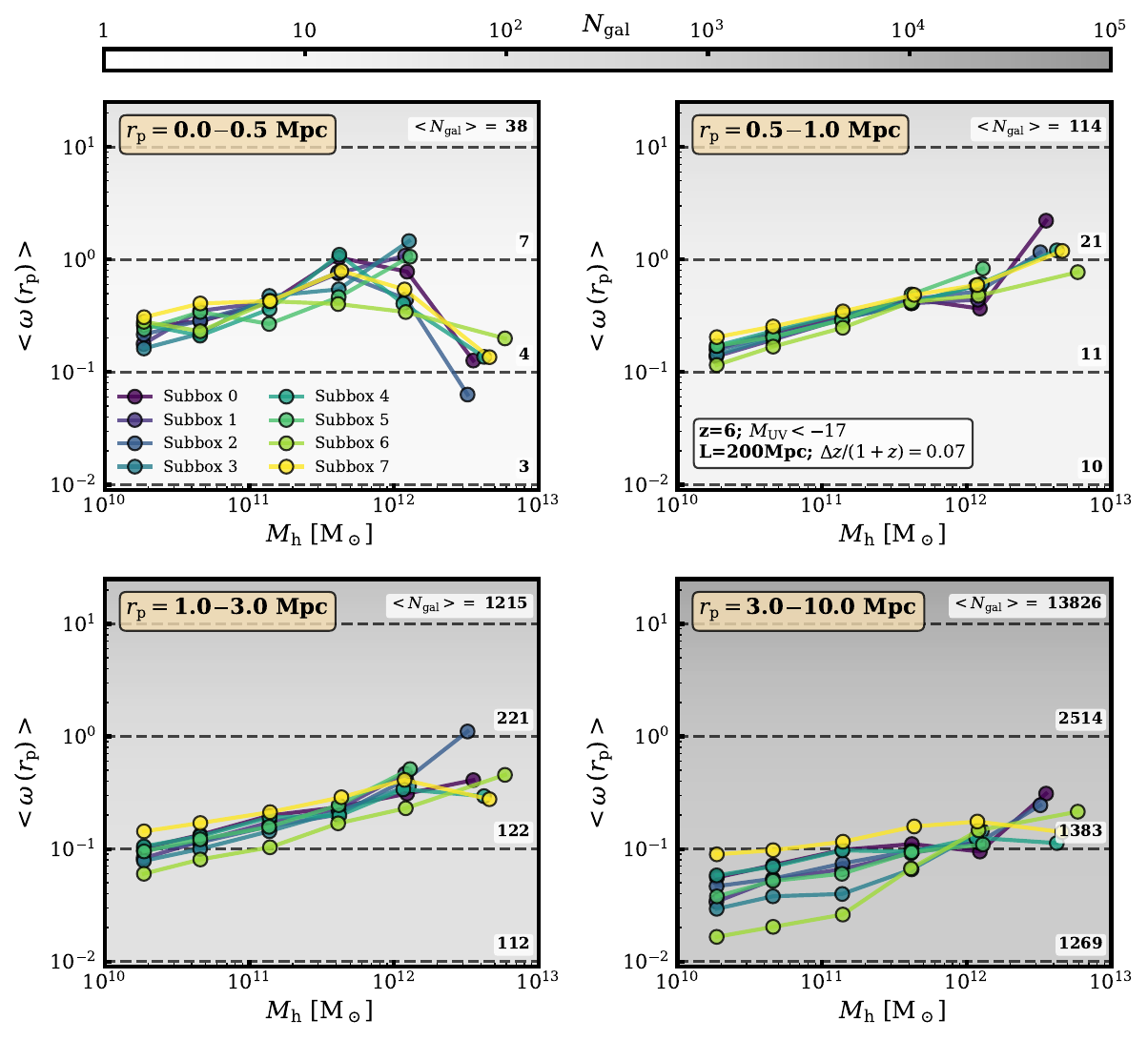}
\vspace{-0.2cm}
\caption{Two-point correlation function and its cosmic variance for mock galaxies across eight survey volumes of $(200\,\text{cMpc})^3$ each, averaged over the projected radial bins indicated in each panel. The mock galaxies are constructed from the COLIBRE simulation (L400m7; Sect.~\ref{ssec_COLIBRE}) combined with the HOMA empirical model (Saleh et al. in prep.; Sect.~\ref{ssec_model}), selected at redshift 6 as an example with UV magnitudes brighter than $M_{\rm UV, cut}=-17$. The measurements are integrated with a redshift uncertainty of $\Delta z/(1+z)=0.07$. The number above each dashed line indicates the average number of mock galaxies corresponding to ${<}\omega{>}$ for the given volume and radial range. Our results demonstrate a significant impact of volume-to-volume variation on clustering measurements (and thus on halo mass estimation), even for survey volumes much larger than those of typical high-$z$ observations currently available. The variance among volumes is smallest for projected distances between 0.5 and 1\,cMpc, offering an optimal range for clustering measurements to minimize bias.}
\label{fig_CV}
\end{figure*}

\subsection{Impact of cosmic variance on SHMR inference}
\label{ssec_CV}

As discussed briefly in Sect.~\ref{ssec_wp_Mh}, cosmic variance emerges as one of the dominant uncertainties in the clustering matching method. Its impact is also scale-dependent, influencing the choice of optimal $r_{\rm p}$ range for measuring clustering strengths while minimizing cosmic variance.

Fig.~\ref{fig_CV} illustrates this effect by showing the average $w_{\rm p}$ as a function of halo mass across a wide range of $r_{\rm p}$ at $z=6$. To obtain these results, we divided the full COLIBRE simulation box (L400m7) into eight subvolumes of $(200$\,cMpc$)^3$ each. Then, for each subvolume, we measured the average clustering strengths for the same halo mass bins used in Sect.~\ref{ssec_SHMR}, adopting a selection of $M_{\rm UV} < -17$ and redshift error $\Delta z/(1+z) = 0.07$ as an example. On small scales of $r_{\rm p}\lesssim 0.5$\,cMpc, the clustering signal is influenced by non-linear effects such as the one-halo term and satellite galaxy distributions, where the linear bias approximation commonly adopted in HOD modelling breaks down as noted in previous studies \citep[e.g.,][]{Jose2017, Shuntov2025b, Weibel2025}. This can lead to systematic differences in inferred halo masses compared to our template-based approach on larger scales. This makes clustering measurements within the virial radius suboptimal for halo mass estimation. At large scales ($r_{\rm p} > 3$ cMpc), cosmic variance becomes severe, with $w_{\rm p}$ varying by nearly an order of magnitude between different subvolumes. This is noteworthy because these eight subvolumes are neighboring regions within the same simulation box and thus share some large-scale density modes, potentially underestimating the full range of cosmic variance. Specifically, from linear theory, the rms variance in the matter fluctuation on the scale of the full simulation box ($400\,\mathrm{cMpc}$) is about one third of that on the scale of each subvolume ($200\,\mathrm{cMpc}$). Assuming a scale-independent linear halo bias, this implies that the volume-to-volume variance estimated from the eight COLIBRE subvolumes is likely underestimated by roughly this factor. This contrasts with, for example, the $1\,\mathrm{cGpc}$ simulation boxes used in \citet{Lim2024, Lim2025b}, which were divided into 1,000 subvolumes of $(100\,\mathrm{cMpc})^3$ each. In that case, the matter variance on the scale of the full box is only about 2 per cent of that on the subvolume scale, and the full representative variance is therefore captured well. We find that the intermediate range $0.5 < r_{\rm p}/\mathrm{cMpc} < 1.0$—our fiducial choice—strikes an optimal balance, minimizing both small-scale nonlinear effects and large-scale cosmic variance. Fig.~\ref{fig_appC} in Appendix~\ref{sec_appC} presents a more direct illustration of this, comparing the performance of our method across different radial bins.

\begin{figure*}
\includegraphics[width=1.\linewidth]{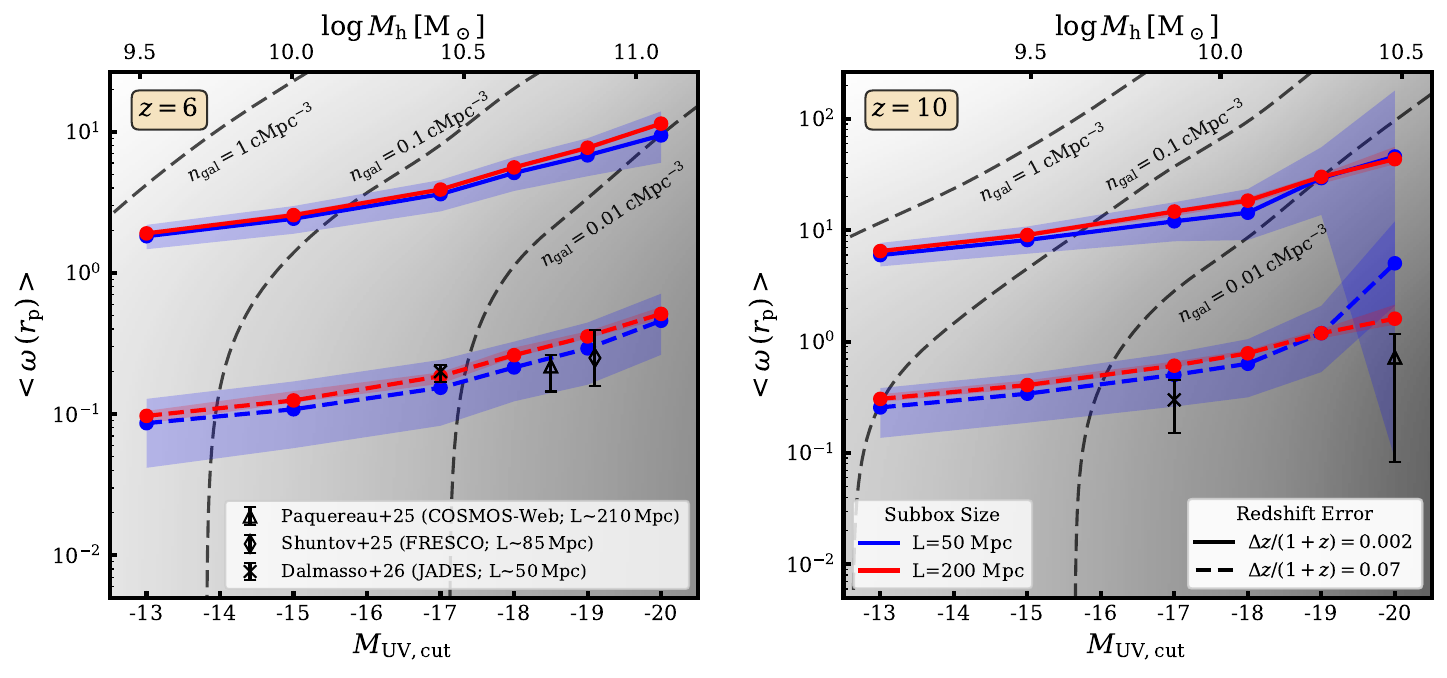}
\vspace{-0.2cm}
\caption{Volume-to-volume variance of the two-point correlation function for mock galaxies, averaged over projected distances between 0.5 and 1\,cMpc, under various magnitude cuts and survey configurations at redshifts 6 (left) and 10 (right). The lines and bands indicate the median and 1-sigma ranges, respectively. The black dashed lines (and the gray color gradient in the background) indicate the comoving number density of mock galaxies corresponding to ${<}\omega{>}$. Our results indicate a potentially significant impact of volume-to-volume variation on clustering measurements and halo mass estimation for surveys with small volume ($\sim 50$\,cMpc) and high redshift error ($\Delta z/(1+z)\simeq 0.07$). For comparison, observational data (\citealt{Paquereau2025, Shuntov2025b, Dalmasso2026}) are presented.}
\label{fig_CV2}
\end{figure*}

Despite the different physical origins of the scatter arising from the one- and two-halo terms, Fig.~\ref{fig_CV} reveals a positive covariance in clustering strengths across scales (for $M_{\rm h}\lesssim 10^{11.5}\,{\rm M}_\odot$). This positive covariance is consistent with the recent findings of \citet{Huang2026}, who showed that volume-averaged correlation functions exhibit significant off-diagonal correlations that increase with halo mass. In practice, this means that large-scale density fluctuations coherently affect multiple radial bins, so treating bins as independent (as Poisson errors do) underestimates the true uncertainty.

This coherent behaviour across scales has direct implications for the interpretation of clustering measurements. The clustering strength at fixed $M_{\rm h}$ within a given volume correlates with the overall number density of selected galaxies in that volume: volumes with higher galaxy number densities exhibit stronger clustering at fixed mass. Recently, \citet{Lim2025b} reported theoretical predictions from the FLAMINGO simulations (\citealt{Schaye2023, Kugel2023}) showing that galaxy formation and abundance at $z \gtrsim 5$ are strongly correlated out to scales $>100$ cMpc. \citet{Huang2026} further found that the impact of cosmic variance (beyond Poisson noise) on galaxy clustering at $z\sim6$ extends to even larger scales, up to $\sim 700$\,cMpc. Our finding of significant clustering variance at $200$\,cMpc scales supports the notion that early galaxy formation is correlated over scales much larger than typical high-redshift survey volumes. 

We find that the impact of cosmic variance on clustering is not strongly dependent on magnitude cut or redshift, whereas survey volume and redshift uncertainty are significant factors, as shown in Fig.~\ref{fig_CV2}. For a volume of $(50$ cMpc$)^3$, $w_{\rm p}$ at fixed mass typically varies by a factor of $\sim 3$ for photometric samples and $\sim 2$ for spectroscopic samples. The level of cosmic variance remains similar across $z \simeq 4$ to $12$ for a given volume and redshift uncertainty. When averaged over our fiducial $0.5 < r_{\rm p} < 1.0$ cMpc range, the \citet{Paquereau2025}, \citet{Shuntov2025b}, and \citet{Dalmasso2026} measurements lie well within the expected cosmic variance (Fig.~\ref{fig_CV2}), indicating that the discrepancy in inferred halo mass is largely attributable to the different radial range combined with cosmic variance.

\subsection{Application to observational data}
\label{ssec_obs}

As demonstrated in Sect.~\ref{ssec_wp_Mh}, applying our clustering matching method with fiducial configuration to the JADES photometric sample at $z \simeq 6$ yields an inferred halo mass of $\log M_{\rm h} /{\rm M}_\odot  = 10.5$, indicating that the galaxies selected with $M_{\rm UV} < -17$ reside, on average, in haloes of that mass. Notably, however, this estimate deviates from the mean halo mass of the same selection calculated directly from the mock catalog, which is $\log M_{\rm h} /{\rm M}_\odot  = 10.72$, or $0.2$ dex higher. Three possibilities may explain this discrepancy.

First, cosmic variance, as discussed in the previous subsection, could play a role. The typical scatter of $\sim 0.25$ dex in $w_{\rm p}$ arising from cosmic variance for the JADES volume with photometric redshift errors (Fig.~\ref{fig_CV2}), combined with the $\sim 0.5$ slope between $\log w_{\rm p}$ and $\log M_{\rm h}$, translates to an estimated halo mass range of $\log M_{\rm h} /{\rm M}_\odot  \simeq 10-11$. If the JADES survey volume is a mildly under-dense region where clustering at fixed halo mass is systematically weaker, the discrepancy could be readily reconciled.

Second, the distribution of clustering strengths for selected samples may not be simply related to the average halo mass; specifically, $\langle w_{\rm p}(M_{\rm h})\rangle \neq w_{\rm p}(\langle M_{\rm h}\rangle)$. We find that the $w_{\rm p}$ distribution at fixed halo mass is skewed towards higher values, with the median lying below the mean. This asymmetry results in halo masses being systematically underestimated when inferred via clustering matching.

\begin{figure*}
\includegraphics[width=1.\linewidth]{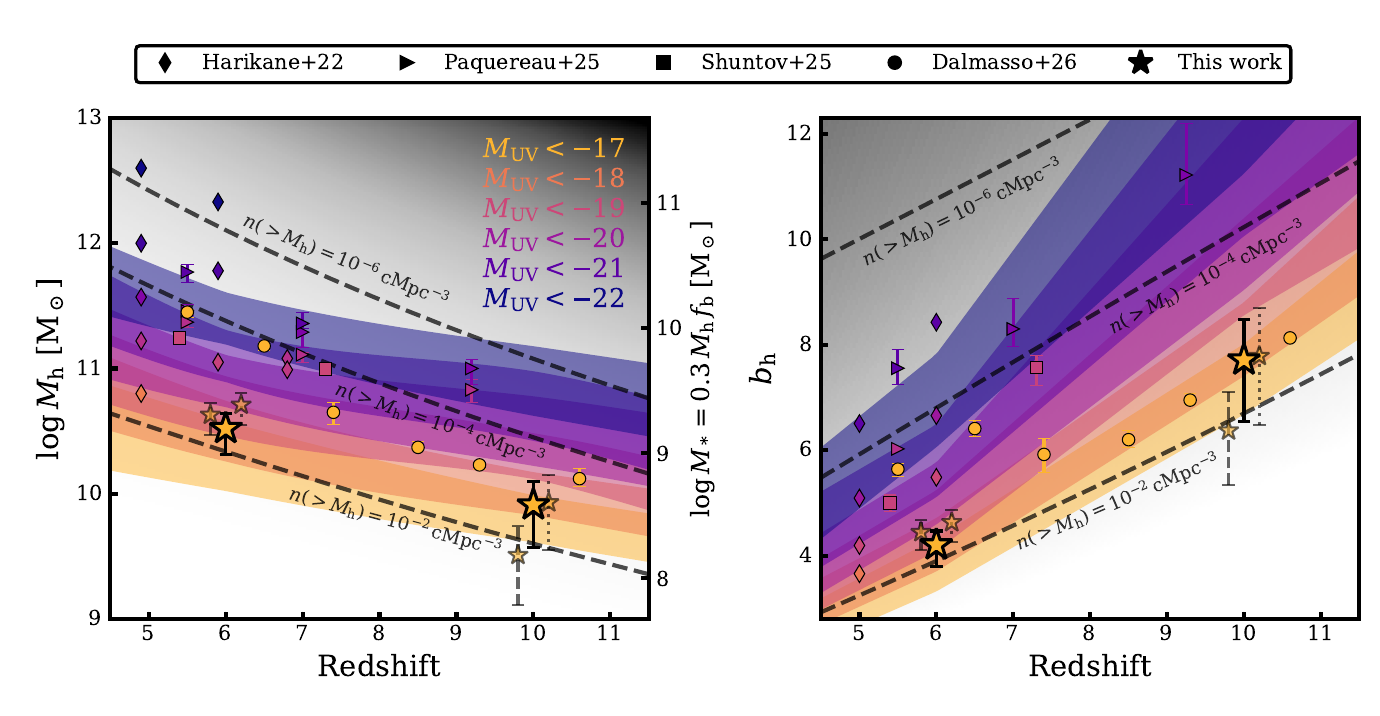}
\vspace{-0.2cm}
\caption{Observational inference of halo mass (left) and linear halo bias (right) using our clustering matching method applied to the JADES photometric samples of \citet{Dalmasso2026}. The linear bias is calculated from the halo mass using the relation of \citet{Tinker2010} (their Eq.~6). The large star symbols with solid errorbars indicate the results obtained using the fiducial HOMA model, while the smaller stars with dashed errorbars represent results from a HOMA variant without the global boost in star formation efficiency, and the smaller stars with dotted errorbars represent results from a variant with increased stochasticity (see Sect.~\ref{ssec_model}). The model variants are horizontally offset by $-0.2$ and $+0.2$\,dex, respectively, for clarity. The colored bands show the 1-sigma percentile range of mock galaxies selected with the given UV magnitude cuts, obtained at each redshift and interpolated in between. The dashed lines and gray color gradient indicate the number of haloes above a given mass. Observational data from \citet{Harikane2022}, \citet{Paquereau2025}, \citet{Shuntov2025b}, and \citet{Dalmasso2026} are shown for comparison. Note that our estimates, despite using the same data, are significantly lower than those of \citet{Dalmasso2026} for reasons discussed in the text.}
\label{fig_obs}
\end{figure*}

Third, the discrepancy may indicate that clustering provides complementary information to luminosity (or stellar mass) functions, helping to tighten model constraints. As explained in Sect.~\ref{ssec_model}, the model used in our analysis was calibrated to the UVLF at $z\simeq6$. An intrinsic disagreement between clustering-inferred halo masses and the representative true halo masses may therefore reflect the ability of clustering to break degeneracies in models constrained solely by abundances. When a model populates haloes with brighter galaxies or adopts a higher SFE, the average halo mass of samples passing a UV magnitude or stellar mass cut decreases, as lower-mass haloes now meet the selection criteria. However, as shown in Fig.~\ref{fig_wp_Mh}, when lower-mass objects enter the sample, the clustering strength at fixed halo mass decreases, leading to a higher halo mass inferred from matching to observational data. Thus, as the true and clustering-inferred halo masses vary in opposite directions with changing the mapping of star formation onto haloes, models can be constrained by requiring the two mass estimates to be consistent.

In Fig.~\ref{fig_obs}, we extend this observational inference to higher redshift, applying our method to the JADES clustering measurement at $z \simeq 10$ from \citet{Dalmasso2026}. The best-fitting result indicates that $z \simeq 10$ galaxies with $M_{\rm UV} < -17$ in the JADES photometric data ($\Delta z \simeq 0.5$) are typically associated with haloes of $\log M_{\rm h}/{\rm M}_\odot  \simeq 9.9$. Again, this is about $0.1$ dex lower than the mean halo mass from the mock catalog ($\log M_{\rm h} /{\rm M}_\odot  \simeq 10$), similar to the offset at $z=6$. The estimate is also below that of \citet{Dalmasso2026} by approximately $0.3$ dex. As discussed in Sect.~\ref{ssec_wp_Mh}, this likely reflects the different $r_{\rm p}$ ranges used for constraining power: the \citet{Dalmasso2026} measurements have smallest uncertainties around $250$ arcsec, which is not only an order of magnitude larger than our fiducial range but also highly sensitive to cosmic variance (Sect.~\ref{ssec_CV}), leading to less accurate halo mass determination. When we fit our clustering template to the same $r_{\rm p}$ range of $\sim 12$\,cMpc (corresponding to $250$ arcsec at $z\simeq 10$), we obtain a much higher inferred halo mass of $\log M_{\rm h}/{\rm M}_\odot \simeq 10.3$, closer to the \citet{Dalmasso2026} estimate. However, constraints derived primarily from such large separations remain highly susceptible to cosmic variance and are therefore suboptimal for inferring associated halo masses. Independent estimations via clustering measurements, such as \citet{Harikane2022}, \citet{Paquereau2025}, and \citet{Shuntov2025b}, have halo masses well aligned with those of \citet{Dalmasso2026} across redshifts. However, most of those observations targeted much brighter samples than $M_{\rm UV}<-17$ used by \citet{Dalmasso2026}.

From the inferred halo masses, we can also compute the linear halo bias $b_{\rm h}$ using, for example, the simulation-calibrated relation from \citet{Tinker2010} (their Eq.~6). Applying this to our best-fit halo masses yields $b_{\rm h} = 4.2$ and $7.7$ for the JADES samples at $z=6$ and $10$, respectively. Compared with the values reported by \citet{Dalmasso2026}, our estimates are lower by approximately a factor of $2$ and $0.5$ at $z \simeq 6$ and $10$, respectively (Fig.~\ref{fig_obs}). As with the halo mass inference, this discrepancy likely stems from the different projected distance ranges used for constraining the clustering signal. We further compare our results with number-weighted linear halo bias measurements from various high-redshift surveys, including \citet{Harikane2022}, \citet{Paquereau2025}, and \citet{Shuntov2025b}. Although most of these observational samples target brighter galaxies than $M_{\rm UV} < -17$, the substantial variation in bias estimates with selection highlights the sensitivity of bias to the adopted magnitude cut. This underscores the importance of accurate modelling and proper accounting for uncertainties, including cosmic variance, when interpreting clustering-based measurements.

Through the analyses presented here and in previous sections, we have demonstrated that our clustering matching technique can infer halo masses for observed galaxies up to $z \simeq 12$ with essentially zero bias and typical uncertainties of $\sim 0.3$ dex, even for photometric data. We have characterized the impact of cosmic variance, identified the optimal radial range for clustering-based inference, and illustrated how the method can be applied to real observational data using the JADES photometric samples as an example.

It is worth noting, however, that the stellar masses used in our analysis are derived assuming a Chabrier IMF, following both the COLIBRE simulation and the HOMA model. The choice of IMF can introduce systematic uncertainties in the inferred stellar masses and, consequently, in the clustering-based halo mass estimates. To illustrate the potential magnitude of this effect, we compare our results with the predictions of the GALFORM semi-analytic model \citep{LuS2025}, which adopts a top-heavy IMF for starburst episodes. At $z\sim10$, for galaxies with $M_{\rm UV}\lesssim-18$, our method infers halo masses of $\log M_{\rm h}/{\rm M}_\odot \sim 10.3$ (see Fig.~\ref{fig_obs}). In contrast, the GALFORM model of \citet{LuS2025} predicts that galaxies of similar UV brightness reside in haloes of $\log M_{\rm h}/{\rm M}_\odot \sim 9.7$. This $\sim0.6$ dex difference arises primarily because the top-heavy IMF produces lower stellar masses for a given UV luminosity. While a detailed comparison with this specific model is beyond the scope of this work, this example highlights the sensitivity of clustering-based halo mass estimates to the assumed IMF, and motivates future work incorporating variable IMFs into both simulations and empirical models.

\subsection{Constraining SFE models using clustering measurements}
\label{ssec_constrain}

As described in Sect.~\ref{sec_intro}, the clustering strength of a galaxy population is closely tied to its bias relative to the underlying dark matter distribution \citep[e.g.,][]{MoWhite1996}, offering a promising avenue for inferring the associated halo mass even at high redshifts. When combined with one-point statistics such as the UVLF and SMF, this two-point statistic may break degeneracies between different models proposed to explain the high abundance of bright galaxies observed in the early Universe \citep[e.g.,][]{Mirocha2020, Sun2025, Munoz2026}.

Recent JWST clustering analyses have provided critical constraints on the SFE--halo mass relation and the relative importance of burstiness versus global SFE enhancement at high redshifts. \citet{Shuntov2025b} simultaneously fitted the two-point correlation functions and UVLFs of H$\alpha$ and [\textsc{Oiii}] emitters at $3.8<z<9$, finding that the SFE peaks at $\sim20\%$ at $M_{\rm h}\sim3\times10^{11}\,{\rm M}_\odot$ and declines towards higher masses. They measured a modest scatter of $\sigma_{\rm UV}\sim0.7$ that remains roughly constant with redshift, indicating limited burstiness in the SFE over this redshift range.

\citet{Paquereau2025} performed a comprehensive HOD analysis of mass-limited clustering in the COSMOS-Web field from $z=0.1$ to $z\sim12$. They found that galaxies at $z\gtrsim8$ are hosted by haloes of $M_{\rm h}\sim10^{10.5}\,{\rm M}_\odot$ with SFE up to 1 dex higher than at lower redshifts. The high galaxy bias ($b_{\rm g}>8$) at $z\gtrsim8$ strongly favours the global SFE boost scenario over stochastic SHMR models, ruling out burstiness as the primary driver of the high abundance of massive high-$z$ galaxies.

Similarly, \citet{Dalmasso2026} performed HOD modelling on $\sim6500$ LBGs at $5\leq z<11$, revealing that galaxies at $z\sim10.6$ reside in haloes of $M_{\rm h}\sim10^{10.12}\,{\rm M}_\odot$ with effective bias reaching $b_{\rm g}=8.13$. This demonstrates that high-$z$ galaxies require a global boost in SFE rather than stochasticity, as the $\sigma_{\rm UV}$ scatter alone cannot explain the observed clustering amplitude.

In Fig.~\ref{fig_obs}, we present the halo mass and bias estimates obtained by fitting the clustering measurements of \citet{Dalmasso2026} with the three HOMA model variants (fiducial, no global boost, and burstier) to explore whether clustering can discriminate between different physical scenarios at high redshifts. The variants yield estimates that agree within uncertainties, indicating that current data cannot distinguish between the models. As discussed in Sect.~\ref{ssec_obs}, a model that associates galaxies of the same brightness with systematically lower- (higher-)mass haloes results in a higher (lower) average halo mass inferred from the same clustering strength. For example, the higher $M_{\rm h}$ values estimated at $z\sim6$ with models of globally less efficient and more stochastic star formation in Fig.~\ref{fig_obs} imply that these models associate mock galaxies with lower-mass haloes.

\begin{figure}
\includegraphics[width=1.\linewidth]{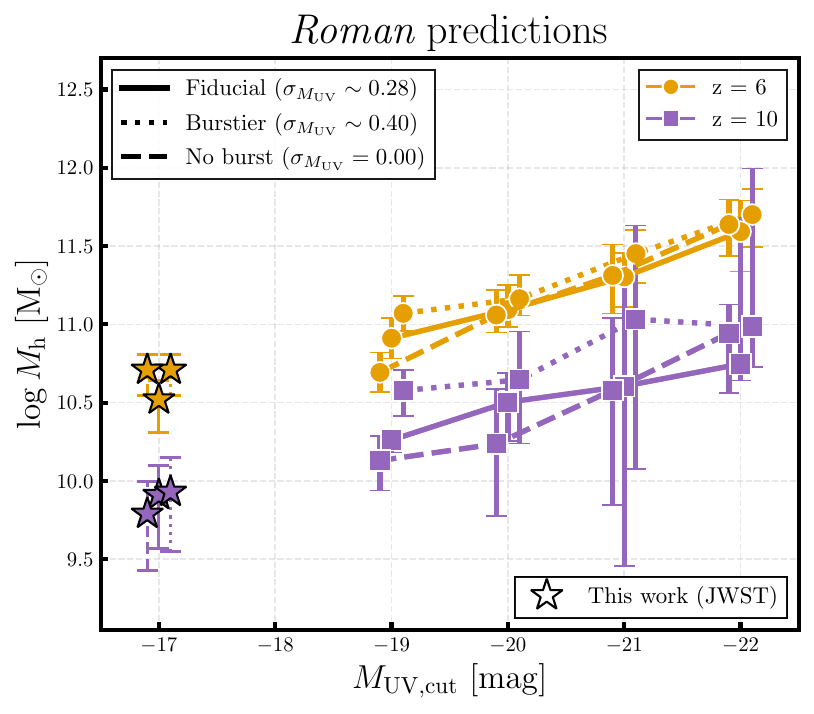}
\vspace{-0.4cm}
\caption{Prediction for the inferred halo mass from the \textit{Roman} Space Telescope using our clustering matching method, for three HOMA model variants with different levels of UV stochasticity (Sect.~\ref{ssec_model}). Results are shown for the Deep Tier fields ($19.2\,{\rm deg}^2$) over a range of $M_{\rm UV}$ cuts, assuming a redshift uncertainty of $\Delta z/(1+z)=0.07$. The 5$\sigma$ point-source depth of 27.5 AB mag in the F158 (H-band) filter corresponds to $M_{\rm UV}\sim-19.9$ at $z\sim6$ and $M_{\rm UV}\sim-21.7$ at $z\sim10$, which we adopt for the galaxy and pair count estimates. The comoving volume is $\simeq1.8\times10^8\,\mathrm{cMpc}^3$. The survey volume is constructed from the COLIBRE simulation as a rectangular box with transverse area corresponding to 19.2 deg$^2$ and line-of-sight depth determined by the comoving volume, with periodic replication when necessary. The HOMA model predicts that the number of pairs within the cylinders used for clustering ($\Delta z$ and $0.5<r_{\rm p}/\mathrm{cMpc}<1.0$) is $\sim10^5$ at $z\sim6$ and $\sim20$ at $z\sim10$ at these depth limits. The halo mass error bars are estimated from the covariance matrix of the clustering measurements across the radial bins, obtained via jackknife resampling of the simulation volume, and propagated to the inferred halo mass through the template matching. For clarity, models with zero (dashed) and enhanced (dotted) star-formation stochasticity are horizontally offset by $0.1$ and $-0.1$\,mag, respectively. For comparison, we overplot our best-fitting results to the JWST measurements of \citet{Dalmasso2026} (star symbols).}
\label{fig_constrain}
\end{figure}

To assess the constraining power of future observations, we use the planned Deep Tier fields of the \textit{Roman} Space Telescope (\citealt{Green2012, Spergel2013, Spergel2015}) High-Latitude Wide Area Survey, which covers a total area of 19.2 deg$^2$. We adopt the 5$\sigma$ point-source depth of 27.5 AB mag in the F158 (H-band) filter. We find that, for a typical star-forming galaxy SED, this corresponds to a rest-frame UV absolute magnitude of $M_{\rm UV}\sim-19.9$ at $z\sim6$ and $M_{\rm UV}\sim-21.7$ at $z\sim10$. We assume a redshift uncertainty of $\Delta z/(1+z)=0.07$. For each redshift bin, the comoving volume is $\simeq 1.8\times10^8\,\mathrm{cMpc}^3$. We construct a rectangular box with a transverse area corresponding to 19.2 deg$^2$ at that redshift and a line-of-sight depth equal to the comoving volume divided by that area, using the COLIBRE simulation. When the required volume exceeds the $(400\,\mathrm{cMpc})^3$ simulation box, we periodically replicate the COLIBRE volume to build the box. For the adopted magnitude limits, the HOMA model predicts $\sim1.6\times10^5$ galaxies at $z\sim6$ ($M_{\rm UV}<-19.9$) and $\sim1.8\times10^3$ galaxies at $z\sim10$ ($M_{\rm UV}<-21.7$) within the Deep Tier volume. When counting pairs within cylinders defined by the redshift uncertainty $\Delta z$ and the radial range $0.5<r_{\rm p}/\mathrm{cMpc}<1.0$ used for our clustering analysis, the expected number of pairs is $\sim10^5$ at $z\sim6$ and $\sim20$ at $z\sim10$. The halo mass uncertainties shown in Fig.~\ref{fig_constrain} are derived from the covariance matrix of the clustering signal across the radial bins used in the fit ($0.5<r_{\rm p}/\mathrm{cMpc}<1.0$), estimated via jackknife resampling of the COLIBRE volume and propagated to the inferred halo mass through the template matching procedure. This accounts for both Poisson noise and cosmic variance, as well as the covariance between different radial bins.

In Fig.~\ref{fig_constrain}, we show the resulting projections, illustrating the constraints expected from the various HOMA model variants. Here, instead of the `no global boost' model, we include another variant that maintains a similar global SFE but assumes no stochasticity, in order to isolate and explore the impact of gradually increasing burstiness. As in Fig.~\ref{fig_obs}, the uncertainties are comparable to or larger than the differences between models. While at higher redshift ($z=10$) and for bright galaxies ($M_{\rm UV}\lesssim-21$) the inferred halo mass shows larger separations between models, the expected number of pairs ($\sim20$) limits the constraining power.

\section[summary]{Summary}
\label{sec_summary}

In this work, we have developed and validated a clustering-based method for inferring halo masses of high-redshift galaxies, and applied it to JWST observations from the JADES survey. Our method matches the projected two-point correlation function of galaxies in stellar mass bins to reference templates constructed from haloes of known mass in the COLIBRE simulation, thereby establishing a direct stellar-to-halo mass relation (SHMR). The templates are built from the largest available volume to minimize cosmic variance, and the matching is performed over an optimal radial range $0.5<r_{\rm p}/\mathrm{cMpc}<1.0$, which we identify as balancing sensitivity to halo mass against non-linear small-scale effects and cosmic variance on large scales. The resulting relation is then applied to individual galaxies to assign clustering-matched halo masses, which we validate against the true halo masses from the simulation. Our approach combines the COLIBRE cosmological hydrodynamical simulation (Sect.~\ref{ssec_COLIBRE}) with the empirical model HOMA calibrated to reproduce the observed UV luminosity function at $z\simeq6$ (Sect.~\ref{ssec_model}), enabling us to construct realistic mock galaxy catalogues (Sect.~\ref{ssec_mock}) and systematically quantify the uncertainties arising from observational configurations (Sect.~\ref{ssec_SHMR}). 

We first established reference clustering templates as a function of halo mass (Sect.~\ref{ssec_wp_Mh}), using the full volume of the L400m7 COLIBRE simulation to minimize cosmic variance. We identified an optimal radial range of $0.5 < r_{\rm p}/\mathrm{cMpc} < 1.0$ for clustering-based halo mass inference (Appendix~\ref{sec_appC}), which balances sensitivity to halo mass against both small-scale nonlinear effects and large-scale cosmic variance. Within this range, the relationship between $\log w_{\rm p}$ and $\log M_{\rm h}$ exhibits a slope of approximately $0.5$, providing a well-defined mapping between clustering strength and halo mass (Fig.~\ref{fig_wp_Mh}).

Validation against true halo masses from the mock catalog (Sect.~\ref{ssec_validation}) demonstrates that our method recovers the underlying stellar mass–halo mass relation with minimal bias and typical uncertainties of $\sim 0.07$ dex under optimal conditions of large volume (Fig.~\ref{fig_validation}). When applied to more realistic survey configurations (Fig.~\ref{fig_validation2}), the combined uncertainty remains below $\sim 0.3$ dex for volumes down to $(50$ cMpc$)^3$ and redshifts out to $z=12$, even for photometric samples with redshift errors of $\Delta z/(1+z) \sim 0.07$.

We systematically explored how survey volume, redshift uncertainty, and UV luminosity threshold affect the accuracy and precision of halo mass inference. Survey volume emerges as the dominant factor (Fig.~\ref{fig_contour1}). For volumes smaller than $10^5$ cMpc$^3$ (roughly $(50$ cMpc$)^3$), the combined uncertainty increases to $>0.5$ dex at $z=6$ and $>0.3$ dex at $z=10$. At fixed volume and redshift error, precision improves with increasing redshift, as brighter cuts at higher redshifts select rarer density peaks with intrinsically stronger clustering (Fig.~\ref{fig_contour2}).

A key result of this work is the identification and characterization of cosmic variance as the dominant uncertainty in clustering-based halo mass estimates (Sect.~\ref{ssec_CV}). We find that for a volume of $(50$ cMpc$)^3$, $w_{\rm p}$ at fixed mass varies by a factor of $\sim 3$ for photometric samples and $\sim 2$ for spectroscopic samples (Fig.~\ref{fig_CV2}), with the level of variance remaining similar across $z \simeq 4$ to $12$. The clustering strength at fixed halo mass correlates with the overall number density of selected galaxies within a volume (Fig.~\ref{fig_CV}), indicating that cosmic variance affects both quantities over similar scales. Our finding of significant clustering variance even at $200$\,cMpc scales is consistent with recent predictions from the FLAMINGO simulations \citep{Lim2025b, Huang2026}, supporting the notion that early galaxy formation is correlated over scales much larger than typical high-redshift survey volumes.

We also demonstrated that incorporating secondary galaxy properties can mitigate assembly bias and environmental effects (Sect.~\ref{ssec_properties}). Using SFR, rest-frame colour $(g-r)$, and stellar age as examples, we showed that splitting the sample by these properties and applying our clustering matching method separately to each subsample effectively removes the systematic offsets introduced by environmental dependencies (Fig.~\ref{fig_SFR100}). At fixed stellar mass, galaxies with different recent star formation histories and stellar populations occupy different halo environments, and accounting for these secondary dependencies corrects for the resulting biases in halo mass estimates. This represents a key advantage over methods such as abundance matching, which rely solely on stellar mass or luminosity and cannot easily account for assembly bias or environmental effects.

Furthermore, we validated our method using galaxy and halo properties taken directly from the COLIBRE simulation (Appendix~\ref{sec_appA}). The COLIBRE SHMR exhibits an intrinsic scatter approximately twice as large as that of the HOMA model at fixed stellar mass. Our method recovers the true halo masses with no significant bias and with dispersions that match the input scatter (Fig.~\ref{fig_AppA}), confirming that the clustering-based inference faithfully traces the underlying SHMR dispersion regardless of its amplitude or the specific assumptions of the galaxy formation model.

Applying our method to the JADES photometric samples from \citet{Dalmasso2026} (Sect.~\ref{ssec_obs}), we infer characteristic halo masses of $\log M_{\rm h} /{\rm M}_\odot \simeq 10.5$ and $9.9$ for galaxies with $M_{\rm UV} < -17$ at $z \simeq 6$ and $10$, respectively (Fig.~\ref{fig_obs}). These estimates are approximately $0.2$ dex lower than the mean halo masses from our mock catalog, a discrepancy that can be explained by a combination of cosmic variance, the skewed distribution of clustering strengths at fixed mass, and the different radial scales used in the analysis. When we fit our clustering templates to the larger scales ($r_{\rm p} \gtrsim 4$ cMpc) that provide the tightest constraints in the \citet{Dalmasso2026} analysis, we obtain halo masses consistent with their HOD-based estimates, but caution that such large scales are highly susceptible to cosmic variance and thus suboptimal for precise halo mass determination. From our inferred halo masses, we derive linear halo biases of $b_{\rm h} = 4.2$ and $7.7$ at $z=6$ and $10$, respectively (Fig.~\ref{fig_obs}).

We further explored the power of clustering to distinguish between different physical scenarios for high-redshift star formation (Sect.~\ref{ssec_constrain}) by fitting the clustering measurements with four HOMA model variants (fiducial, no global boost, and less and more burstiness). The variants yield estimates that agree within uncertainties, indicating that current data cannot yet discriminate between these models. However, our forecasts for the \textit{Roman} Deep Tier show that at $z\sim10$, the inferred halo masses for $M_{\rm UV}\lesssim-21$ galaxies differ by $\sim0.5$ dex between bursty and non-bursty models, while the expected number of pairs ($\sim20$) limits the constraining power (Fig.~\ref{fig_constrain}).

In conclusion, our clustering-based method provides a robust framework for inferring halo masses of high-redshift galaxies, with well-characterized uncertainties that account for survey volume, redshift errors, and cosmic variance. The method's application to JWST data yields estimates that are broadly consistent with independent HOD analyses, while offering the advantages of direct empirical calibration and the ability to incorporate secondary galaxy properties. A limitation of our approach is that it does not distinguish between centrals and satellites when assigning halo masses, as it relies on a single SHMR derived from clustering matching. For applications where the central/satellite distinction is important—particularly for small-scale clustering studies where satellite contributions are significant—a more detailed treatment would be required. We have also shown that while current clustering measurements cannot yet distinguish between competing physical models of early star formation, future surveys with reduced errors could break these degeneracies and provide valuable constraints on the nature of star formation and galaxy growth in the early Universe. As JWST continues to deliver deep imaging and spectroscopy of early galaxy populations, our approach will enable increasingly precise constraints on the galaxy–halo connection at the highest redshifts, informing models of galaxy formation and evolution during the epoch of reionization. Future work will extend this analysis to larger samples, incorporate additional galaxy properties, and apply the method to other JWST fields to further constrain cosmic variance and its impact on our understanding of the early Universe. 

\section*{ACKNOWLEDGEMENTS}

SL and RM acknowledge support by the Science and Technology Facilities Council (STFC) and by the UKRI Frontier Research grant RISEandFALL. RM also acknowledges funding from a research professorship from the Royal Society. EC acknowledges support from STFC consolidated grant ST/X001075/1. FH acknowledges funding from the Netherlands Organization for Scientific Research (NWO) through research programme Athena 184.034.002. SL also thanks Kareem El-Badry for insightful comments and discussion. This work used the DiRAC@Durham facility managed by the Institute for Computational Cosmology on behalf of the STFC DiRAC HPC Facility (\url{www.dirac.ac.uk}). The equipment was funded by BEIS capital funding via STFC capital grants ST/K00042X/1, ST/P002293/1, ST/R002371/1 and ST/S002502/1, Durham University and STFC operations grant ST/R000832/1. DiRAC is part of the National e-Infrastructure. This work is partly funded by research programme Athena 184.034.002 from the NWO.

\section*{DATA AVAILABILITY}

The data underlying this article will be shared on reasonable request to the corresponding author.

\bibliographystyle{mnras}
\bibliography{GalEnv.bib}

@BOOK{MovdBWhite2010,
       author = {{Mo}, Houjun and {van den Bosch}, Frank C. and {White}, Simon},
        title = "{Galaxy Formation and Evolution}",
         year = 2010,
       adsurl = {https://ui.adsabs.harvard.edu/abs/2010gfe..book.....M}
}

@ARTICLE{Helton2024b,
       author = {{Helton}, Jakob M. and {Sun}, Fengwu and {Woodrum}, Charity and {Hainline}, Kevin N. and {Willmer}, Christopher N.~A. and {Rieke}, Marcia J. and {Rieke}, George H. and {Alberts}, Stacey and {Eisenstein}, Daniel J. and {Tacchella}, Sandro and {Robertson}, Brant and {Johnson}, Benjamin D. and {Baker}, William M. and {Bhatawdekar}, Rachana and {Bunker}, Andrew J. and {Chen}, Zuyi and {Egami}, Eiichi and {Ji}, Zhiyuan and {Maiolino}, Roberto and {Willott}, Chris and {Witstok}, Joris},
        title = "{Identification of High-redshift Galaxy Overdensities in GOODS-N and GOODS-S}",
      journal = {\apj},
         year = 2024,
        month = oct,
       volume = {974},
       number = {1},
          eid = {41},
        pages = {41},
          doi = {10.3847/1538-4357/ad6867},
archivePrefix = {arXiv},
       eprint = {2311.04270},
 primaryClass = {astro-ph.GA},
       adsurl = {https://ui.adsabs.harvard.edu/abs/2024ApJ...974...41H}
}

@ARTICLE{Schaye2023,
       author = {{Schaye}, Joop and {Kugel}, Roi and {Schaller}, Matthieu and {Helly}, John C. and {Braspenning}, Joey and {Elbers}, Willem and {McCarthy}, Ian G. and {van Daalen}, Marcel P. and {Vandenbroucke}, Bert and {Frenk}, Carlos S. and {Kwan}, Juliana and {Salcido}, Jaime and {Bah{\'e}}, Yannick M. and {Borrow}, Josh and {Chaikin}, Evgenii and {Hahn}, Oliver and {Hu{\v{s}}ko}, Filip and {Jenkins}, Adrian and {Lacey}, Cedric G. and {Nobels}, Folkert S.~J.},
        title = "{The FLAMINGO project: cosmological hydrodynamical simulations for large-scale structure and galaxy cluster surveys}",
      journal = {\mnras},
         year = 2023,
        month = dec,
       volume = {526},
       number = {4},
        pages = {4978-5020},
          doi = {10.1093/mnras/stad2419},
archivePrefix = {arXiv},
       eprint = {2306.04024},
 primaryClass = {astro-ph.CO},
       adsurl = {https://ui.adsabs.harvard.edu/abs/2023MNRAS.526.4978S}
}

@ARTICLE{Kugel2023,
       author = {{Kugel}, Roi and {Schaye}, Joop and {Schaller}, Matthieu and {Helly}, John C. and {Braspenning}, Joey and {Elbers}, Willem and {Frenk}, Carlos S. and {McCarthy}, Ian G. and {Kwan}, Juliana and {Salcido}, Jaime and {van Daalen}, Marcel P. and {Vandenbroucke}, Bert and {Bah{\'e}}, Yannick M. and {Borrow}, Josh and {Chaikin}, Evgenii and {Hu{\v{s}}ko}, Filip and {Jenkins}, Adrian and {Lacey}, Cedric G. and {Nobels}, Folkert S.~J. and {Vernon}, Ian},
        title = "{FLAMINGO: calibrating large cosmological hydrodynamical simulations with machine learning}",
      journal = {\mnras},
         year = 2023,
        month = dec,
       volume = {526},
       number = {4},
        pages = {6103-6127},
          doi = {10.1093/mnras/stad2540},
archivePrefix = {arXiv},
       eprint = {2306.05492},
 primaryClass = {astro-ph.CO},
       adsurl = {https://ui.adsabs.harvard.edu/abs/2023MNRAS.526.6103K}
}

@ARTICLE{Abbott2022,
       author = {{Abbott}, T.~M.~C. and {Aguena}, M. and {Alarcon}, A. and {Allam}, S. and {Alves}, O. and {Amon}, A. and {Andrade-Oliveira}, F. and {Annis}, J. and {Avila}, S. and {Bacon}, D. and {Baxter}, E. and {Bechtol}, K. and {Becker}, M.~R. and {Bernstein}, G.~M. and {Bhargava}, S. and {Birrer}, S. and {Blazek}, J. and {Brandao-Souza}, A. and {Bridle}, S.~L. and {Brooks}, D. and {Buckley-Geer}, E. and {Burke}, D.~L. and {Camacho}, H. and {Campos}, A. and {Carnero Rosell}, A. and {Carrasco Kind}, M. and {Carretero}, J. and {Castander}, F.~J. and {Cawthon}, R. and {Chang}, C. and {Chen}, A. and {Chen}, R. and {Choi}, A. and {Conselice}, C. and {Cordero}, J. and {Costanzi}, M. and {Crocce}, M. and {da Costa}, L.~N. and {da Silva Pereira}, M.~E. and {Davis}, C. and {Davis}, T.~M. and {De Vicente}, J. and {DeRose}, J. and {Desai}, S. and {Di Valentino}, E. and {Diehl}, H.~T. and {Dietrich}, J.~P. and {Dodelson}, S. and {Doel}, P. and {Doux}, C. and {Drlica-Wagner}, A. and {Eckert}, K. and {Eifler}, T.~F. and {Elsner}, F. and {Elvin-Poole}, J. and {Everett}, S. and {Evrard}, A.~E. and {Fang}, X. and {Farahi}, A. and {Fernandez}, E. and {Ferrero}, I. and {Fert{\'e}}, A. and {Fosalba}, P. and {Friedrich}, O. and {Frieman}, J. and {Garc{\'\i}a-Bellido}, J. and {Gatti}, M. and {Gaztanaga}, E. and {Gerdes}, D.~W. and {Giannantonio}, T. and {Giannini}, G. and {Gruen}, D. and {Gruendl}, R.~A. and {Gschwend}, J. and {Gutierrez}, G. and {Harrison}, I. and {Hartley}, W.~G. and {Herner}, K. and {Hinton}, S.~R. and {Hollowood}, D.~L. and {Honscheid}, K. and {Hoyle}, B. and {Huff}, E.~M. and {Huterer}, D. and {Jain}, B. and {James}, D.~J. and {Jarvis}, M. and {Jeffrey}, N. and {Jeltema}, T. and {Kovacs}, A. and {Krause}, E. and {Kron}, R. and {Kuehn}, K. and {Kuropatkin}, N. and {Lahav}, O. and {Leget}, P. -F. and {Lemos}, P. and {Liddle}, A.~R. and {Lidman}, C. and {Lima}, M. and {Lin}, H. and {MacCrann}, N. and {Maia}, M.~A.~G. and {Marshall}, J.~L. and {Martini}, P. and {McCullough}, J. and {Melchior}, P. and {Mena-Fern{\'a}ndez}, J. and {Menanteau}, F. and {Miquel}, R. and {Mohr}, J.~J. and {Morgan}, R. and {Muir}, J. and {Myles}, J. and {Nadathur}, S. and {Navarro-Alsina}, A. and {Nichol}, R.~C. and {Ogando}, R.~L.~C. and {Omori}, Y. and {Palmese}, A. and {Pandey}, S. and {Park}, Y. and {Paz-Chinch{\'o}n}, F. and {Petravick}, D. and {Pieres}, A. and {Plazas Malag{\'o}n}, A.~A. and {Porredon}, A. and {Prat}, J. and {Raveri}, M. and {Rodriguez-Monroy}, M. and {Rollins}, R.~P. and {Romer}, A.~K. and {Roodman}, A. and {Rosenfeld}, R. and {Ross}, A.~J. and {Rykoff}, E.~S. and {Samuroff}, S. and {S{\'a}nchez}, C. and {Sanchez}, E. and {Sanchez}, J. and {Sanchez Cid}, D. and {Scarpine}, V. and {Schubnell}, M. and {Scolnic}, D. and {Secco}, L.~F. and {Serrano}, S. and {Sevilla-Noarbe}, I. and {Sheldon}, E. and {Shin}, T. and {Smith}, M. and {Soares-Santos}, M. and {Suchyta}, E. and {Swanson}, M.~E.~C. and {Tabbutt}, M. and {Tarle}, G. and {Thomas}, D. and {To}, C. and {Troja}, A. and {Troxel}, M.~A. and {Tucker}, D.~L. and {Tutusaus}, I. and {Varga}, T.~N. and {Walker}, A.~R. and {Weaverdyck}, N. and {Wechsler}, R. and {Weller}, J. and {Yanny}, B. and {Yin}, B. and {Zhang}, Y. and {Zuntz}, J. and {DES Collaboration}},
        title = "{Dark Energy Survey Year 3 results: Cosmological constraints from galaxy clustering and weak lensing}",
      journal = {\prd},
         year = 2022,
        month = jan,
       volume = {105},
       number = {2},
          eid = {023520},
        pages = {023520},
          doi = {10.1103/PhysRevD.105.023520},
archivePrefix = {arXiv},
       eprint = {2105.13549},
 primaryClass = {astro-ph.CO},
       adsurl = {https://ui.adsabs.harvard.edu/abs/2022PhRvD.105b3520A}
}

@ARTICLE{Schaller2024,
       author = {{Schaller}, Matthieu and {Borrow}, Josh and {Draper}, Peter W. and {Ivkovic}, Mladen and {McAlpine}, Stuart and {Vandenbroucke}, Bert and {Bah{\'e}}, Yannick and {Chaikin}, Evgenii and {Chalk}, Aidan B.~G. and {Chan}, Tsang Keung and {Correa}, Camila and {van Daalen}, Marcel and {Elbers}, Willem and {Gonnet}, Pedro and {Hausammann}, Lo{\"\i}c and {Helly}, John and {Hu{\v{s}}ko}, Filip and {Kegerreis}, Jacob A. and {Nobels}, Folkert S.~J. and {Ploeckinger}, Sylvia and {Revaz}, Yves and {Roper}, William J. and {Ruiz-Bonilla}, Sergio and {Sandnes}, Thomas D. and {Uyttenhove}, Yolan and {Willis}, James S. and {Xiang}, Zhen},
        title = "{SWIFT: A modern highly-parallel gravity and smoothed particle hydrodynamics solver for astrophysical and cosmological applications}",
      journal = {\mnras},
         year = 2024,
        month = may,
       volume = {530},
       number = {2},
        pages = {2378-2419},
          doi = {10.1093/mnras/stae922},
archivePrefix = {arXiv},
       eprint = {2305.13380},
 primaryClass = {astro-ph.IM},
       adsurl = {https://ui.adsabs.harvard.edu/abs/2024MNRAS.530.2378S}
}

@ARTICLE{Chabrier2003,
       author = {{Chabrier}, Gilles},
        title = "{Galactic Stellar and Substellar Initial Mass Function}",
      journal = {\pasp},
         year = 2003,
        month = jul,
       volume = {115},
       number = {809},
        pages = {763-795},
          doi = {10.1086/376392},
archivePrefix = {arXiv},
       eprint = {astro-ph/0304382},
 primaryClass = {astro-ph},
       adsurl = {https://ui.adsabs.harvard.edu/abs/2003PASP..115..763C}
}

@ARTICLE{BoothSchaye2009,
       author = {{Booth}, C.~M. and {Schaye}, Joop},
        title = "{Cosmological simulations of the growth of supermassive black holes and feedback from active galactic nuclei: method and tests}",
      journal = {\mnras},
         year = 2009,
        month = sep,
       volume = {398},
       number = {1},
        pages = {53-74},
          doi = {10.1111/j.1365-2966.2009.15043.x},
archivePrefix = {arXiv},
       eprint = {0904.2572},
 primaryClass = {astro-ph.CO},
       adsurl = {https://ui.adsabs.harvard.edu/abs/2009MNRAS.398...53B}
}

@ARTICLE{Bahe2022,
       author = {{Bah{\'e}}, Yannick M. and {Schaye}, Joop and {Schaller}, Matthieu and {Bower}, Richard G. and {Borrow}, Josh and {Chaikin}, Evgenii and {Kugel}, Roi and {Nobels}, Folkert and {Ploeckinger}, Sylvia},
        title = "{The importance of black hole repositioning for galaxy formation simulations}",
      journal = {\mnras},
         year = 2022,
        month = oct,
       volume = {516},
       number = {1},
        pages = {167-184},
          doi = {10.1093/mnras/stac1339},
archivePrefix = {arXiv},
       eprint = {2109.01489},
 primaryClass = {astro-ph.GA},
       adsurl = {https://ui.adsabs.harvard.edu/abs/2022MNRAS.516..167B}
}

@ARTICLE{Davis1985,
       author = {{Davis}, M. and {Efstathiou}, G. and {Frenk}, C.~S. and {White}, S.~D.~M.},
        title = "{The evolution of large-scale structure in a universe dominated by cold dark matter}",
      journal = {\apj},
         year = 1985,
        month = may,
       volume = {292},
        pages = {371-394},
          doi = {10.1086/163168},
       adsurl = {https://ui.adsabs.harvard.edu/abs/1985ApJ...292..371D}
}

@ARTICLE{Oesch2023, 
       author = {{Oesch}, P.~A. and {Brammer}, G. and {Naidu}, R.~P. and {Bouwens}, R.~J. and {Chisholm}, J. and {Illingworth}, G.~D. and {Matthee}, J. and {Nelson}, E. and {Qin}, Y. and {Reddy}, N. and {Shapley}, A. and {Shivaei}, I. and {van Dokkum}, P. and {Weibel}, A. and {Whitaker}, K. and {Wuyts}, S. and {Covelo-Paz}, A. and {Endsley}, R. and {Fudamoto}, Y. and {Giovinazzo}, E. and {Herard-Demanche}, T. and {Kerutt}, J. and {Kramarenko}, I. and {Labbe}, I. and {Leonova}, E. and {Lin}, J. and {Magee}, D. and {Marchesini}, D. and {Maseda}, M. and {Mason}, C. and {Matharu}, J. and {Meyer}, R.~A. and {Neufeld}, C. and {Prieto Lyon}, G. and {Schaerer}, D. and {Sharma}, R. and {Shuntov}, M. and {Smit}, R. and {Stefanon}, M. and {Wyithe}, J.~S.~B. and {Xiao}, M.},
        title = "{The JWST FRESCO survey: legacy NIRCam/grism spectroscopy and imaging in the two GOODS fields}",
      journal = {\mnras},
         year = 2023,
        month = oct,
       volume = {525},
       number = {2},
        pages = {2864-2874},
          doi = {10.1093/mnras/stad2411},
archivePrefix = {arXiv},
       eprint = {2304.02026},
 primaryClass = {astro-ph.GA},
       adsurl = {https://ui.adsabs.harvard.edu/abs/2023MNRAS.525.2864O}
}

@ARTICLE{Behroozi2019,
       author = {{Behroozi}, Peter and {Wechsler}, Risa H. and {Hearin}, Andrew P. and {Conroy}, Charlie},
        title = "{UNIVERSEMACHINE: The correlation between galaxy growth and dark matter halo assembly from z = 0-10}",
      journal = {\mnras},
         year = 2019,
        month = sep,
       volume = {488},
       number = {3},
        pages = {3143-3194},
          doi = {10.1093/mnras/stz1182},
archivePrefix = {arXiv},
       eprint = {1806.07893},
 primaryClass = {astro-ph.GA},
       adsurl = {https://ui.adsabs.harvard.edu/abs/2019MNRAS.488.3143B}
}

@ARTICLE{Boylan-Kolchin2023,
       author = {{Boylan-Kolchin}, Michael},
        title = "{Stress testing {\ensuremath{\Lambda}}CDM with high-redshift galaxy candidates}",
      journal = {Nature Astronomy},
         year = 2023,
        month = jun,
       volume = {7},
        pages = {731-735},
          doi = {10.1038/s41550-023-01937-7},
archivePrefix = {arXiv},
       eprint = {2208.01611},
 primaryClass = {astro-ph.CO},
       adsurl = {https://ui.adsabs.harvard.edu/abs/2023NatAs...7..731B}
}

@ARTICLE{WhiteFrenk1991,
       author = {{White}, Simon D.~M. and {Frenk}, Carlos S.},
        title = "{Galaxy Formation through Hierarchical Clustering}",
      journal = {\apj},
         year = 1991,
        month = sep,
       volume = {379},
        pages = {52},
          doi = {10.1086/170483},
       adsurl = {https://ui.adsabs.harvard.edu/abs/1991ApJ...379...52W}
}

@ARTICLE{WhiteRees1978,
       author = {{White}, S.~D.~M. and {Rees}, M.~J.},
        title = "{Core condensation in heavy halos: a two-stage theory for galaxy formation and clustering.}",
      journal = {\mnras},
         year = 1978,
        month = may,
       volume = {183},
        pages = {341-358},
          doi = {10.1093/mnras/183.3.341},
       adsurl = {https://ui.adsabs.harvard.edu/abs/1978MNRAS.183..341W}
}

@ARTICLE{Carnall2023,
       author = {{Carnall}, A.~C. and {McLeod}, D.~J. and {McLure}, R.~J. and {Dunlop}, J.~S. and {Begley}, R. and {Cullen}, F. and {Donnan}, C.~T. and {Hamadouche}, M.~L. and {Jewell}, S.~M. and {Jones}, E.~W. and {Pollock}, C.~L. and {Wild}, V.},
        title = "{A surprising abundance of massive quiescent galaxies at 3 < z < 5 in the first data from JWST CEERS}",
      journal = {\mnras},
         year = 2023,
        month = apr,
       volume = {520},
       number = {3},
        pages = {3974-3985},
          doi = {10.1093/mnras/stad369},
archivePrefix = {arXiv},
       eprint = {2208.00986},
 primaryClass = {astro-ph.GA},
       adsurl = {https://ui.adsabs.harvard.edu/abs/2023MNRAS.520.3974C}
}

@ARTICLE{Harikane2023a,
       author = {{Harikane}, Yuichi and {Ouchi}, Masami and {Oguri}, Masamune and {Ono}, Yoshiaki and {Nakajima}, Kimihiko and {Isobe}, Yuki and {Umeda}, Hiroya and {Mawatari}, Ken and {Zhang}, Yechi},
        title = "{A Comprehensive Study of Galaxies at z   9-16 Found in the Early JWST Data: Ultraviolet Luminosity Functions and Cosmic Star Formation History at the Pre-reionization Epoch}",
      journal = {\apjs},
         year = 2023,
        month = mar,
       volume = {265},
       number = {1},
          eid = {5},
        pages = {5},
          doi = {10.3847/1538-4365/acaaa9},
archivePrefix = {arXiv},
       eprint = {2208.01612},
 primaryClass = {astro-ph.GA},
       adsurl = {https://ui.adsabs.harvard.edu/abs/2023ApJS..265....5H}
}

@ARTICLE{Harikane2023b,
       author = {{Harikane}, Yuichi and {Zhang}, Yechi and {Nakajima}, Kimihiko and {Ouchi}, Masami and {Isobe}, Yuki and {Ono}, Yoshiaki and {Hatano}, Shun and {Xu}, Yi and {Umeda}, Hiroya},
        title = "{A JWST/NIRSpec First Census of Broad-line AGNs at z = 4-7: Detection of 10 Faint AGNs with M $_{BH}$ {}10$^{6}$-{}10$^{8}$ M $_{{\ensuremath{\odot}}}$ and Their Host Galaxy Properties}",
      journal = {\apj},
         year = 2023,
        month = dec,
       volume = {959},
       number = {1},
          eid = {39},
        pages = {39},
          doi = {10.3847/1538-4357/ad029e},
archivePrefix = {arXiv},
       eprint = {2303.11946},
 primaryClass = {astro-ph.GA},
       adsurl = {https://ui.adsabs.harvard.edu/abs/2023ApJ...959...39H}
}

@ARTICLE{Carnall2024,
       author = {{Carnall}, A.~C. and {Cullen}, F. and {McLure}, R.~J. and {McLeod}, D.~J. and {Begley}, R. and {Donnan}, C.~T. and {Dunlop}, J.~S. and {Shapley}, A.~E. and {Rowlands}, K. and {Almaini}, O. and {Arellano-C{\'o}rdova}, K.~Z. and {Barrufet}, L. and {Cimatti}, A. and {Ellis}, R.~S. and {Grogin}, N.~A. and {Hamadouche}, M.~L. and {Illingworth}, G.~D. and {Koekemoer}, A.~M. and {Leung}, H. -H. and {Lovell}, C.~C. and {P{\'e}rez-Gonz{\'a}lez}, P.~G. and {Santini}, P. and {Stanton}, T.~M. and {Wild}, V.},
        title = "{The JWST EXCELS survey: too much, too young, too fast? Ultra-massive quiescent galaxies at 3 < z < 5}",
      journal = {\mnras},
         year = 2024,
        month = oct,
       volume = {534},
       number = {1},
        pages = {325-348},
          doi = {10.1093/mnras/stae2092},
archivePrefix = {arXiv},
       eprint = {2405.02242},
 primaryClass = {astro-ph.GA},
       adsurl = {https://ui.adsabs.harvard.edu/abs/2024MNRAS.534..325C}
}

@ARTICLE{Lim2016,
       author = {{Lim}, S.~H. and {Mo}, H.~J. and {Wang}, Huiyuan and {Yang}, Xiaohu},
        title = "{An observational proxy of halo assembly time and its correlation with galaxy properties}",
      journal = {\mnras},
         year = 2016,
        month = jan,
       volume = {455},
       number = {1},
        pages = {499-510},
          doi = {10.1093/mnras/stv2282},
archivePrefix = {arXiv},
       eprint = {1502.01256},
 primaryClass = {astro-ph.GA},
       adsurl = {https://ui.adsabs.harvard.edu/abs/2016MNRAS.455..499L}
}

@ARTICLE{Chaikin2023,
       author = {{Chaikin}, Evgenii and {Schaye}, Joop and {Schaller}, Matthieu and {Ben{\'\i}tez-Llambay}, Alejandro and {Nobels}, Folkert S.~J. and {Ploeckinger}, Sylvia},
        title = "{A thermal-kinetic subgrid model for supernova feedback in simulations of galaxy formation}",
      journal = {\mnras},
         year = 2023,
        month = aug,
       volume = {523},
       number = {3},
        pages = {3709-3731},
          doi = {10.1093/mnras/stad1626},
archivePrefix = {arXiv},
       eprint = {2211.04619},
 primaryClass = {astro-ph.GA},
       adsurl = {https://ui.adsabs.harvard.edu/abs/2023MNRAS.523.3709C}
}

@ARTICLE{Glazebrook2024,
       author = {{Glazebrook}, Karl and {Nanayakkara}, Themiya and {Schreiber}, Corentin and {Lagos}, Claudia and {Kawinwanichakij}, Lalitwadee and {Jacobs}, Colin and {Chittenden}, Harry and {Brammer}, Gabriel and {Kacprzak}, Glenn G. and {Labbe}, Ivo and {Marchesini}, Danilo and {Marsan}, Z. Cemile and {Oesch}, Pascal A. and {Papovich}, Casey and {Remus}, Rhea-Silvia and {Tran}, Kim-Vy H. and {Esdaile}, James and {Chandro-Gomez}, Angel},
        title = "{A massive galaxy that formed its stars at z {\ensuremath{\approx}} 11}",
      journal = {\nat},
         year = 2024,
        month = apr,
       volume = {628},
       number = {8007},
        pages = {277-281},
          doi = {10.1038/s41586-024-07191-9},
archivePrefix = {arXiv},
       eprint = {2308.05606},
 primaryClass = {astro-ph.GA},
       adsurl = {https://ui.adsabs.harvard.edu/abs/2024Natur.628..277G}
}

@ARTICLE{Dekel2023,
       author = {{Dekel}, Avishai and {Sarkar}, Kartick C. and {Birnboim}, Yuval and {Mandelker}, Nir and {Li}, Zhaozhou},
        title = "{Efficient formation of massive galaxies at cosmic dawn by feedback-free starbursts}",
      journal = {\mnras},
         year = 2023,
        month = aug,
       volume = {523},
       number = {3},
        pages = {3201-3218},
          doi = {10.1093/mnras/stad1557},
archivePrefix = {arXiv},
       eprint = {2303.04827},
 primaryClass = {astro-ph.GA},
       adsurl = {https://ui.adsabs.harvard.edu/abs/2023MNRAS.523.3201D}
}

@ARTICLE{Shen2023,
       author = {{Shen}, Xuejian and {Vogelsberger}, Mark and {Boylan-Kolchin}, Michael and {Tacchella}, Sandro and {Kannan}, Rahul},
        title = "{The impact of UV variability on the abundance of bright galaxies at z {\ensuremath{\geq}} 9}",
      journal = {\mnras},
         year = 2023,
        month = nov,
       volume = {525},
       number = {3},
        pages = {3254-3261},
          doi = {10.1093/mnras/stad2508},
archivePrefix = {arXiv},
       eprint = {2305.05679},
 primaryClass = {astro-ph.GA},
       adsurl = {https://ui.adsabs.harvard.edu/abs/2023MNRAS.525.3254S}
}

@ARTICLE{Sun2023,
       author = {{Sun}, Guochao and {Faucher-Gigu{\`e}re}, Claude-Andr{\'e} and {Hayward}, Christopher C. and {Shen}, Xuejian and {Wetzel}, Andrew and {Cochrane}, Rachel K.},
        title = "{Bursty Star Formation Naturally Explains the Abundance of Bright Galaxies at Cosmic Dawn}",
      journal = {\apjl},
         year = 2023,
        month = oct,
       volume = {955},
       number = {2},
          eid = {L35},
        pages = {L35},
          doi = {10.3847/2041-8213/acf85a},
archivePrefix = {arXiv},
       eprint = {2307.15305},
 primaryClass = {astro-ph.GA},
       adsurl = {https://ui.adsabs.harvard.edu/abs/2023ApJ...955L..35S}
}

@ARTICLE{KravtsovBelokurov2024,
       author = {{Kravtsov}, Andrey and {Belokurov}, Vasily},
        title = "{Stochastic star formation and the abundance of $z>10$ UV-bright galaxies}",
      journal = {arXiv e-prints},
         year = 2024,
        month = may,
          eid = {arXiv:2405.04578},
        pages = {arXiv:2405.04578},
          doi = {10.48550/arXiv.2405.04578},
archivePrefix = {arXiv},
       eprint = {2405.04578},
 primaryClass = {astro-ph.GA},
       adsurl = {https://ui.adsabs.harvard.edu/abs/2024arXiv240504578K}
}

@ARTICLE{Nanayakkara2024,
       author = {{Nanayakkara}, Themiya and {Glazebrook}, Karl and {Jacobs}, Colin and {Kawinwanichakij}, Lalitwadee and {Schreiber}, Corentin and {Brammer}, Gabriel and {Esdaile}, James and {Kacprzak}, Glenn G. and {Labbe}, Ivo and {Lagos}, Claudia and {Marchesini}, Danilo and {Marsan}, Z. Cemile and {Oesch}, Pascal A. and {Papovich}, Casey and {Remus}, Rhea-Silvia and {Tran}, Kim-Vy H.},
        title = "{A population of faint, old, and massive quiescent galaxies at 3 <z <4 revealed by JWST NIRSpec Spectroscopy}",
      journal = {Scientific Reports},
         year = 2024,
        month = feb,
       volume = {14},
          eid = {3724},
        pages = {3724},
          doi = {10.1038/s41598-024-52585-4},
archivePrefix = {arXiv},
       eprint = {2212.11638},
 primaryClass = {astro-ph.GA},
       adsurl = {https://ui.adsabs.harvard.edu/abs/2024NatSR..14.3724N}
}

@ARTICLE{Weibel2024,
       author = {{Weibel}, Andrea and {Oesch}, Pascal A. and {Barrufet}, Laia and {Gottumukkala}, Rashmi and {Ellis}, Richard S. and {Santini}, Paola and {Weaver}, John R. and {Allen}, Natalie and {Bouwens}, Rychard and {Bowler}, Rebecca A.~A. and {Brammer}, Gabe and {Carnall}, Adam C. and {Cullen}, Fergus and {Dayal}, Pratika and {Dickinson}, Mark and {Donnan}, Callum T. and {Dunlop}, James S. and {Giavalisco}, Mauro and {Grogin}, Norman A. and {Illingworth}, Garth D. and {Koekemoer}, Anton M. and {Labbe}, Ivo and {Marchesini}, Danilo and {McLeod}, Derek J. and {McLure}, Ross J. and {Naidu}, Rohan P. and {P{\'e}rez-Gonz{\'a}lez}, Pablo G. and {Shuntov}, Marko and {Stefanon}, Mauro and {Toft}, Sune and {Xiao}, Mengyuan},
        title = "{Galaxy build-up in the first 1.5 Gyr of cosmic history: insights from the stellar mass function at z   4-9 from JWST NIRCam observations}",
      journal = {\mnras},
         year = 2024,
        month = sep,
       volume = {533},
       number = {2},
        pages = {1808-1838},
          doi = {10.1093/mnras/stae1891},
archivePrefix = {arXiv},
       eprint = {2403.08872},
 primaryClass = {astro-ph.GA},
       adsurl = {https://ui.adsabs.harvard.edu/abs/2024MNRAS.533.1808W}
}

@ARTICLE{Harvey2025,
       author = {{Harvey}, Thomas and {Conselice}, Christopher J. and {Adams}, Nathan J. and {Austin}, Duncan and {Juod{\v{z}}balis}, Ignas and {Trussler}, James and {Li}, Qiong and {Ormerod}, Katherine and {Ferreira}, Leonardo and {Lovell}, Christopher C. and {Duan}, Qiao and {Westcott}, Lewi and {Harris}, Honor and {Bhatawdekar}, Rachana and {Coe}, Dan and {Cohen}, Seth H. and {Caruana}, Joseph and {Cheng}, Cheng and {Driver}, Simon P. and {Frye}, Brenda and {Furtak}, Lukas J. and {Grogin}, Norman A. and {Hathi}, Nimish P. and {Holwerda}, Benne W. and {Jansen}, Rolf A. and {Koekemoer}, Anton M. and {Marshall}, Madeline A. and {Nonino}, Mario and {Vijayan}, Aswin P. and {Wilkins}, Stephen M. and {Windhorst}, Rogier and {Willmer}, Christopher N.~A. and {Yan}, Haojing and {Zitrin}, Adi},
        title = "{EPOCHS. IV. SED Modeling Assumptions and Their Impact on the Stellar Mass Function at 6.5 {\ensuremath{\leq}} z {\ensuremath{\leq}} 13.5 Using PEARLS and Public JWST Observations}",
      journal = {\apj},
         year = 2025,
        month = jan,
       volume = {978},
       number = {1},
          eid = {89},
        pages = {89},
          doi = {10.3847/1538-4357/ad8c29},
archivePrefix = {arXiv},
       eprint = {2403.03908},
 primaryClass = {astro-ph.GA},
       adsurl = {https://ui.adsabs.harvard.edu/abs/2025ApJ...978...89H}
}

@ARTICLE{Jespersen2025a,
       author = {{Jespersen}, Christian Kragh and {Steinhardt}, Charles L. and {Somerville}, Rachel S. and {Lovell}, Christopher C.},
        title = "{On the Significance of Rare Objects at High Redshift: The Impact of Cosmic Variance}",
      journal = {\apj},
         year = 2025,
        month = mar,
       volume = {982},
       number = {1},
          eid = {23},
        pages = {23},
          doi = {10.3847/1538-4357/adb422},
archivePrefix = {arXiv},
       eprint = {2403.00050},
 primaryClass = {astro-ph.GA},
       adsurl = {https://ui.adsabs.harvard.edu/abs/2025ApJ...982...23J}
}

@ARTICLE{Jespersen2025b,
       author = {{Jespersen}, Christian Kragh and {Carnall}, Adam C. and {Lovell}, Christopher C.},
        title = "{Explaining Ultramassive Quiescent Galaxies at 3 < z < 5 in the Context of Their Environments}",
      journal = {\apjl},
         year = 2025,
        month = jul,
       volume = {988},
       number = {1},
          eid = {L19},
        pages = {L19},
          doi = {10.3847/2041-8213/adeb7c},
archivePrefix = {arXiv},
       eprint = {2507.05340},
 primaryClass = {astro-ph.GA},
       adsurl = {https://ui.adsabs.harvard.edu/abs/2025ApJ...988L..19J}
}

@ARTICLE{Lim2024,
       author = {{Lim}, Seunghwan and {Tacchella}, Sandro and {Schaye}, Joop and {Schaller}, Matthieu and {Helton}, Jakob M. and {Kugel}, Roi and {Maiolino}, Roberto},
        title = "{The FLAMINGO simulation view of cluster progenitors observed in the epoch of reionization with JWST}",
      journal = {\mnras},
         year = 2024,
        month = aug,
       volume = {532},
       number = {4},
        pages = {4551-4569},
          doi = {10.1093/mnras/stae1790},
archivePrefix = {arXiv},
       eprint = {2402.17819},
 primaryClass = {astro-ph.GA},
       adsurl = {https://ui.adsabs.harvard.edu/abs/2024MNRAS.532.4551L}
}

@ARTICLE{Lovell2023,
       author = {{Lovell}, Christopher C. and {Harrison}, Ian and {Harikane}, Yuichi and {Tacchella}, Sandro and {Wilkins}, Stephen M.},
        title = "{Extreme value statistics of the halo and stellar mass distributions at high redshift: are JWST results in tension with {\ensuremath{\Lambda}}CDM?}",
      journal = {\mnras},
         year = 2023,
        month = jan,
       volume = {518},
       number = {2},
        pages = {2511-2520},
          doi = {10.1093/mnras/stac3224},
archivePrefix = {arXiv},
       eprint = {2208.10479},
 primaryClass = {astro-ph.GA},
       adsurl = {https://ui.adsabs.harvard.edu/abs/2023MNRAS.518.2511L}
}

@ARTICLE{Lim2025,
       author = {{Lim}, Seunghwan and {Tacchella}, Sandro and {Maiolino}, Roberto and {Schaye}, Joop and {Schaller}, Matthieu},
        title = "{In-situ vs. ex-situ drivers of galaxy quenching: ubiquity of main sequence and critical black hole mass from the FLAMINGO simulation}",
      journal = {arXiv e-prints},
         year = 2025,
        month = apr,
          eid = {arXiv:2504.02027},
        pages = {arXiv:2504.02027},
          doi = {10.48550/arXiv.2504.02027},
archivePrefix = {arXiv},
       eprint = {2504.02027},
 primaryClass = {astro-ph.GA},
       adsurl = {https://ui.adsabs.harvard.edu/abs/2025arXiv250402027L}
}

@ARTICLE{Turner2025,
       author = {{Turner}, Crispin and {Tacchella}, Sandro and {D'Eugenio}, Francesco and {Carniani}, Stefano and {Curti}, Mirko and {Glazebrook}, Karl and {Johnson}, Benjamin D. and {Lim}, Seunghwan and {Looser}, Tobias and {Maiolino}, Roberto and {Nanayakkara}, Themiya and {Wan}, Jenny},
        title = "{Age-dating early quiescent galaxies: high star formation efficiency, but consistent with direct, higher-redshift observations}",
      journal = {\mnras},
         year = 2025,
        month = feb,
       volume = {537},
       number = {2},
        pages = {1826-1848},
          doi = {10.1093/mnras/staf128},
archivePrefix = {arXiv},
       eprint = {2410.05377},
 primaryClass = {astro-ph.GA},
       adsurl = {https://ui.adsabs.harvard.edu/abs/2025MNRAS.537.1826T}
}

@ARTICLE{Tacchella2018,
       author = {{Tacchella}, Sandro and {Bose}, Sownak and {Conroy}, Charlie and {Eisenstein}, Daniel J. and {Johnson}, Benjamin D.},
        title = "{A Redshift-independent Efficiency Model: Star Formation and Stellar Masses in Dark Matter Halos at z {\ensuremath{\gtrsim}} 4}",
      journal = {\apj},
         year = 2018,
        month = dec,
       volume = {868},
       number = {2},
          eid = {92},
        pages = {92},
          doi = {10.3847/1538-4357/aae8e0},
archivePrefix = {arXiv},
       eprint = {1806.03299},
 primaryClass = {astro-ph.GA},
       adsurl = {https://ui.adsabs.harvard.edu/abs/2018ApJ...868...92T}
}

@ARTICLE{Shuntov2025,
       author = {{Shuntov}, M. and {Ilbert}, O. and {Toft}, S. and {Arango-Toro}, R.~C. and {Akins}, H.~B. and {Casey}, C.~M. and {Franco}, M. and {Harish}, S. and {Kartaltepe}, J.~S. and {Koekemoer}, A.~M. and {McCracken}, H.~J. and {Paquereau}, L. and {Laigle}, C. and {Bethermin}, M. and {Dubois}, Y. and {Drakos}, N.~E. and {Faisst}, A. and {Gozaliasl}, G. and {Gillman}, S. and {Hayward}, C.~C. and {Hirschmann}, M. and {Huertas-Company}, M. and {Jespersen}, C.~K. and {Jin}, S. and {Kokorev}, V. and {Lambrides}, E. and {Le Borgne}, D. and {Liu}, D. and {Magdis}, G. and {Massey}, R. and {McPartland}, C.~J.~R. and {Mercier}, W. and {McCleary}, J.~E. and {McKinney}, J. and {Oesch}, P.~A. and {Renzini}, A. and {Rhodes}, J.~D. and {Rich}, R.~M. and {Robertson}, B.~E. and {Sanders}, D. and {Trebitsch}, M. and {Tresse}, L. and {Valentino}, F. and {Vijayan}, A.~P. and {Weaver}, J.~R. and {Weibel}, A. and {Wilkins}, S.~M. and {Yang}, L.},
        title = "{COSMOS-Web: Stellar mass assembly in relation to dark matter halos across 0.2 < z < 12 of cosmic history}",
      journal = {\aap},
         year = 2025,
        month = mar,
       volume = {695},
          eid = {A20},
        pages = {A20},
          doi = {10.1051/0004-6361/202452570},
archivePrefix = {arXiv},
       eprint = {2410.08290},
 primaryClass = {astro-ph.GA},
       adsurl = {https://ui.adsabs.harvard.edu/abs/2025A&A...695A..20S}
}

@ARTICLE{Harikane2018,
       author = {{Harikane}, Yuichi and {Ouchi}, Masami and {Ono}, Yoshiaki and {Saito}, Shun and {Behroozi}, Peter and {More}, Surhud and {Shimasaku}, Kazuhiro and {Toshikawa}, Jun and {Lin}, Yen-Ting and {Akiyama}, Masayuki and {Coupon}, Jean and {Komiyama}, Yutaka and {Konno}, Akira and {Lin}, Sheng-Chieh and {Miyazaki}, Satoshi and {Nishizawa}, Atsushi J. and {Shibuya}, Takatoshi and {Silverman}, John},
        title = "{GOLDRUSH. II. Clustering of galaxies at z {\ensuremath{\sim}} 4-6 revealed with the half-million dropouts over the 100 deg$^{2}$ area corresponding to 1 Gpc$^{3}$}",
      journal = {\pasj},
         year = 2018,
        month = jan,
       volume = {70},
          eid = {S11},
        pages = {S11},
          doi = {10.1093/pasj/psx097},
archivePrefix = {arXiv},
       eprint = {1704.06535},
 primaryClass = {astro-ph.GA},
       adsurl = {https://ui.adsabs.harvard.edu/abs/2018PASJ...70S..11H}
}

@ARTICLE{Casey2023,
       author = {{Casey}, Caitlin M. and {Kartaltepe}, Jeyhan S. and {Drakos}, Nicole E. and {Franco}, Maximilien and {Harish}, Santosh and {Paquereau}, Louise and {Ilbert}, Olivier and {Rose}, Caitlin and {Cox}, Isabella G. and {Nightingale}, James W. and {Robertson}, Brant E. and {Silverman}, John D. and {Koekemoer}, Anton M. and {Massey}, Richard and {McCracken}, Henry Joy and {Rhodes}, Jason and {Akins}, Hollis B. and {Allen}, Natalie and {Amvrosiadis}, Aristeidis and {Arango-Toro}, Rafael C. and {Bagley}, Micaela B. and {Bongiorno}, Angela and {Capak}, Peter L. and {Champagne}, Jaclyn B. and {Chartab}, Nima and {Ch{\'a}vez Ortiz}, {\'O}scar A. and {Chworowsky}, Katherine and {Cooke}, Kevin C. and {Cooper}, Olivia R. and {Darvish}, Behnam and {Ding}, Xuheng and {Faisst}, Andreas L. and {Finkelstein}, Steven L. and {Fujimoto}, Seiji and {Gentile}, Fabrizio and {Gillman}, Steven and {Gould}, Katriona M.~L. and {Gozaliasl}, Ghassem and {Hayward}, Christopher C. and {He}, Qiuhan and {Hemmati}, Shoubaneh and {Hirschmann}, Michaela and {Jahnke}, Knud and {Jin}, Shuowen and {Khostovan}, Ali Ahmad and {Kokorev}, Vasily and {Lambrides}, Erini and {Laigle}, Clotilde and {Larson}, Rebecca L. and {Leung}, Gene C.~K. and {Liu}, Daizhong and {Liaudat}, Tobias and {Long}, Arianna S. and {Magdis}, Georgios and {Mahler}, Guillaume and {Mainieri}, Vincenzo and {Manning}, Sinclaire M. and {Maraston}, Claudia and {Martin}, Crystal L. and {McCleary}, Jacqueline E. and {McKinney}, Jed and {McPartland}, Conor J.~R. and {Mobasher}, Bahram and {Pattnaik}, Rohan and {Renzini}, Alvio and {Rich}, R. Michael and {Sanders}, David B. and {Sattari}, Zahra and {Scognamiglio}, Diana and {Scoville}, Nick and {Sheth}, Kartik and {Shuntov}, Marko and {Sparre}, Martin and {Suzuki}, Tomoko L. and {Talia}, Margherita and {Toft}, Sune and {Trakhtenbrot}, Benny and {Urry}, C. Megan and {Valentino}, Francesco and {Vanderhoof}, Brittany N. and {Vardoulaki}, Eleni and {Weaver}, John R. and {Whitaker}, Katherine E. and {Wilkins}, Stephen M. and {Yang}, Lilan and {Zavala}, Jorge A.},
        title = "{COSMOS-Web: An Overview of the JWST Cosmic Origins Survey}",
      journal = {\apj},
         year = 2023,
        month = sep,
       volume = {954},
       number = {1},
          eid = {31},
        pages = {31},
          doi = {10.3847/1538-4357/acc2bc},
archivePrefix = {arXiv},
       eprint = {2211.07865},
 primaryClass = {astro-ph.GA},
       adsurl = {https://ui.adsabs.harvard.edu/abs/2023ApJ...954...31C}
}

@ARTICLE{Schaye2026, 
       author = {{Schaye}, Joop and {Chaikin}, Evgenii and {Schaller}, Matthieu and {Ploeckinger}, Sylvia and {Hu{\v{s}}ko}, Filip and {McGibbon}, Robert J. and {Trayford}, James W. and {Ben{\'\i}tez-Llambay}, Alejandro and {Correa}, Camila and {Frenk}, Carlos S. and {Richings}, Alexander J. and {Forouhar Moreno}, Victor J. and {Bah{\'e}}, Yannick M. and {Borrow}, Josh and {Durrant}, Anna and {Gebek}, Andrea and {Helly}, John C. and {Jenkins}, Adrian and {Lacey}, Cedric G. and {Ludlow}, Aaron and {Nobels}, Folkert S.~J.},
        title = "{The COLIBRE project: cosmological hydrodynamical simulations of galaxy formation and evolution}",
      journal = {\mnras},
         year = 2026,
        month = may,
       volume = {548},
       number = {1},
          eid = {stag375},
        pages = {stag375},
          doi = {10.1093/mnras/stag375},
archivePrefix = {arXiv},
       eprint = {2508.21126},
 primaryClass = {astro-ph.GA},
       adsurl = {https://ui.adsabs.harvard.edu/abs/2026MNRAS.548ag375S}
}

@ARTICLE{Chaikin2026, 
       author = {{Chaikin}, Evgenii and {Schaye}, Joop and {Schaller}, Matthieu and {Ploeckinger}, Sylvia and {Bah{\'e}}, Yannick M. and {Ben{\'\i}tez-Llambay}, Alejandro and {Correa}, Camila and {Forouhar Moreno}, Victor J. and {Frenk}, Carlos S. and {Hu{\v{s}}ko}, Filip and {Kugel}, Roi and {McGibbon}, Robert and {Richings}, Alexander J. and {Trayford}, James W. and {Borrow}, Josh and {Crain}, Robert A. and {Helly}, John C. and {Lacey}, Cedric G. and {Ludlow}, Aaron and {Nobels}, Folkert S.~J.},
        title = "{COLIBRE: calibrating subgrid feedback in cosmological simulations that include a cold gas phase}",
      journal = {\mnras},
         year = 2026,
        month = may,
       volume = {548},
       number = {1},
          eid = {stag300},
        pages = {stag300},
          doi = {10.1093/mnras/stag300},
archivePrefix = {arXiv},
       eprint = {2509.04067},
 primaryClass = {astro-ph.GA},
       adsurl = {https://ui.adsabs.harvard.edu/abs/2026MNRAS.548ag300C}
}

@ARTICLE{McGibbon2025,
       author = {{McGibbon}, Robert and {Helly}, John and {Schaye}, Joop and {Schaller}, Matthieu and {Vandenbroucke}, Bert},
        title = "{SOAP: A Python Package for Calculating the Properties of Galaxies and Halos Formed in Cosmological Simulations}",
      journal = {The Journal of Open Source Software},
         year = 2025,
        month = jul,
       volume = {10},
       number = {111},
          eid = {8252},
        pages = {8252},
          doi = {10.21105/joss.08252},
archivePrefix = {arXiv},
       eprint = {2507.22669},
 primaryClass = {astro-ph.IM},
       adsurl = {https://ui.adsabs.harvard.edu/abs/2025JOSS...10.8252M}
}

@ARTICLE{Conroy2009,
       author = {{Conroy}, Charlie and {Gunn}, James E. and {White}, Martin},
        title = "{The Propagation of Uncertainties in Stellar Population Synthesis Modeling. I. The Relevance of Uncertain Aspects of Stellar Evolution and the Initial Mass Function to the Derived Physical Properties of Galaxies}",
      journal = {\apj},
         year = 2009,
        month = jul,
       volume = {699},
       number = {1},
        pages = {486-506},
          doi = {10.1088/0004-637X/699/1/486},
archivePrefix = {arXiv},
       eprint = {0809.4261},
 primaryClass = {astro-ph},
       adsurl = {https://ui.adsabs.harvard.edu/abs/2009ApJ...699..486C}
}

@ARTICLE{Dalmasso2026,
       author = {{Dalmasso}, Nicol{\`o} and {Ferrami}, Giovanni and {Leethochawalit}, Nicha and {Ventura}, Emanuele M. and {Trenti}, Michele},
        title = "{Accelerated evolution of galaxy host halo masses during Cosmic Dawn from deep JWST clustering}",
      journal = {\mnras},
         year = 2026,
        month = feb,
       volume = {546},
       number = {2},
          eid = {stag001},
        pages = {stag001},
          doi = {10.1093/mnras/stag001},
archivePrefix = {arXiv},
       eprint = {2601.01697},
 primaryClass = {astro-ph.GA},
       adsurl = {https://ui.adsabs.harvard.edu/abs/2026MNRAS.546ag001D}
}

@ARTICLE{Paquereau2025,
       author = {{Paquereau}, L. and {Laigle}, C. and {McCracken}, H.~J. and {Shuntov}, M. and {Ilbert}, O. and {Akins}, H.~B. and {Allen}, N. and {Arango-Togo}, R. and {Berman}, E.~M. and {B{\'e}thermin}, M. and {Casey}, C.~M. and {McCleary}, J. and {Dubois}, Y. and {Drakos}, N.~E. and {Faisst}, A.~L. and {Franco}, M. and {Harish}, S. and {Jespersen}, C.~K. and {Kartaltepe}, J.~S. and {Koekemoer}, A.~M. and {Kokorev}, V. and {Lambrides}, E. and {Larson}, R. and {Liu}, D. and {Le Borgne}, D. and {Lewis}, J.~S.~W. and {McKinney}, J. and {Mercier}, W. and {Rhodes}, J.~D. and {Robertson}, B.~E. and {Toft}, S. and {Trebitsch}, M. and {Tresse}, L. and {Weaver}, J.~R.},
        title = "{Tracing the galaxy-halo connection with galaxy clustering in COSMOS-Web from z = 0.1 to z {\ensuremath{\sim}} 12}",
      journal = {\aap},
         year = 2025,
        month = oct,
       volume = {702},
          eid = {A163},
        pages = {A163},
          doi = {10.1051/0004-6361/202553828},
archivePrefix = {arXiv},
       eprint = {2501.11674},
 primaryClass = {astro-ph.GA},
       adsurl = {https://ui.adsabs.harvard.edu/abs/2025A&A...702A.163P}
}

@ARTICLE{Munoz2023,
       author = {{Mu{\~n}oz}, Julian B. and {Mirocha}, Jordan and {Furlanetto}, Steven and {Sabti}, Nashwan},
        title = "{Breaking degeneracies in the first galaxies with clustering}",
      journal = {\mnras},
         year = 2023,
        month = nov,
       volume = {526},
       number = {1},
        pages = {L47-L55},
          doi = {10.1093/mnrasl/slad115},
archivePrefix = {arXiv},
       eprint = {2306.09403},
 primaryClass = {astro-ph.CO},
       adsurl = {https://ui.adsabs.harvard.edu/abs/2023MNRAS.526L..47M}
}

@ARTICLE{Sun2025,
       author = {{Sun}, Guochao and {Mu{\~n}oz}, Julian B. and {Mirocha}, Jordan and {Faucher-Gigu{\`e}re}, Claude-Andr{\'e}},
        title = "{Constraining bursty star formation histories with galaxy UV and H{\ensuremath{\alpha}} luminosity functions and clustering}",
      journal = {\jcap},
         year = 2025,
        month = apr,
       volume = {2025},
       number = {4},
          eid = {034},
        pages = {034},
          doi = {10.1088/1475-7516/2025/04/034},
archivePrefix = {arXiv},
       eprint = {2410.21409},
 primaryClass = {astro-ph.GA},
       adsurl = {https://ui.adsabs.harvard.edu/abs/2025JCAP...04..034S}
}

@ARTICLE{Munoz2026,
       author = {{Mu{\~n}oz}, Julian B. and {Chisholm}, John and {Sun}, Guochao and {Samuel}, Jenna and {Mirocha}, Jordan and {Bregou}, Emily and {Venditti}, Alessandra and {Qezlou}, Mahdi and {Simmonds}, Charlotte and {Endsley}, Ryan},
        title = "{Relatively Fast and Reasonably Furious: Evidence for Increased Burstiness in Smaller Halos at Cosmic Dawn}",
      journal = {\mnras},
         year = 2026,
        month = mar,
          doi = {10.1093/mnras/stag415},
archivePrefix = {arXiv},
       eprint = {2601.07912},
 primaryClass = {astro-ph.GA},
       adsurl = {https://ui.adsabs.harvard.edu/abs/2026MNRAS.tmp..405M}
}

@ARTICLE{Harikane2022,
       author = {{Harikane}, Yuichi and {Ono}, Yoshiaki and {Ouchi}, Masami and {Liu}, Chengze and {Sawicki}, Marcin and {Shibuya}, Takatoshi and {Behroozi}, Peter S. and {He}, Wanqiu and {Shimasaku}, Kazuhiro and {Arnouts}, Stephane and {Coupon}, Jean and {Fujimoto}, Seiji and {Gwyn}, Stephen and {Huang}, Jiasheng and {Inoue}, Akio K. and {Kashikawa}, Nobunari and {Komiyama}, Yutaka and {Matsuoka}, Yoshiki and {Willott}, Chris J.},
        title = "{GOLDRUSH. IV. Luminosity Functions and Clustering Revealed with  4,000,000 Galaxies at z   2-7: Galaxy-AGN Transition, Star Formation Efficiency, and Implication for Evolution at z > 10}",
      journal = {\apjs},
         year = 2022,
        month = mar,
       volume = {259},
       number = {1},
          eid = {20},
        pages = {20},
          doi = {10.3847/1538-4365/ac3dfc},
archivePrefix = {arXiv},
       eprint = {2108.01090},
 primaryClass = {astro-ph.GA},
       adsurl = {https://ui.adsabs.harvard.edu/abs/2022ApJS..259...20H}
}

@ARTICLE{Shuntov2025b, 
       author = {{Shuntov}, Marko and {Oesch}, Pascal A. and {Toft}, Sune and {Meyer}, Romain A. and {Covelo-Paz}, Alba and {Paquereau}, Louise and {Bouwens}, Rychard and {Brammer}, Gabriel and {Gelli}, Viola and {Giovinazzo}, Emma and {Herard-Demanche}, Thomas and {Illingworth}, Garth D. and {Mason}, Charlotte and {Naidu}, Rohan P. and {Weibel}, Andrea and {Xiao}, Mengyuan},
        title = "{Constraints on the early Universe star formation efficiency from galaxy clustering and halo modeling of H{\ensuremath{\alpha}} and [O III] emitters}",
      journal = {\aap},
         year = 2025,
        month = jul,
       volume = {699},
          eid = {A231},
        pages = {A231},
          doi = {10.1051/0004-6361/202554618},
archivePrefix = {arXiv},
       eprint = {2503.14280},
 primaryClass = {astro-ph.GA},
       adsurl = {https://ui.adsabs.harvard.edu/abs/2025A&A...699A.231S}
}

@ARTICLE{WechslerTinker2018,
       author = {{Wechsler}, Risa H. and {Tinker}, Jeremy L.},
        title = "{The Connection Between Galaxies and Their Dark Matter Halos}",
      journal = {\araa},
         year = 2018,
        month = sep,
       volume = {56},
        pages = {435-487},
          doi = {10.1146/annurev-astro-081817-051756},
archivePrefix = {arXiv},
       eprint = {1804.03097},
 primaryClass = {astro-ph.GA},
       adsurl = {https://ui.adsabs.harvard.edu/abs/2018ARA&A..56..435W}
}

@ARTICLE{ConroyWechsler2009,
       author = {{Conroy}, Charlie and {Wechsler}, Risa H.},
        title = "{Connecting Galaxies, Halos, and Star Formation Rates Across Cosmic Time}",
      journal = {\apj},
         year = 2009,
        month = may,
       volume = {696},
       number = {1},
        pages = {620-635},
          doi = {10.1088/0004-637X/696/1/620},
archivePrefix = {arXiv},
       eprint = {0805.3346},
 primaryClass = {astro-ph},
       adsurl = {https://ui.adsabs.harvard.edu/abs/2009ApJ...696..620C}
}

@ARTICLE{Baker2025b,
       author = {{Baker}, William M. and {Valentino}, Francesco and {Lagos}, Claudia del P. and {Ito}, Kei and {Jespersen}, Christian Kragh and {Gottumukkala}, Rashmi and {Hjorth}, Jens and {Langeroodi}, Danial and {Sedgewick}, Aidan},
        title = "{Exploring over 700 massive quiescent galaxies at z = 2─7: Demographics and stellar mass functions}",
      journal = {\aap},
         year = 2025,
        month = oct,
       volume = {702},
          eid = {A270},
        pages = {A270},
          doi = {10.1051/0004-6361/202555829},
archivePrefix = {arXiv},
       eprint = {2506.04119},
 primaryClass = {astro-ph.GA},
       adsurl = {https://ui.adsabs.harvard.edu/abs/2025A&A...702A.270B}
}

@ARTICLE{Maiolino2024b,
       author = {{Maiolino}, Roberto and {Scholtz}, Jan and {Curtis-Lake}, Emma and {Carniani}, Stefano and {Baker}, William and {de Graaff}, Anna and {Tacchella}, Sandro and {{\"U}bler}, Hannah and {D'Eugenio}, Francesco and {Witstok}, Joris and {Curti}, Mirko and {Arribas}, Santiago and {Bunker}, Andrew J. and {Charlot}, St{\'e}phane and {Chevallard}, Jacopo and {Eisenstein}, Daniel J. and {Egami}, Eiichi and {Ji}, Zhiyuan and {Jones}, Gareth C. and {Lyu}, Jianwei and {Rawle}, Tim and {Robertson}, Brant and {Rujopakarn}, Wiphu and {Perna}, Michele and {Sun}, Fengwu and {Venturi}, Giacomo and {Williams}, Christina C. and {Willott}, Chris},
        title = "{JADES: The diverse population of infant black holes at 4 < z < 11: Merging, tiny, poor, but mighty}",
      journal = {\aap},
         year = 2024,
        month = nov,
       volume = {691},
          eid = {A145},
        pages = {A145},
          doi = {10.1051/0004-6361/202347640},
archivePrefix = {arXiv},
       eprint = {2308.01230},
 primaryClass = {astro-ph.GA},
       adsurl = {https://ui.adsabs.harvard.edu/abs/2024A&A...691A.145M}
}

@ARTICLE{Kocevski2025,
       author = {{Kocevski}, Dale D. and {Finkelstein}, Steven L. and {Barro}, Guillermo and {Taylor}, Anthony J. and {Calabr{\`o}}, Antonello and {Laloux}, Brivael and {Buchner}, Johannes and {Trump}, Jonathan R. and {Leung}, Gene C.~K. and {Yang}, Guang and {Dickinson}, Mark and {P{\'e}rez-Gonz{\'a}lez}, Pablo G. and {Pacucci}, Fabio and {Inayoshi}, Kohei and {Somerville}, Rachel S. and {McGrath}, Elizabeth J. and {Akins}, Hollis B. and {Bagley}, Micaela B. and {Bowler}, Rebecca A.~A. and {Bisigello}, Laura and {Carnall}, Adam and {Casey}, Caitlin M. and {Cheng}, Yingjie and {Cleri}, Nikko J. and {Costantin}, Luca and {Cullen}, Fergus and {Davis}, Kelcey and {Donnan}, Callum T. and {Dunlop}, James S. and {Ellis}, Richard S. and {Ferguson}, Henry C. and {Fujimoto}, Seiji and {Fontana}, Adriano and {Giavalisco}, Mauro and {Grazian}, Andrea and {Grogin}, Norman A. and {Hathi}, Nimish P. and {Hirschmann}, Michaela and {Huertas-Company}, Marc and {Holwerda}, Benne W. and {Illingworth}, Garth and {Juneau}, St{\'e}phanie and {Kartaltepe}, Jeyhan S. and {Koekemoer}, Anton M. and {Li}, Wenxiu and {Lucas}, Ray A. and {Magee}, Dan and {Mason}, Charlotte and {McLeod}, Derek J. and {McLure}, Ross J. and {Napolitano}, Lorenzo and {Papovich}, Casey and {Pirzkal}, Nor and {Rodighiero}, Giulia and {Santini}, Paola and {Wilkins}, Stephen M. and {Yung}, L.~Y. Aaron},
        title = "{The Rise of Faint, Red Active Galactic Nuclei at z > 4: A Sample of Little Red Dots in the JWST Extragalactic Legacy Fields}",
      journal = {\apj},
         year = 2025,
        month = jun,
       volume = {986},
       number = {2},
          eid = {126},
        pages = {126},
          doi = {10.3847/1538-4357/adbc7d},
archivePrefix = {arXiv},
       eprint = {2404.03576},
 primaryClass = {astro-ph.GA},
       adsurl = {https://ui.adsabs.harvard.edu/abs/2025ApJ...986..126K}
}

@ARTICLE{Donnan2023,
       author = {{Donnan}, C.~T. and {McLeod}, D.~J. and {Dunlop}, J.~S. and {McLure}, R.~J. and {Carnall}, A.~C. and {Begley}, R. and {Cullen}, F. and {Hamadouche}, M.~L. and {Bowler}, R.~A.~A. and {Magee}, D. and {McCracken}, H.~J. and {Milvang-Jensen}, B. and {Moneti}, A. and {Targett}, T.},
        title = "{The evolution of the galaxy UV luminosity function at redshifts z ≃ 8 - 15 from deep JWST and ground-based near-infrared imaging}",
      journal = {\mnras},
         year = 2023,
        month = feb,
       volume = {518},
       number = {4},
        pages = {6011-6040},
          doi = {10.1093/mnras/stac3472},
archivePrefix = {arXiv},
       eprint = {2207.12356},
 primaryClass = {astro-ph.GA},
       adsurl = {https://ui.adsabs.harvard.edu/abs/2023MNRAS.518.6011D}
}

@ARTICLE{Finkelstein2024,
       author = {{Finkelstein}, Steven L. and {Leung}, Gene C.~K. and {Bagley}, Micaela B. and {Dickinson}, Mark and {Ferguson}, Henry C. and {Papovich}, Casey and {Akins}, Hollis B. and {Arrabal Haro}, Pablo and {Dav{\'e}}, Romeel and {Dekel}, Avishai and {Kartaltepe}, Jeyhan S. and {Kocevski}, Dale D. and {Koekemoer}, Anton M. and {Pirzkal}, Nor and {Somerville}, Rachel S. and {Yung}, L.~Y. Aaron and {Amor{\'\i}n}, Ricardo O. and {Backhaus}, Bren E. and {Behroozi}, Peter and {Bisigello}, Laura and {Bromm}, Volker and {Casey}, Caitlin M. and {Ch{\'a}vez Ortiz}, {\'O}scar A. and {Cheng}, Yingjie and {Chworowsky}, Katherine and {Cleri}, Nikko J. and {Cooper}, M.~C. and {Davis}, Kelcey and {de la Vega}, Alexander and {Elbaz}, David and {Franco}, Maximilien and {Fontana}, Adriano and {Fujimoto}, Seiji and {Giavalisco}, Mauro and {Grogin}, Norman A. and {Holwerda}, Benne W. and {Huertas-Company}, Marc and {Hirschmann}, Michaela and {Iyer}, Kartheik G. and {Jogee}, Shardha and {Jung}, Intae and {Larson}, Rebecca L. and {Lucas}, Ray A. and {Mobasher}, Bahram and {Morales}, Alexa M. and {Morley}, Caroline V. and {Mukherjee}, Sagnick and {P{\'e}rez-Gonz{\'a}lez}, Pablo G. and {Ravindranath}, Swara and {Rodighiero}, Giulia and {Rowland}, Melanie J. and {Tacchella}, Sandro and {Taylor}, Anthony J. and {Trump}, Jonathan R. and {Wilkins}, Stephen M.},
        title = "{The Complete CEERS Early Universe Galaxy Sample: A Surprisingly Slow Evolution of the Space Density of Bright Galaxies at z {\ensuremath{\sim}} 8.5{\textendash}14.5}",
      journal = {\apjl},
         year = 2024,
        month = jul,
       volume = {969},
       number = {1},
          eid = {L2},
        pages = {L2},
          doi = {10.3847/2041-8213/ad4495},
archivePrefix = {arXiv},
       eprint = {2311.04279},
 primaryClass = {astro-ph.GA},
       adsurl = {https://ui.adsabs.harvard.edu/abs/2024ApJ...969L...2F}
}

@ARTICLE{Gelli2024,
       author = {{Gelli}, Viola and {Mason}, Charlotte and {Hayward}, Christopher C.},
        title = "{The Impact of Mass-dependent Stochasticity at Cosmic Dawn}",
      journal = {\apj},
         year = 2024,
        month = nov,
       volume = {975},
       number = {2},
          eid = {192},
        pages = {192},
          doi = {10.3847/1538-4357/ad7b36},
archivePrefix = {arXiv},
       eprint = {2405.13108},
 primaryClass = {astro-ph.GA},
       adsurl = {https://ui.adsabs.harvard.edu/abs/2024ApJ...975..192G}
}

@MISC{Egami2023,
       author = {{Egami}, Eiichi and {Sun}, Fengwu and {Alberts}, Stacey and {Baum}, Stefi A. and {Boyett}, Kristan and {Bunker}, Andrew and {Cameron}, Alex James and {Carniani}, Stefano and {Charlot}, Stephane and {Chen}, Zuyi and {Chevallard}, Jacopo and {Curti}, Mirko and {D'Eugenio}, Francesco and {Danhaive}, Lola and {DeCoursey}, Christa Noel and {Dudzeviciute}, Ugne and {Eisenstein}, Daniel J. and {Hainline}, Kevin and {Helton}, Jakob and {Ji}, Zhiyuan and {Johnson}, Benjamin D. and {Kumari}, Nimisha and {Looser}, Tobias Jakob and {Lyu}, Jianwei and {Ma}, Zheng and {Maiolino}, Roberto and {Maseda}, Michael and {Nelson}, Erica and {Rawle}, Tim and {Rieke}, Marcia J. and {Robertson}, Brant and {Sandles}, Lester and {Shivaei}, Irene and {Smit}, Renske and {Suess}, Katherine and {Tacchella}, Sandro and {Uebler}, Hannah and {Whitler}, Lily and {Williams}, Christina C. and {Willmer}, Christopher Nicholas Andrew and {Willott}, Chris J. and {Witstok}, Joris and {de Graaff}, Anna G.},
        title = "{Complete NIRCam Grism Redshift Survey (CONGRESS)}",
 howpublished = {JWST Proposal. Cycle 2, ID. \#3577},
         year = 2023,
        month = aug,
        pages = {3577},
       adsurl = {https://ui.adsabs.harvard.edu/abs/2023jwst.prop.3577E}
}

@ARTICLE{Mirocha2020,
       author = {{Mirocha}, Jordan},
        title = "{Prospects for distinguishing galaxy evolution models with surveys at redshifts z {\ensuremath{\gtrsim}} 4}",
      journal = {\mnras},
         year = 2020,
        month = dec,
       volume = {499},
       number = {3},
        pages = {4534-4544},
          doi = {10.1093/mnras/staa3150},
archivePrefix = {arXiv},
       eprint = {2008.04322},
 primaryClass = {astro-ph.GA},
       adsurl = {https://ui.adsabs.harvard.edu/abs/2020MNRAS.499.4534M}
}

@ARTICLE{MoWhite1996,
       author = {{Mo}, H.~J. and {White}, S.~D.~M.},
        title = "{An analytic model for the spatial clustering of dark matter haloes}",
      journal = {\mnras},
         year = 1996,
        month = sep,
       volume = {282},
       number = {2},
        pages = {347-361},
          doi = {10.1093/mnras/282.2.347},
archivePrefix = {arXiv},
       eprint = {astro-ph/9512127},
 primaryClass = {astro-ph},
       adsurl = {https://ui.adsabs.harvard.edu/abs/1996MNRAS.282..347M}
}

@ARTICLE{Endsley2020,
       author = {{Endsley}, Ryan and {Behroozi}, Peter and {Stark}, Daniel P. and {Williams}, Christina C. and {Robertson}, Brant E. and {Rieke}, Marcia and {Gottl{\"o}ber}, Stefan and {Yepes}, Gustavo},
        title = "{Clustering with JWST: Constraining galaxy host halo masses, satellite quenching efficiencies, and merger rates at z = 4-10}",
      journal = {\mnras},
         year = 2020,
        month = mar,
       volume = {493},
       number = {1},
        pages = {1178-1196},
          doi = {10.1093/mnras/staa324},
archivePrefix = {arXiv},
       eprint = {1907.02546},
 primaryClass = {astro-ph.GA},
       adsurl = {https://ui.adsabs.harvard.edu/abs/2020MNRAS.493.1178E}
}

@ARTICLE{Tinker2010,
       author = {{Tinker}, Jeremy L. and {Robertson}, Brant E. and {Kravtsov}, Andrey V. and {Klypin}, Anatoly and {Warren}, Michael S. and {Yepes}, Gustavo and {Gottl{\"o}ber}, Stefan},
        title = "{The Large-scale Bias of Dark Matter Halos: Numerical Calibration and Model Tests}",
      journal = {\apj},
         year = 2010,
        month = dec,
       volume = {724},
       number = {2},
        pages = {878-886},
          doi = {10.1088/0004-637X/724/2/878},
archivePrefix = {arXiv},
       eprint = {1001.3162},
 primaryClass = {astro-ph.CO},
       adsurl = {https://ui.adsabs.harvard.edu/abs/2010ApJ...724..878T}
}

@ARTICLE{Jose2017,
       author = {{Jose}, Charles and {Baugh}, Carlton M. and {Lacey}, Cedric G. and {Subramanian}, Kandaswamy},
        title = "{Understanding the non-linear clustering of high-redshift galaxies}",
      journal = {\mnras},
         year = 2017,
        month = aug,
       volume = {469},
       number = {4},
        pages = {4428-4436},
          doi = {10.1093/mnras/stx1014},
archivePrefix = {arXiv},
       eprint = {1702.00853},
 primaryClass = {astro-ph.CO},
       adsurl = {https://ui.adsabs.harvard.edu/abs/2017MNRAS.469.4428J}
}

@ARTICLE{Jose2016,
       author = {{Jose}, Charles and {Lacey}, Cedric G. and {Baugh}, Carlton M.},
        title = "{The clustering of dark matter haloes: scale-dependent bias on quasi-linear scales}",
      journal = {\mnras},
         year = 2016,
        month = nov,
       volume = {463},
       number = {1},
        pages = {270-281},
          doi = {10.1093/mnras/stw1702},
archivePrefix = {arXiv},
       eprint = {1509.06715},
 primaryClass = {astro-ph.CO},
       adsurl = {https://ui.adsabs.harvard.edu/abs/2016MNRAS.463..270J}
}

@ARTICLE{Jose2013,
       author = {{Jose}, Charles and {Subramanian}, Kandaswamy and {Srianand}, Raghunathan and {Samui}, Saumyadip},
        title = "{Spatial clustering of high-redshift Lyman-break galaxies}",
      journal = {\mnras},
         year = 2013,
        month = mar,
       volume = {429},
       number = {3},
        pages = {2333-2350},
          doi = {10.1093/mnras/sts503},
archivePrefix = {arXiv},
       eprint = {1208.2097},
 primaryClass = {astro-ph.CO},
       adsurl = {https://ui.adsabs.harvard.edu/abs/2013MNRAS.429.2333J}
}

@ARTICLE{Dalmasso2024,
       author = {{Dalmasso}, Nicol{\`o} and {Leethochawalit}, Nicha and {Trenti}, Michele and {Boyett}, Kristan},
        title = "{Galaxy clustering at cosmic dawn from JWST/NIRCam observations to redshift z 11}",
      journal = {\mnras},
         year = 2024,
        month = sep,
       volume = {533},
       number = {2},
        pages = {2391-2398},
          doi = {10.1093/mnras/stae2006},
archivePrefix = {arXiv},
       eprint = {2402.18052},
 primaryClass = {astro-ph.GA},
       adsurl = {https://ui.adsabs.harvard.edu/abs/2024MNRAS.533.2391D}
}

@ARTICLE{Lim2025b,
       author = {{Lim}, Seunghwan and {Tacchella}, Sandro and {Maiolino}, Roberto and {Lovell}, Christopher C. and {Schaye}, Joop},
        title = "{Think inside the box: cosmic variance and large-scale conformity of high-redshift massive galaxies in the FLAMINGO simulations}",
      journal = {arXiv e-prints},
         year = 2025,
        month = nov,
          eid = {arXiv:2511.09618},
        pages = {arXiv:2511.09618},
          doi = {10.48550/arXiv.2511.09618},
archivePrefix = {arXiv},
       eprint = {2511.09618},
 primaryClass = {astro-ph.GA},
       adsurl = {https://ui.adsabs.harvard.edu/abs/2025arXiv251109618L}
}

@ARTICLE{Weibel2025,
       author = {{Weibel}, Andrea and {Jespersen}, Christian Kragh and {Oesch}, Pascal A. and {Williams}, Christina C. and {Bezanson}, Rachel and {Brammer}, Gabriel and {Cloonan}, Aidan P. and {Dayal}, Pratika and {Hutter}, Anne and {Ji}, Zhiyuan and {Maseda}, Michael V. and {Shuntov}, Marko and {Whitaker}, Katherine E.},
        title = "{Exploring Cosmic Dawn with PANORAMIC II: Cosmic Variance and Galaxy Clustering at $z\sim10$}",
      journal = {arXiv e-prints},
         year = 2025,
        month = dec,
          eid = {arXiv:2512.14212},
        pages = {arXiv:2512.14212},
          doi = {10.48550/arXiv.2512.14212},
archivePrefix = {arXiv},
       eprint = {2512.14212},
 primaryClass = {astro-ph.GA},
       adsurl = {https://ui.adsabs.harvard.edu/abs/2025arXiv251214212W}
}

@ARTICLE{Mead2015,
       author = {{Mead}, A.~J. and {Peacock}, J.~A. and {Heymans}, C. and {Joudaki}, S. and {Heavens}, A.~F.},
        title = "{An accurate halo model for fitting non-linear cosmological power spectra and baryonic feedback models}",
      journal = {\mnras},
         year = 2015,
        month = dec,
       volume = {454},
       number = {2},
        pages = {1958-1975},
          doi = {10.1093/mnras/stv2036},
archivePrefix = {arXiv},
       eprint = {1505.07833},
 primaryClass = {astro-ph.CO},
       adsurl = {https://ui.adsabs.harvard.edu/abs/2015MNRAS.454.1958M}
}

@ARTICLE{LandySzalay1993,
       author = {{Landy}, Stephen D. and {Szalay}, Alexander S.},
        title = "{Bias and Variance of Angular Correlation Functions}",
      journal = {\apj},
         year = 1993,
        month = jul,
       volume = {412},
        pages = {64},
          doi = {10.1086/172900},
       adsurl = {https://ui.adsabs.harvard.edu/abs/1993ApJ...412...64L}
}

@ARTICLE{Borrow2022, 
       author = {{Borrow}, Josh and {Schaller}, Matthieu and {Bower}, Richard G. and {Schaye}, Joop},
        title = "{SPHENIX: smoothed particle hydrodynamics for the next generation of galaxy formation simulations}",
      journal = {\mnras},
         year = 2022,
        month = apr,
       volume = {511},
       number = {2},
        pages = {2367-2389},
          doi = {10.1093/mnras/stab3166},
archivePrefix = {arXiv},
       eprint = {2012.03974},
 primaryClass = {astro-ph.GA},
       adsurl = {https://ui.adsabs.harvard.edu/abs/2022MNRAS.511.2367B}
}

@ARTICLE{Ludlow2019,
       author = {{Ludlow}, Aaron D. and {Schaye}, Joop and {Schaller}, Matthieu and {Richings}, Jack},
        title = "{Energy equipartition between stellar and dark matter particles in cosmological simulations results in spurious growth of galaxy sizes}",
      journal = {\mnras},
         year = 2019,
        month = sep,
       volume = {488},
       number = {1},
        pages = {L123-L128},
          doi = {10.1093/mnrasl/slz110},
archivePrefix = {arXiv},
       eprint = {1903.10110},
 primaryClass = {astro-ph.GA},
       adsurl = {https://ui.adsabs.harvard.edu/abs/2019MNRAS.488L.123L}
}

@ARTICLE{Richings2014b,
       author = {{Richings}, A.~J. and {Schaye}, J. and {Oppenheimer}, B.~D.},
        title = "{Non-equilibrium chemistry and cooling in the diffuse interstellar medium - II. Shielded gas}",
      journal = {\mnras},
         year = 2014,
        month = aug,
       volume = {442},
       number = {3},
        pages = {2780-2796},
          doi = {10.1093/mnras/stu1046},
archivePrefix = {arXiv},
       eprint = {1403.6155},
 primaryClass = {astro-ph.GA},
       adsurl = {https://ui.adsabs.harvard.edu/abs/2014MNRAS.442.2780R}
}

@ARTICLE{Richings2014a,
       author = {{Richings}, A.~J. and {Schaye}, J. and {Oppenheimer}, B.~D.},
        title = "{Non-equilibrium chemistry and cooling in the diffuse interstellar medium - I. Optically thin regime}",
      journal = {\mnras},
         year = 2014,
        month = jun,
       volume = {440},
       number = {4},
        pages = {3349-3369},
          doi = {10.1093/mnras/stu525},
archivePrefix = {arXiv},
       eprint = {1401.4719},
 primaryClass = {astro-ph.GA},
       adsurl = {https://ui.adsabs.harvard.edu/abs/2014MNRAS.440.3349R}
}

@ARTICLE{Ploeckinger2025,
       author = {{Ploeckinger}, Sylvia and {Richings}, Alexander J. and {Schaye}, Joop and {Trayford}, James W. and {Schaller}, Matthieu and {Chaikin}, Evgenii},
        title = "{HYBRID-CHIMES: a model for radiative cooling and the abundances of ions and molecules in simulations of galaxy formation}",
      journal = {\mnras},
         year = 2025,
        month = oct,
       volume = {543},
       number = {2},
        pages = {891-916},
          doi = {10.1093/mnras/staf1402},
archivePrefix = {arXiv},
       eprint = {2506.15773},
 primaryClass = {astro-ph.GA},
       adsurl = {https://ui.adsabs.harvard.edu/abs/2025MNRAS.543..891P}
}

@ARTICLE{ForouharMoreno2025,
       author = {{Forouhar Moreno}, Victor J. and {Helly}, John and {McGibbon}, Robert and {Schaye}, Joop and {Schaller}, Matthieu and {Han}, Jiaxin and {Kugel}, Roi and {Bah{\'e}}, Yannick M.},
        title = "{Assessing subhalo finders in cosmological hydrodynamical simulations}",
      journal = {\mnras},
         year = 2025,
        month = oct,
       volume = {543},
       number = {2},
        pages = {1339-1372},
          doi = {10.1093/mnras/staf1478},
archivePrefix = {arXiv},
       eprint = {2502.06932},
 primaryClass = {astro-ph.CO},
       adsurl = {https://ui.adsabs.harvard.edu/abs/2025MNRAS.543.1339F}
}

@ARTICLE{Nobels2024,
       author = {{Nobels}, Folkert S.~J. and {Schaye}, Joop and {Schaller}, Matthieu and {Ploeckinger}, Sylvia and {Chaikin}, Evgenii and {Richings}, Alexander J.},
        title = "{Tests of subgrid models for star formation using simulations of isolated disc galaxies}",
      journal = {\mnras},
         year = 2024,
        month = aug,
       volume = {532},
       number = {3},
        pages = {3299-3321},
          doi = {10.1093/mnras/stae1390},
archivePrefix = {arXiv},
       eprint = {2309.13750},
 primaryClass = {astro-ph.GA},
       adsurl = {https://ui.adsabs.harvard.edu/abs/2024MNRAS.532.3299N}
}

@ARTICLE{Ludlow2021,
       author = {{Ludlow}, Aaron D. and {Fall}, S. Michael and {Schaye}, Joop and {Obreschkow}, Danail},
        title = "{Spurious heating of stellar motions in simulated galactic discs by dark matter halo particles}",
      journal = {\mnras},
         year = 2021,
        month = dec,
       volume = {508},
       number = {4},
        pages = {5114-5137},
          doi = {10.1093/mnras/stab2770},
archivePrefix = {arXiv},
       eprint = {2105.03561},
 primaryClass = {astro-ph.GA},
       adsurl = {https://ui.adsabs.harvard.edu/abs/2021MNRAS.508.5114L}
}

@ARTICLE{Fakhouri2010,
       author = {{Fakhouri}, Onsi and {Ma}, Chung-Pei and {Boylan-Kolchin}, Michael},
        title = "{The merger rates and mass assembly histories of dark matter haloes in the two Millennium simulations}",
      journal = {\mnras},
         year = 2010,
        month = aug,
       volume = {406},
       number = {4},
        pages = {2267-2278},
          doi = {10.1111/j.1365-2966.2010.16859.x},
archivePrefix = {arXiv},
       eprint = {1001.2304},
 primaryClass = {astro-ph.CO},
       adsurl = {https://ui.adsabs.harvard.edu/abs/2010MNRAS.406.2267F}
}

@ARTICLE{Boylan-Kolchin2025,
       author = {{Boylan-Kolchin}, Michael},
        title = "{Accelerated by dark matter: a high-redshift pathway to efficient galaxy-scale star formation}",
      journal = {\mnras},
         year = 2025,
        month = apr,
       volume = {538},
       number = {4},
        pages = {3210-3218},
          doi = {10.1093/mnras/staf471},
archivePrefix = {arXiv},
       eprint = {2407.10900},
 primaryClass = {astro-ph.GA},
       adsurl = {https://ui.adsabs.harvard.edu/abs/2025MNRAS.538.3210B}
}

@ARTICLE{Bouwens2021,
       author = {{Bouwens}, R.~J. and {Oesch}, P.~A. and {Stefanon}, M. and {Illingworth}, G. and {Labb{\'e}}, I. and {Reddy}, N. and {Atek}, H. and {Montes}, M. and {Naidu}, R. and {Nanayakkara}, T. and {Nelson}, E. and {Wilkins}, S.},
        title = "{New Determinations of the UV Luminosity Functions from z   9 to 2 Show a Remarkable Consistency with Halo Growth and a Constant Star Formation Efficiency}",
      journal = {\aj},
         year = 2021,
        month = aug,
       volume = {162},
       number = {2},
          eid = {47},
        pages = {47},
          doi = {10.3847/1538-3881/abf83e},
archivePrefix = {arXiv},
       eprint = {2102.07775},
 primaryClass = {astro-ph.GA},
       adsurl = {https://ui.adsabs.harvard.edu/abs/2021AJ....162...47B}
}

@ARTICLE{SinhaGarrison2020,
    author = {{Sinha}, Manodeep and {Garrison}, Lehman H.},
    title = "{CORRFUNC - a suite of blazing fast correlation functions on
    the CPU}",
    journal = {\mnras},
    year = "2020",
    month = "Jan",
    volume = {491},
    number = {2},
    pages = {3022-3041},
    doi = {10.1093/mnras/stz3157},
    adsurl =
    {https://ui.adsabs.harvard.edu/abs/2020MNRAS.491.3022S}
}

@ARTICLE{Robertson2024,
       author = {{Robertson}, Brant and {Johnson}, Benjamin D. and {Tacchella}, Sandro and {Eisenstein}, Daniel J. and {Hainline}, Kevin and {Arribas}, Santiago and {Baker}, William M. and {Bunker}, Andrew J. and {Carniani}, Stefano and {Cargile}, Phillip A. and {Carreira}, Courtney and {Charlot}, Stephane and {Chevallard}, Jacopo and {Curti}, Mirko and {Curtis-Lake}, Emma and {D'Eugenio}, Francesco and {Egami}, Eiichi and {Hausen}, Ryan and {Helton}, Jakob M. and {Jakobsen}, Peter and {Ji}, Zhiyuan and {Jones}, Gareth C. and {Maiolino}, Roberto and {Maseda}, Michael V. and {Nelson}, Erica and {P{\'e}rez-Gonz{\'a}lez}, Pablo G. and {Pusk{\'a}s}, D{\'a}vid and {Rieke}, Marcia and {Smit}, Renske and {Sun}, Fengwu and {{\"U}bler}, Hannah and {Whitler}, Lily and {Williams}, Christina C. and {Willmer}, Christopher N.~A. and {Willott}, Chris and {Witstok}, Joris},
        title = "{Earliest Galaxies in the JADES Origins Field: Luminosity Function and Cosmic Star Formation Rate Density 300 Myr after the Big Bang}",
      journal = {\apj},
         year = 2024,
        month = jul,
       volume = {970},
       number = {1},
          eid = {31},
        pages = {31},
          doi = {10.3847/1538-4357/ad463d},
archivePrefix = {arXiv},
       eprint = {2312.10033},
 primaryClass = {astro-ph.GA},
       adsurl = {https://ui.adsabs.harvard.edu/abs/2024ApJ...970...31R}
}

@ARTICLE{Whitler2025,
       author = {{Whitler}, Lily and {Stark}, Daniel P. and {Topping}, Michael W. and {Robertson}, Brant and {Rieke}, Marcia and {Hainline}, Kevin N. and {Endsley}, Ryan and {Chen}, Zuyi and {Baker}, William M. and {Bhatawdekar}, Rachana and {Bunker}, Andrew J. and {Carniani}, Stefano and {Charlot}, St{\'e}phane and {Chevallard}, Jacopo and {Curtis-Lake}, Emma and {Egami}, Eiichi and {Eisenstein}, Daniel J. and {Helton}, Jakob M. and {Ji}, Zhiyuan and {Johnson}, Benjamin D. and {P{\'e}rez-Gonz{\'a}lez}, Pablo G. and {Rinaldi}, Pierluigi and {Tacchella}, Sandro and {Williams}, Christina C. and {Willmer}, Christopher N.~A. and {Willott}, Chris and {Witstok}, Joris},
        title = "{The z {\ensuremath{\gtrsim}} 9 Galaxy UV Luminosity Function from the JWST Advanced Deep Extragalactic Survey: Insights into Early Galaxy Evolution and Reionization}",
      journal = {\apj},
         year = 2025,
        month = oct,
       volume = {992},
       number = {1},
          eid = {63},
        pages = {63},
          doi = {10.3847/1538-4357/adfddc},
archivePrefix = {arXiv},
       eprint = {2501.00984},
 primaryClass = {astro-ph.GA},
       adsurl = {https://ui.adsabs.harvard.edu/abs/2025ApJ...992...63W}
}

@ARTICLE{Huang2026,
       author = {{Huang}, Jiamu and {Pizzati}, Elia and {Hennawi}, Joseph F. and {Schaye}, Joop and {Schaller}, Matthieu and {Snyder}, Benjamin and {Kang}, Yi},
        title = "{The Impact of Cosmic Variance and Satellites on JWST Clustering Measurements at Redshift around 6}",
      journal = {arXiv e-prints},
         year = 2026,
        month = may,
          eid = {arXiv:2605.11077},
        pages = {arXiv:2605.11077},
          doi = {10.48550/arXiv.2605.11077},
archivePrefix = {arXiv},
       eprint = {2605.11077},
 primaryClass = {astro-ph.CO},
       adsurl = {https://ui.adsabs.harvard.edu/abs/2026arXiv260511077H}
}

@ARTICLE{Shuntov2025C,
       author = {{Shuntov}, Marko and {Akins}, Hollis B. and {Paquereau}, Louise and {Casey}, Caitlin M. and {Ilbert}, Olivier and {Arango-Toro}, Rafael C. and {McCracken}, Henry Joy and {Franco}, Maximilien and {Harish}, Santosh and {Kartaltepe}, Jeyhan S. and {Koekemoer}, Anton M. and {Yang}, Lilan and {Huertas-Company}, Marc and {Berman}, Edward M. and {McCleary}, Jacqueline E. and {Toft}, Sune and {Gavazzi}, Rapha{\"e}l and {Achenbach}, Mark J. and {Bertin}, Emmanuel and {Brinch}, Malte and {Champagne}, Jackie and {Chartab}, Nima and {Drakos}, Nicole E. and {Egami}, Eiichi and {Endsley}, Ryan and {Faisst}, Andreas L. and {Fan}, Xiaohui and {Flayhart}, Carter and {Hartley}, William G. and {Hatamnia}, Hossein and {Gozaliasl}, Ghassem and {Gentile}, Fabrizio and {Jermann}, Iris and {Jin}, Shuowen and {Kakiichi}, Koki and {Khostovan}, Ali Ahmad and {K{\"u}mmel}, Martin and {Laigle}, Clotilde and {Laishram}, Ronaldo and {Lambrides}, Erini and {Liu}, Daizhong and {Lyu}, Jianwei and {Magdis}, Georgios and {Mobasher}, Bahram and {Moutard}, Thibaud and {Renzini}, Alvio and {Rich}, R. Michael and {Sanders}, David B. and {Sattari}, Zahra and {Robertson}, Brant E. and {Schefer}, Marc and {Scognamiglio}, Diana and {Scoville}, Nick and {Silverman}, John D. and {Taamoli}, Sina and {Trakhtenbrot}, Benny and {Valentino}, Francesco and {Wang}, Feige and {Weaver}, John R. and {Yang}, Jinyi},
        title = "{COSMOS2025: The COSMOS-Web galaxy catalog of photometry, morphology, redshifts, and physical parameters from JWST, HST, and ground-based imaging}",
      journal = {\aap},
         year = 2025,
        month = dec,
       volume = {704},
          eid = {A339},
        pages = {A339},
          doi = {10.1051/0004-6361/202555799},
archivePrefix = {arXiv},
       eprint = {2506.03243},
 primaryClass = {astro-ph.GA},
       adsurl = {https://ui.adsabs.harvard.edu/abs/2025A&A...704A.339S}
}

@ARTICLE{Spergel2013, 
       author = {{Spergel}, D. and {Gehrels}, N. and {Breckinridge}, J. and {Donahue}, M. and {Dressler}, A. and {Gaudi}, B.~S. and {Greene}, T. and {Guyon}, O. and {Hirata}, C. and {Kalirai}, J. and {Kasdin}, N.~J. and {Moos}, W. and {Perlmutter}, S. and {Postman}, M. and {Rauscher}, B. and {Rhodes}, J. and {Wang}, Y. and {Weinberg}, D. and {Centrella}, J. and {Traub}, W. and {Baltay}, C. and {Colbert}, J. and {Bennett}, D. and {Kiessling}, A. and {Macintosh}, B. and {Merten}, J. and {Mortonson}, M. and {Penny}, M. and {Rozo}, E. and {Savransky}, D. and {Stapelfeldt}, K. and {Zu}, Y. and {Baker}, C. and {Cheng}, E. and {Content}, D. and {Dooley}, J. and {Foote}, M. and {Goullioud}, R. and {Grady}, K. and {Jackson}, C. and {Kruk}, J. and {Levine}, M. and {Melton}, M. and {Peddie}, C. and {Ruffa}, J. and {Shaklan}, S.},
        title = "{Wide-Field InfraRed Survey Telescope-Astrophysics Focused Telescope Assets WFIRST-AFTA Final Report}",
      journal = {arXiv e-prints},
         year = 2013,
        month = may,
          eid = {arXiv:1305.5422},
        pages = {arXiv:1305.5422},
          doi = {10.48550/arXiv.1305.5422},
archivePrefix = {arXiv},
       eprint = {1305.5422},
 primaryClass = {astro-ph.IM},
       adsurl = {https://ui.adsabs.harvard.edu/abs/2013arXiv1305.5422S}
}

@ARTICLE{Spergel2015, 
       author = {{Spergel}, D. and {Gehrels}, N. and {Baltay}, C. and {Bennett}, D. and {Breckinridge}, J. and {Donahue}, M. and {Dressler}, A. and {Gaudi}, B.~S. and {Greene}, T. and {Guyon}, O. and {Hirata}, C. and {Kalirai}, J. and {Kasdin}, N.~J. and {Macintosh}, B. and {Moos}, W. and {Perlmutter}, S. and {Postman}, M. and {Rauscher}, B. and {Rhodes}, J. and {Wang}, Y. and {Weinberg}, D. and {Benford}, D. and {Hudson}, M. and {Jeong}, W.-S. and {Mellier}, Y. and {Traub}, W. and {Yamada}, T. and {Capak}, P. and {Colbert}, J. and {Masters}, D. and {Penny}, M. and {Savransky}, D. and {Stern}, D. and {Zimmerman}, N. and {Barry}, R. and {Bartusek}, L. and {Carpenter}, K. and {Cheng}, E. and {Content}, D. and {Dekens}, F. and {Demers}, R. and {Grady}, K. and {Jackson}, C. and {Kuan}, G. and {Kruk}, J. and {Melton}, M. and {Nemati}, B. and {Parvin}, B. and {Poberezhskiy}, I. and {Peddie}, C. and {Ruffa}, J. and {Wallace}, J.~K. and {Whipple}, A. and {Wollack}, E. and {Zhao}, F.},
        title = "{Wide-Field InfrarRed Survey Telescope-Astrophysics Focused Telescope Assets WFIRST-AFTA 2015 Report}",
      journal = {arXiv e-prints},
         year = 2015,
        month = mar,
          eid = {arXiv:1503.03757},
        pages = {arXiv:1503.03757},
          doi = {10.48550/arXiv.1503.03757},
archivePrefix = {arXiv},
       eprint = {1503.03757},
 primaryClass = {astro-ph.IM},
       adsurl = {https://ui.adsabs.harvard.edu/abs/2015arXiv150303757S}
}

@ARTICLE{Green2012, 
       author = {{Green}, J. and {Schechter}, P. and {Baltay}, C. and {Bean}, R. and {Bennett}, D. and {Brown}, R. and {Conselice}, C. and {Donahue}, M. and {Fan}, X. and {Gaudi}, B.~S. and {Hirata}, C. and {Kalirai}, J. and {Lauer}, T. and {Nichol}, B. and {Padmanabhan}, N. and {Perlmutter}, S. and {Rauscher}, B. and {Rhodes}, J. and {Roellig}, T. and {Stern}, D. and {Sumi}, T. and {Tanner}, A. and {Wang}, Y. and {Weinberg}, D. and {Wright}, E. and {Gehrels}, N. and {Sambruna}, R. and {Traub}, W. and {Anderson}, J. and {Cook}, K. and {Garnavich}, P. and {Hillenbrand}, L. and {Ivezic}, Z. and {Kerins}, E. and {Lunine}, J. and {McDonald}, P. and {Penny}, M. and {Phillips}, M. and {Rieke}, G. and {Riess}, A. and {van der Marel}, R. and {Barry}, R.~K. and {Cheng}, E. and {Content}, D. and {Cutri}, R. and {Goullioud}, R. and {Grady}, K. and {Helou}, G. and {Jackson}, C. and {Kruk}, J. and {Melton}, M. and {Peddie}, C. and {Rioux}, N. and {Seiffert}, M.},
        title = "{Wide-Field InfraRed Survey Telescope (WFIRST) Final Report}",
      journal = {arXiv e-prints},
         year = 2012,
        month = aug,
          eid = {arXiv:1208.4012},
        pages = {arXiv:1208.4012},
          doi = {10.48550/arXiv.1208.4012},
archivePrefix = {arXiv},
       eprint = {1208.4012},
 primaryClass = {astro-ph.IM},
       adsurl = {https://ui.adsabs.harvard.edu/abs/2012arXiv1208.4012G}
}

@ARTICLE{Michaux2021,
       author = {{Michaux}, Micha{\"e}l and {Hahn}, Oliver and {Rampf}, Cornelius and {Angulo}, Raul E.},
        title = "{Accurate initial conditions for cosmological N-body simulations: minimizing truncation and discreteness errors}",
      journal = {\mnras},
         year = 2021,
        month = jan,
       volume = {500},
       number = {1},
        pages = {663-683},
          doi = {10.1093/mnras/staa3149},
archivePrefix = {arXiv},
       eprint = {2008.09588},
 primaryClass = {astro-ph.CO},
       adsurl = {https://ui.adsabs.harvard.edu/abs/2021MNRAS.500..663M}
}

@software{Hahn2020,
       author = {{Hahn}, Oliver and {Michaux}, Micha{\"e}l and {Rampf}, Cornelius and {Uhlemann}, Cora and {Angulo}, Raul E.},
        title = "{MUSIC2-monofonIC: 3LPT initial condition generator}",
 howpublished = {Astrophysics Source Code Library, record ascl:2008.024},
         year = 2020,
        month = aug,
          eid = {ascl:2008.024},
archivePrefix = {ascl},
       eprint = {2008.024},
       adsurl = {https://ui.adsabs.harvard.edu/abs/2020ascl.soft08024H}
}

@ARTICLE{AnguloPontzen2016,
       author = {{Angulo}, Raul E. and {Pontzen}, Andrew},
        title = "{Cosmological N-body simulations with suppressed variance}",
      journal = {\mnras},
         year = 2016,
        month = oct,
       volume = {462},
       number = {1},
        pages = {L1-L5},
          doi = {10.1093/mnrasl/slw098},
archivePrefix = {arXiv},
       eprint = {1603.05253},
 primaryClass = {astro-ph.CO},
       adsurl = {https://ui.adsabs.harvard.edu/abs/2016MNRAS.462L...1A}
}

@ARTICLE{vanDaalen2014,
       author = {{van Daalen}, Marcel P. and {Schaye}, Joop and {McCarthy}, Ian G. and {Booth}, C.~M. and {Dalla Vecchia}, Claudio},
        title = "{The impact of baryonic processes on the two-point correlation functions of galaxies, subhaloes and matter}",
      journal = {\mnras},
         year = 2014,
        month = jun,
       volume = {440},
       number = {4},
        pages = {2997-3010},
          doi = {10.1093/mnras/stu482},
archivePrefix = {arXiv},
       eprint = {1310.7571},
 primaryClass = {astro-ph.CO},
       adsurl = {https://ui.adsabs.harvard.edu/abs/2014MNRAS.440.2997V}
}

@ARTICLE{Gao2005,
       author = {{Gao}, Liang and {Springel}, Volker and {White}, Simon D.~M.},
        title = "{The age dependence of halo clustering}",
      journal = {\mnras},
         year = 2005,
        month = oct,
       volume = {363},
       number = {1},
        pages = {L66-L70},
          doi = {10.1111/j.1745-3933.2005.00084.x},
archivePrefix = {arXiv},
       eprint = {astro-ph/0506510},
 primaryClass = {astro-ph},
       adsurl = {https://ui.adsabs.harvard.edu/abs/2005MNRAS.363L..66G}
}

@ARTICLE{Wechsler2006,
       author = {{Wechsler}, Risa H. and {Zentner}, Andrew R. and {Bullock}, James S. and {Kravtsov}, Andrey V. and {Allgood}, Brandon},
        title = "{The Dependence of Halo Clustering on Halo Formation History, Concentration, and Occupation}",
      journal = {\apj},
         year = 2006,
        month = nov,
       volume = {652},
       number = {1},
        pages = {71-84},
          doi = {10.1086/507120},
archivePrefix = {arXiv},
       eprint = {astro-ph/0512416},
 primaryClass = {astro-ph},
       adsurl = {https://ui.adsabs.harvard.edu/abs/2006ApJ...652...71W}
}

@ARTICLE{WuZ2026,
       author = {{Wu}, Zihao and {Eisenstein}, Daniel J. and {Johnson}, Benjamin D. and {Hainline}, Kevin and {Baker}, William M. and {Bunker}, Andrew J. and {Cameron}, Alex J. and {Curtis-Lake}, Emma and {Danhaive}, A. Lola and {Hausen}, Ryan and {Helton}, Jakob M. and {Ji}, Zhiyuan and {Looser}, Tobias J. and {Maiolino}, Roberto and {Mengistu}, Petra and {Rinaldi}, Pierluigi and {Robertson}, Brant E. and {Sun}, Fengwu and {Tacchella}, Sandro and {Trussler}, James A.~A. and {Williams}, Christina C. and {Willmer}, Christopher N.~A. and {Witstok}, Joris},
        title = "{JADES: A Prominent Galaxy Overdensity Candidate within the First 500 Myr}",
      journal = {arXiv e-prints},
         year = 2026,
        month = jan,
          eid = {arXiv:2601.15960},
        pages = {arXiv:2601.15960},
          doi = {10.48550/arXiv.2601.15960},
archivePrefix = {arXiv},
       eprint = {2601.15960},
 primaryClass = {astro-ph.GA},
       adsurl = {https://ui.adsabs.harvard.edu/abs/2026arXiv260115960W}
}

@ARTICLE{TrentiStiavelli2008,
       author = {{Trenti}, M. and {Stiavelli}, M.},
        title = "{Cosmic Variance and Its Effect on the Luminosity Function Determination in Deep High-z Surveys}",
      journal = {\apj},
         year = 2008,
        month = apr,
       volume = {676},
       number = {2},
        pages = {767-780},
          doi = {10.1086/528674},
archivePrefix = {arXiv},
       eprint = {0712.0398},
 primaryClass = {astro-ph},
       adsurl = {https://ui.adsabs.harvard.edu/abs/2008ApJ...676..767T}
}

@ARTICLE{Somerville2004, 
       author = {{Somerville}, Rachel S. and {Lee}, Kyoungsoo and {Ferguson}, Henry C. and {Gardner}, Jonathan P. and {Moustakas}, Leonidas A. and {Giavalisco}, Mauro},
        title = "{Cosmic Variance in the Great Observatories Origins Deep Survey}",
      journal = {\apjl},
         year = 2004,
        month = jan,
       volume = {600},
       number = {2},
        pages = {L171-L174},
          doi = {10.1086/378628},
archivePrefix = {arXiv},
       eprint = {astro-ph/0309071},
 primaryClass = {astro-ph},
       adsurl = {https://ui.adsabs.harvard.edu/abs/2004ApJ...600L.171S}
}

@ARTICLE{NewmanDavis2002,
       author = {{Newman}, Jeffrey A. and {Davis}, Marc},
        title = "{Measuring the Cosmic Equation of State with Counts of Galaxies. II. Error Budget for the DEEP2 Redshift Survey}",
      journal = {\apj},
         year = 2002,
        month = jan,
       volume = {564},
       number = {2},
        pages = {567-575},
          doi = {10.1086/324148},
archivePrefix = {arXiv},
       eprint = {astro-ph/0109130},
 primaryClass = {astro-ph},
       adsurl = {https://ui.adsabs.harvard.edu/abs/2002ApJ...564..567N}
}

@ARTICLE{Robertson2010,
       author = {{Robertson}, Brant E.},
        title = "{A Method for Measuring the Bias of High-redshift Galaxies from Cosmic Variance}",
      journal = {\apjl},
         year = 2010,
        month = jun,
       volume = {716},
       number = {2},
        pages = {L229-L234},
          doi = {10.1088/2041-8205/716/2/L229},
archivePrefix = {arXiv},
       eprint = {1005.4927},
 primaryClass = {astro-ph.CO},
       adsurl = {https://ui.adsabs.harvard.edu/abs/2010ApJ...716L.229R}
}

@ARTICLE{Trayford2026, 
       author = {{Trayford}, James W. and {Schaye}, Joop and {Correa}, Camila and {Ploeckinger}, Sylvia and {Richings}, Alexander J. and {Chaikin}, Evgenii and {Schaller}, Matthieu and {Ben{\'\i}tez-Llambay}, Alejandro and {Frenk}, Carlos and {Hu{\v{s}}ko}, Filip},
        title = "{Modelling the evolution and influence of dust in cosmological simulations that include the cold phase of the interstellar medium}",
      journal = {\mnras},
         year = 2026,
        month = feb,
       volume = {545},
       number = {4},
          eid = {staf2040},
        pages = {staf2040},
          doi = {10.1093/mnras/staf2040},
archivePrefix = {arXiv},
       eprint = {2505.13056},
 primaryClass = {astro-ph.GA},
       adsurl = {https://ui.adsabs.harvard.edu/abs/2026MNRAS.545f2040T}
}

@ARTICLE{DallaVecchiaSchaye2012,
       author = {{Dalla Vecchia}, Claudio and {Schaye}, Joop},
        title = "{Simulating galactic outflows with thermal supernova feedback}",
      journal = {\mnras},
         year = 2012,
        month = oct,
       volume = {426},
       number = {1},
        pages = {140-158},
          doi = {10.1111/j.1365-2966.2012.21704.x},
archivePrefix = {arXiv},
       eprint = {1203.5667},
 primaryClass = {astro-ph.GA},
       adsurl = {https://ui.adsabs.harvard.edu/abs/2012MNRAS.426..140D}
}

@ARTICLE{Correa2026, 
       author = {{Correa}, Camila A. and {Schaye}, Joop and {Schaller}, Matthieu and {Trayford}, James W. and {Chaikin}, Evgenii and {Ben{\'\i}tez-Llambay}, Alejandro and {Frenk}, Carlos S. and {Ploeckinger}, Sylvia and {Richings}, Alexander J.},
        title = "{A subgrid model for chemical enrichment in cosmological simulations}",
      journal = {\mnras},
         year = 2026,
        month = may,
       volume = {548},
       number = {3},
          eid = {stag645},
        pages = {stag645},
          doi = {10.1093/mnras/stag645},
archivePrefix = {arXiv},
       eprint = {2604.00980},
 primaryClass = {astro-ph.GA},
       adsurl = {https://ui.adsabs.harvard.edu/abs/2026MNRAS.548ag645C}
}

@ARTICLE{Benitez-Llambay2026, 
       author = {{Ben{\'\i}tez-Llambay}, Alejandro and {Ploeckinger}, Sylvia and {Schaye}, Joop and {Richings}, Alexander J. and {Chaikin}, Evgenii and {Schaller}, Matthieu and {Trayford}, James W. and {Frenk}, Carlos S. and {Hu{\v{s}}ko}, Filip and {Correa}, Camila},
        title = "{Non-explosive pre-supernova feedback in the COLIBRE model of galaxy formation}",
      journal = {\mnras},
         year = 2026,
        month = mar,
       volume = {546},
       number = {4},
          eid = {stag268},
        pages = {stag268},
          doi = {10.1093/mnras/stag268},
archivePrefix = {arXiv},
       eprint = {2509.25309},
 primaryClass = {astro-ph.GA},
       adsurl = {https://ui.adsabs.harvard.edu/abs/2026MNRAS.546ag268B}
}

@ARTICLE{BruzualCharlot2003, 
       author = {{Bruzual}, G. and {Charlot}, S.},
        title = "{Stellar population synthesis at the resolution of 2003}",
      journal = {\mnras},
         year = 2003,
        month = oct,
       volume = {344},
       number = {4},
        pages = {1000-1028},
          doi = {10.1046/j.1365-8711.2003.06897.x},
archivePrefix = {arXiv},
       eprint = {astro-ph/0309134},
 primaryClass = {astro-ph},
       adsurl = {https://ui.adsabs.harvard.edu/abs/2003MNRAS.344.1000B}
}

@ARTICLE{Limber1953,
       author = {{Limber}, D. Nelson},
        title = "{The Analysis of Counts of the Extragalactic Nebulae in Terms of a Fluctuating Density Field.}",
      journal = {\apj},
         year = 1953,
        month = jan,
       volume = {117},
        pages = {134},
          doi = {10.1086/145672},
       adsurl = {https://ui.adsabs.harvard.edu/abs/1953ApJ...117..134L}
}

@ARTICLE{Cowley2018, 
       author = {{Cowley}, William I. and {Baugh}, Carlton M. and {Cole}, Shaun and {Frenk}, Carlos S. and {Lacey}, Cedric G.},
        title = "{Predictions for deep galaxy surveys with JWST from {\ensuremath{\Lambda}}CDM}",
      journal = {\mnras},
         year = 2018,
        month = feb,
       volume = {474},
       number = {2},
        pages = {2352-2372},
          doi = {10.1093/mnras/stx2897},
archivePrefix = {arXiv},
       eprint = {1702.02146},
 primaryClass = {astro-ph.GA},
       adsurl = {https://ui.adsabs.harvard.edu/abs/2018MNRAS.474.2352C}
}

@ARTICLE{LuS2025,
       author = {{Lu}, Shengdong and {Frenk}, Carlos S. and {Bose}, Sownak and {Lacey}, Cedric G. and {Cole}, Shaun and {Baugh}, Carlton M. and {Helly}, John C.},
        title = "{A comparison of pre-existing {\ensuremath{\Lambda}}CDM predictions with the abundance of JWST galaxies at high redshift}",
      journal = {\mnras},
         year = 2025,
        month = jan,
       volume = {536},
       number = {1},
        pages = {1018-1034},
          doi = {10.1093/mnras/stae2646},
archivePrefix = {arXiv},
       eprint = {2406.02672},
 primaryClass = {astro-ph.GA},
       adsurl = {https://ui.adsabs.harvard.edu/abs/2025MNRAS.536.1018L}
}

@ARTICLE{LuS2026,
       author = {{Lu}, Shengdong and {Frenk}, Carlos S. and {Lacey}, Cedric G. and {Gebek}, Andrea and {Schaye}, Joop and {Cole}, Shaun and {Bose}, Sownak and {Durrant}, Anna and {Andreadis}, Nick and {Baes}, Maarten and {Ben{\'\i}tez-Llambay}, Alejandro and {Chaikin}, Evgenii and {Correa}, Camila and {Crain}, Robert A. and {Hu{\v{s}}ko}, Filip and {McGibbon}, Robert J. and {Ploeckinger}, Sylvia and {Richings}, Alexander J. and {Schaller}, Matthieu and {Trayford}, James W.},
        title = "{The galaxy ultraviolet luminosity function from $z=7$ to $15$ in the COLIBRE simulations}",
      journal = {arXiv e-prints},
         year = 2026,
        month = may,
          eid = {arXiv:2605.06782},
        pages = {arXiv:2605.06782},
          doi = {10.48550/arXiv.2605.06782},
archivePrefix = {arXiv},
       eprint = {2605.06782},
 primaryClass = {astro-ph.GA},
       adsurl = {https://ui.adsabs.harvard.edu/abs/2026arXiv260506782L}
}

@ARTICLE{Baugh2005, 
       author = {{Baugh}, C.~M. and {Lacey}, C.~G. and {Frenk}, C.~S. and {Granato}, G.~L. and {Silva}, L. and {Bressan}, A. and {Benson}, A.~J. and {Cole}, S.},
        title = "{Can the faint submillimetre galaxies be explained in the {\ensuremath{\Lambda}} cold dark matter model?}",
      journal = {\mnras},
         year = 2005,
        month = jan,
       volume = {356},
       number = {3},
        pages = {1191-1200},
          doi = {10.1111/j.1365-2966.2004.08553.x},
archivePrefix = {arXiv},
       eprint = {astro-ph/0406069},
 primaryClass = {astro-ph},
       adsurl = {https://ui.adsabs.harvard.edu/abs/2005MNRAS.356.1191B}
}

@article{Conroy2010,
    author = {Conroy, Charlie and Gunn, James E.},
    title = {The Propagation of Uncertainties in Stellar Population Synthesis Modeling. III. Model Calibration, Comparison, and Evaluation},
    journal = {ApJ},
    volume = {712},
    pages = {833},
    year = {2010},
    doi = {10.1088/0004-637X/712/2/833}
}

\appendix

\section{Validation of clustering-based halo mass inference using direct COLIBRE properties}
\label{sec_appA}

\begin{figure*}
\includegraphics[width=1.0\linewidth]{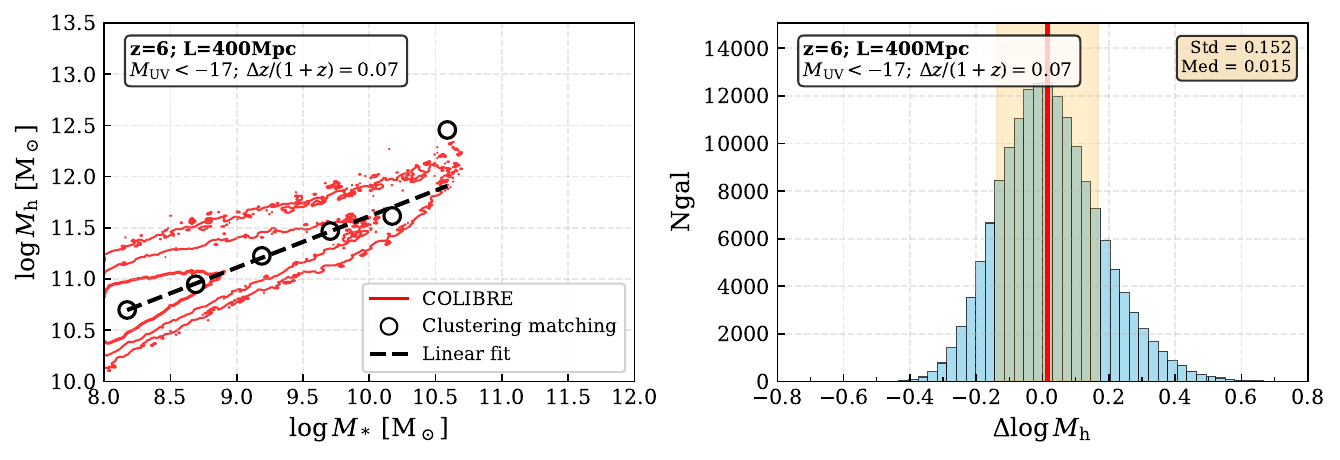}
\vspace{-0.5cm}
\caption{Validation of clustering-based halo mass inference using galaxy and halo properties taken directly from the COLIBRE simulation, with UV magnitudes from HOMA for selection. \textit{Left:} Stellar-to-halo mass relation (SHMR) inferred by our clustering matching method (circles; power-law fit shown as black dashed line) for stellar mass bins of width $0.5$ dex at redshift $z=6$, for a $(400$ cMpc$)^3$ volume, a UV magnitude cut of $M_{\rm UV, cut}=-17$, and a redshift uncertainty of $\Delta z/(1+z)=0.07$. The red contours indicate the true SHMR distribution of the COLIBRE simulation (68\%, 95\%, and 99.7\% contours). \textit{Right:} Distribution of residuals $\Delta \log M_{\rm h} = \log M_{\rm h,true} - \log M_{\rm h,CM}$ between the true halo mass and that inferred from clustering matching (using the power-law fit from the left panel). The red vertical line and orange band indicate the median and $1\sigma$ dispersion of the distribution, respectively. The combined uncertainty $\sigma_{\rm comb} = \sqrt{\mathrm{median}(\Delta \log M_{\rm h})^2 + \sigma_{\Delta \log M_{\rm h}}^2}$ is consistent with the intrinsic scatter of the COLIBRE simulation, confirming that our method recovers the underlying SHMR dispersion without introducing bias or additional uncertainty.}
\label{fig_AppA}
\end{figure*}

In Sect.~\ref{ssec_validation}, we validated our clustering-based halo mass inference using mock catalogues constructed by assigning HOMA-predicted galaxy properties to COLIBRE haloes. Here, we present an independent cross-validation using galaxy and halo properties taken directly from the COLIBRE hydrodynamical simulation for stellar masses and halo masses, while still relying on HOMA for the UV magnitudes needed to apply the selection cuts. This tests whether our method performs equally well when applied to a different baryonic model with different intrinsic scatter. It also confirms that our conclusions are not contingent on the specific assumptions of the HOMA model.

Fig.~\ref{fig_AppA} shows the same validation diagnostics as in Fig.~\ref{fig_validation}, but for the direct COLIBRE sample. The COLIBRE SHMR exhibits an intrinsic scatter approximately twice as large as that of the HOMA model at fixed stellar mass. Despite this larger scatter, our clustering matching recovers the true mean relation with deviations comparable to those found in the HOMA-based validation. The combined uncertainty $\sigma_{\rm comb} = \sqrt{\mathrm{median}(\Delta \log M_{\rm h})^2 + \sigma_{\Delta \log M_{\rm h}}^2}$ is consistent with the intrinsic scatter of the COLIBRE simulation, confirming that our method faithfully traces the underlying dispersion without introducing bias or artificially inflating the scatter.

This cross-validation demonstrates that the clustering matching technique robustly recovers the true halo masses and SHMR for different galaxy formation models with varying levels of intrinsic scatter, and that the performance metrics derived from the HOMA mock catalogues in Sect.~\ref{ssec_validation} are representative of the method's general capabilities.

\section{Impact of satellite assignment on clustering-based halo mass inference}
\label{sec_appB}

\begin{figure}
\includegraphics[width=0.98\linewidth]{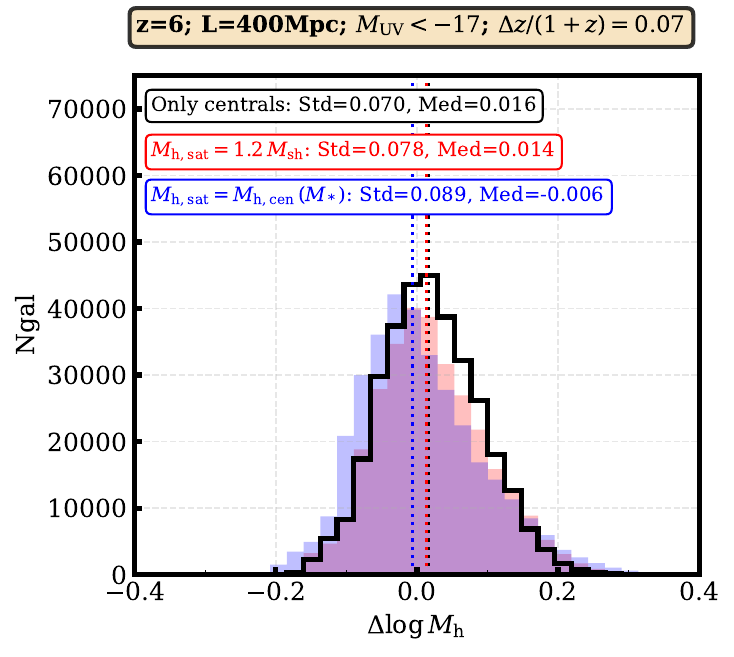}
\vspace{-0.2cm}
\caption{Comparison of clustering-based halo mass inference for different satellite assignment approaches. The histograms show the distribution of residuals $\Delta\log M_{\rm h} = \log M_{\rm h,true} - \log M_{\rm h,CM}$ at $z=6$ for a $(400\,\mathrm{cMpc})^{3}$ volume, $M_{\rm UV}<-17$, and $\Delta z/(1+z)=0.07$. The black histogram shows the centrals-only case (fiducial), while the red histogram shows the case where subhalo masses are scaled by a factor of 1.2 to account for the typical mass loss of satellites after accretion. The blue histogram shows the case where satellites are directly assigned the halo mass that would match centrals of the same stellar mass. Vertical dotted lines indicate the median of each distribution. The standard deviations and medians are reported in the legend. All three cases yield nearly identical distributions, demonstrating that our method is robust to the treatment of satellite galaxies.}
\label{fig_appB}
\end{figure}

In Sect.~\ref{ssec_mock}, we described two approaches for assigning halo masses to satellite galaxies: (i) scaling their bound subhalo masses by a factor of 1.2 to account for the typical mass loss of satellites after accretion, and (ii) directly assigning them the halo mass that would match centrals of the same stellar mass. Here we present the results of these tests.

Fig.~\ref{fig_appB} shows the distribution of residuals $\Delta\log M_{\rm h} = \log M_{\rm h,true} - \log M_{\rm h,CM}$ for the three cases: centrals-only (our fiducial approach), scaled bound mass, and centric-equivalent. The three distributions are nearly indistinguishable. This confirms that the inclusion of satellites with either assignment approach does not introduce significant bias or additional uncertainty in the inferred halo masses.

\section{Performance of the method as a function of fitting scale}
\label{sec_appC}

\begin{figure}
\includegraphics[width=1.0\linewidth]{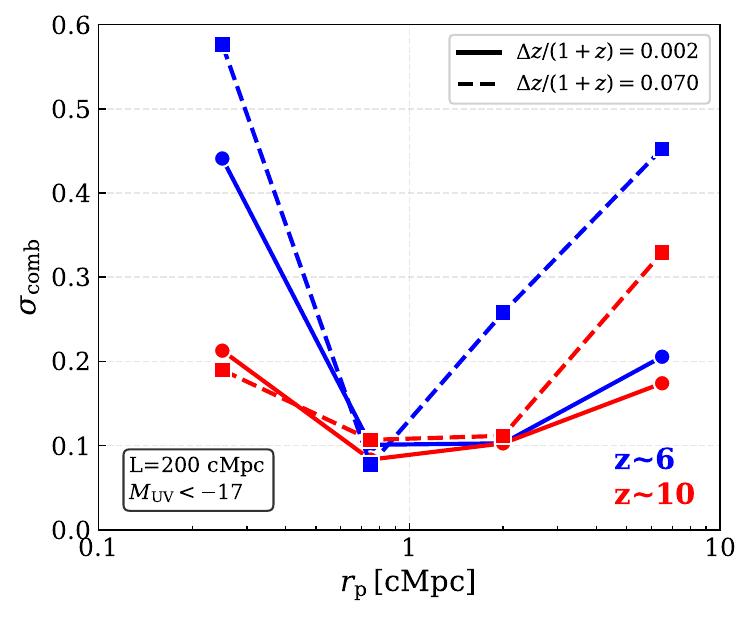}
\vspace{-0.5cm}
\caption{Performance of the clustering-based halo mass assignment as a function of the projected radial bin $r_{\rm p}$ used for fitting. The performance is quantified by $\sigma_{\rm comb} = \sqrt{\mathrm{median}(\Delta \log M_{\rm h})^2 + \sigma_{\Delta \log M_{\rm h}}^2}$, which combines the bias (median offset) and scatter (standard deviation) in the inferred halo mass relative to the true mass. Results are shown for two redshifts, $z\sim6$ (blue) and $z\sim10$ (red), and two levels of redshift uncertainty: $\Delta z/(1+z)=0.002$ (solid) and $0.070$ (dashed). The box size and UV magnitude cut are fixed at $200\,\mathrm{cMpc}$ and $M_{\rm UV}<-17$, respectively, for illustration, although the same qualitative trends hold for other survey configurations. The strong dependence on radial scale, with $\sigma_{\rm comb}$ varying by factors of $>5$, reflects the dominance of different physical effects: non-linear object-to-object scatter on small scales ($r_{\rm p}\lesssim0.5$\,cMpc) and cosmic variance on large scales ($r_{\rm p}\gtrsim3$\,cMpc). The minimum near $0.5<r_{\rm p}/\mathrm{cMpc}<1.0$ indicates that this radial range provides the optimal balance, minimizing both sources of uncertainty.}
\label{fig_appC}
\end{figure}

Fig.~\ref{fig_appC} demonstrates the performance of our clustering-based matching method, using the combined metric $\sigma_{\rm comb} = \sqrt{\mathrm{median}(\Delta \log M_{\rm h})^2 + \sigma_{\Delta \log M_{\rm h}}^2}$ as a proxy for the bias and dispersion in inferred halo masses relative to the true masses, evaluated across different projected radial bins used in the SHMR fitting. The results reveal a strong dependence on the chosen radial scale, with $\sigma_{\rm comb}$ varying by factors of $>5$ between the smallest and largest scales. This sensitivity is driven by two distinct physical effects that dominate at different radial ranges.

On small scales ($r_{\rm p} \lesssim 0.5$\,cMpc), the large values of $\sigma_{\rm comb}$ arise primarily from non-linear effects in the one-halo regime. At these separations, the clustering signal is sensitive to the detailed internal structure of haloes, including the distribution and occupation of satellite galaxies, as well as object-by-object variations in the galaxy--halo connection. These effects introduce significant scatter in the measured clustering strength for a given halo mass, which propagates into large uncertainties in the inferred SHMR. Importantly, this scatter is largely stochastic and reflects intrinsic variations in how galaxies populate their haloes rather than systematic offsets in the mean clustering amplitude.

On large scales ($r_{\rm p} \gtrsim 3$\,cMpc), the increase in $\sigma_{\rm comb}$ is instead driven by cosmic variance (see Sect.~\ref{ssec_CV}). At these separations, the clustering signal probes the large-scale density field, which varies substantially between different survey volumes. As shown in Fig.~\ref{fig_CV}, the volume-to-volume variance in $w_{\rm p}$ can reach nearly an order of magnitude at $r_{\rm p}\gtrsim3$\,cMpc, even for subvolumes of $(200$\,cMpc$)^3$. This cosmic variance affects the measured clustering amplitude systematically across an entire subvolume, leading to biases in the inferred SHMR that are coherent across all galaxies within that volume.

The contrasting nature of these two sources of uncertainty can be illustrated by comparing Fig.~\ref{fig_appC} with Fig.~\ref{fig_CV} for the same configuration: $z=6$, subvolume of ($200$\,cMpc)$^3$, $\Delta z/(1+z)=0.07$, and $M_{\rm UV}<-17$. On small scales ($r_{\rm p}<0.5$\,cMpc), the values of $\sigma_{\rm comb}$ in Fig.~\ref{fig_appC} are significantly larger than the cosmic variance contribution inferred from Fig.~\ref{fig_CV}. This indicates that on these scales, the dominant source of uncertainty is not the volume-to-volume variation in the mean clustering amplitude, but rather the object-to-object scatter arising from non-linear halo physics. Conversely, on large scales ($r_{\rm p}>3$\,cMpc), the two metrics become comparable, consistent with cosmic variance being the primary driver of the uncertainty.

The intermediate range $0.5<r_{\rm p}/\mathrm{cMpc}<1.0$ strikes an optimal balance, minimizing both the non-linear object-to-object scatter that plagues small scales and the cosmic variance that dominates on large scales. This range lies in the transition between the one-halo and two-halo regimes, where the clustering signal is robustly determined by the underlying halo mass function and is relatively insensitive to both the detailed internal structure of haloes and the large-scale density fluctuations. Consequently, this is the range where our method achieves the highest precision and accuracy, as reflected by the minimum in $\sigma_{\rm comb}$ around $r_{\rm p}\sim0.75$\,cMpc. Although the results shown here are for a fixed box size of $200\,\mathrm{cMpc}$ and a magnitude cut of $M_{\rm UV}<-17$, we find consistent conclusions across other survey configurations, confirming that $0.5<r_{\rm p}/\mathrm{cMpc}<1.0$ provides the optimal clustering scale for robust halo mass inference.

\section{Tabulated cross-correlation values}
\label{sec_appD}

\begin{table*}
 \renewcommand{\arraystretch}{1.4} 
 \centering
  \begin{minipage}{151mm}
\caption{Cross-correlation values at $r_{\rm p}=0.75$\,cMpc, the median of our fiducial radial range $0.5<r_{\rm p}/\mathrm{cMpc}<1.0$, as a function of halo mass at $z=6$ and $12$ for redshift uncertainties of $\Delta z/(1+z)=0.002$ and $0.07$. The values are derived from measurements across 64 subvolumes of $(100$ cMpc$)^3$ each, and the quoted errors represent the 16th--84th percentile range of the subvolume distribution.}
  \begin{tabular}{lcccccc}
\hline
 & $\log M_{\rm h}/{\rm M}_\odot$ & $M_{\rm UV}<-13$ & $M_{\rm UV}<-15$ & $M_{\rm UV}<-17$ & $M_{\rm UV}<-19$ & $M_{\rm UV}<-21$ \\ 
\hline
\hline
& & & \hspace{0.7cm} $\Delta z/(1+z)=0.002$ & & & \\
\hline
$z=6$ & $10.2$ & $1.9_{-0.15}^{+0.17}$ & $2.5_{-0.30}^{+0.21}$ & $3.5_{-0.44}^{+0.30}$ & $5.1_{-1.8}^{+1.5}$ & $6.7_{-6.7}^{+7.8}$ \\
 & $10.7$ & $2.5_{-0.22}^{+0.17}$ & $3.5_{-0.35}^{+0.23}$ & $4.9_{-0.44}^{+0.52}$ & $7.2_{-1.1}^{+1.4}$ & $15_{-5.1}^{+10}$ \\
 & $11.2$ & $3.5_{-0.30}^{+0.24}$ & $5.1_{-0.57}^{+0.53}$ & $7.7_{-0.98}^{+0.63}$ & $11_{-2.3}^{+1.9}$ & $28_{-12}^{+14}$ \\
 & $11.7$ & $4.9_{-0.58}^{+0.58}$ & $7.3_{-1.1}^{+1.1}$ & $12_{-1.1}^{+2.3}$ & $19_{-4.2}^{+4.5}$ & $38_{-38}^{+100}$ \\
 & $12.1$ & $7.2_{-3.0}^{+2.3}$ & $11_{-3.8}^{+4.9}$ & $17_{-6.5}^{+12}$ & $31_{-29}^{+36}$ & $77_{-77}^{+180}$ \\
 & $12.6$ & $13_{-3.2}^{+6.3}$ & $20_{-3.8}^{+12}$ & $36_{-4.6}^{+12}$ & $78_{-3.5}^{+14}$ & $-$ \\
$z=12$ & $10.1$ & $13_{-1.9}^{+3.4}$ & $17_{-2.9}^{+4.5}$ & $27_{-7.9}^{+8.2}$ & $47_{-48}^{+35}$ & $-$ \\
 & $10.6$ & $21_{-8.3}^{+14}$ & $28_{-12}^{+27}$ & $44_{-44}^{+80}$ & $-$ & $-$ \\
 & $11.1$ & $30_{-9.2}^{+9.2}$ & $52_{-24}^{+24}$ & $93_{-11}^{+11}$ & $-$ & $-$ \\
\hline
\hline
& & & \hspace{0.6cm} $\Delta z/(1+z)=0.07$ & & & \\
\hline
$z=6$ & $10.2$ & $0.10_{-0.04}^{+0.03}$ & $0.12_{-0.05}^{+0.03}$ & $0.16_{-0.05}^{+0.05}$ & $0.22_{-0.13}^{+0.15}$ & $0.47_{-0.65}^{+0.60}$ \\
 & $10.7$ & $0.12_{-0.04}^{+0.03}$ & $0.16_{-0.05}^{+0.04}$ & $0.23_{-0.07}^{+0.06}$ & $0.34_{-0.12}^{+0.08}$ & $0.78_{-0.44}^{+0.49}$ \\
 & $11.2$ & $0.17_{-0.04}^{+0.05}$ & $0.24_{-0.07}^{+0.05}$ & $0.36_{-0.08}^{+0.06}$ & $0.53_{-0.19}^{+0.11}$ & $1.2_{-0.54}^{+0.75}$ \\
 & $11.7$ & $0.24_{-0.10}^{+0.08}$ & $0.34_{-0.10}^{+0.12}$ & $0.50_{-0.08}^{+0.17}$ & $0.81_{-0.23}^{+0.33}$ & $1.8_{-2.1}^{+4.9}$ \\
 & $12.1$ & $0.33_{-0.31}^{+0.46}$ & $0.57_{-0.45}^{+0.34}$ & $0.73_{-0.47}^{+0.61}$ & $1.3_{-1.2}^{+1.2}$ & $1.9_{-2.0}^{+7.2}$ \\
 & $12.6$ & $0.31_{-0.25}^{+0.91}$ & $1.1_{-0.44}^{+0.20}$ & $1.2_{-0.11}^{+1.0}$ & $2.3_{-0.89}^{+0.64}$ & $-$ \\
$z=12$ & $10.1$ & $0.60_{-0.20}^{+0.14}$ & $0.75_{-0.20}^{+0.20}$ & $1.1_{-0.44}^{+0.47}$ & $1.8_{-1.7}^{+2.1}$ & $-$ \\
 & $10.6$ & $0.86_{-0.50}^{+0.63}$ & $1.1_{-0.57}^{+1.1}$ & $1.7_{-2.0}^{+3.5}$ & $-$ & $-$ \\
 & $11.1$ & $0.96_{-0.64}^{+0.64}$ & $2.0_{-0.74}^{+0.74}$ & $2.9_{-1.0}^{+1.0}$ & $-$ & $-$ \\
\hline
\\
\end{tabular}
\label{tab_corr}
\end{minipage}
\vspace{-0.2cm}
\end{table*}

In Sect.~\ref{ssec_SHMR}, we use the cross-correlation between the full galaxy sample and haloes in bins of halo mass as reference templates for our clustering matching method. To facilitate reproducibility and to provide a quantitative reference for the clustering amplitudes, we tabulate the cross-correlation values at $r_{\rm p}=0.75$\,cMpc, the median of our fiducial radial range $0.5<r_{\rm p}/\mathrm{cMpc}<1.0$.

The values presented in Table~\ref{tab_corr} are derived from measurements across 64 subvolumes of $(100$ cMpc$)^3$ each, obtained by dividing the full volume of the L400m7 COLIBRE simulation box. Results are shown for redshifts $z=6$ and $12$, for redshift uncertainties of $\Delta z/(1+z)=0.002$ and $0.07$, and for six halo mass bins spanning $\log M_{\rm h}/{\rm M}_\odot = 10$ to $13$. For the cross-correlation, we show measurements for five UV magnitude cuts from $M_{\rm UV}<-13$ to $M_{\rm UV}<-21$. For each entry, we report the median value and the asymmetric errors corresponding to the 16th and 84th percentiles of the distribution across the 64 subvolumes. The errors thus reflect the subvolume-to-subvolume variance, providing a measure of the cosmic variance expected for a survey volume of $(100$ cMpc$)^3$.

\label{lastpage}

\end{document}